\documentclass[%
 reprint,
superscriptaddress,
 amsmath,amssymb,
 aps,
 prb,
 floatfix,
]{revtex4-2}

\usepackage{amsmath}
\usepackage{array}
\usepackage{dsfont}
\usepackage{mathtools}
\usepackage{graphicx}
\usepackage{dcolumn}
\usepackage{bm}
\usepackage{hyperref}
\usepackage{physics}
\usepackage{empheq}
\usepackage{color}
\usepackage{multirow}
\usepackage{afterpage}
\newcommand\inv[1]{#1\raisebox{1.15ex}{$\scriptscriptstyle-\!1$}}
\newcommand{\RE}[1]{\mathrm{Re}\left( #1 \right)}
\newcommand{\IM}[1]{\mathrm{Im}\left( #1 \right)}

\begin{document}

\preprint{APS/123-QED}

\title{Competing \texorpdfstring{$U(1)$}{U(1)} and \texorpdfstring{$\mathbb{Z}_2$}{Z2} dipolar-octupolar quantum spin liquids on the pyrochlore lattice: application to \texorpdfstring{Ce$_2$Zr$_2$O$_7$}{Ce2Zr2O7}}

\author{F\'elix Desrochers}
\email{felix.desrochers@mail.utoronto.ca}
\affiliation{%
 Department of Physics, University of Toronto, Toronto, Ontario M5S 1A7, Canada
}%
\author{Li Ern Chern}%
\affiliation{%
 T.C.M. Group, Cavendish Laboratory, JJ Thomson Avenue, Cambridge CB3 0HE, United Kingdom
}%
\author{Yong Baek Kim}%
\affiliation{%
 Department of Physics, University of Toronto, Toronto, Ontario M5S 1A7, Canada
}%
\affiliation{%
 School of Physics, Korea Institute for Advanced Study, Seoul 130-722, Korea
}%

\date{\today}

\begin{abstract}
Recent experiments on the dipolar-octupolar pyrochlore compound Ce$_2$Zr$_2$O$_7$ indicate that it may realize a three-dimensional quantum spin liquid (QSL). In particular, the analyses of available data suggest that the system is in a region of parameter space, where the so-called $\pi$-flux $U(1)$ octupolar quantum spin ice (QSI) with an emergent photon could be realized. On the other hand, because the system is far from the perturbative classical spin ice regime, it is unclear whether the quantum ground state is primarily a coherent superposition of 2-in-2-out configurations as in the canonical QSI phases. In this work, we explore other possible competing quantum spin liquid states beyond QSI using the Schwinger boson parton construction of the dipolar-octupolar pseudospin-1/2 model. After classifying all symmetric $U(1)$ and $\mathbb{Z}_2$ QSLs using the projective symmetry group (PSG), we construct a mean-field phase diagram that possesses a number of important features observed in an earlier exact diagonalization study. In the experimentally relevant frustrated region of the phase diagram, we find two closely competing gapped $\mathbb{Z}_2$ QSLs with a narrow spinon dispersion. The equal-time and dynamical spin structure factors of these states show key features similar to what was reported in neutron scattering experiments. Hence, these QSLs are closely competing ground states in addition to the $\pi$-flux $U(1)$ QSI, and they should be taken into account on equal footing for the interpretation of experiments on Ce$_2$Zr$_2$O$_7$.
\end{abstract}

\maketitle

\section{\label{sec: Introduction} Introduction}

Quantum spin liquids (QSLs) are quantum paramagnetic ground states characterized by a long-range entanglement (LRE) \cite{wen2004quantum, knolle2019field, savary2016quantum, zhou2017quantum, balents2010spin}, namely they cannot be transformed into direct product states via any set of finite local unitary transformations \cite{chen2010local}, and therefore have a non-vanishing topological entanglement entropy \cite{kitaev2006topological}. The entanglement in the ground state leads to deconfined fractionalized excitations that couple to emergent gauge fields. However, the discovery of a material that hosts a stable QSL combined with a convincing theoretical understanding of both the microscopic origin and characteristics of its ground state is an objective that has yet to be achieved.

The dipolar-octupolar (DO) pyrochlore Ce$_2$Zr$_2$O$_7$ provides a unique opportunity in this regard. From a theoretical point of view, its low-energy behavior is described by the remarkably simple $XYZ$ model with pseudospin-1/2 degrees of freedom that have two components transforming as dipoles, and one as an octupole \cite{rau2019frustrated, huang2014quantum}. This description with only a few parameters enables a tractable analysis of its ground state. Experimentally, investigations have shown strong evidence for QSL physics \cite{gaudet2019quantum, gao2019experimental, sibille2020quantum, smith2021case, bhardwaj2021sleuthing}. It lacks any magnetic ordering down to the lowest temperature, and it was recently determined that the leading pseudospin exchange coupling is antiferromagnetic and associated with the octupolar pseudospin component. Both transverse (dipolar) couplings are also frustrated \cite{bhardwaj2021sleuthing, smith2021case}. That region of parameters space is theoretically suggested to host a $\pi$-flux $U(1)$ octupolar quantum spin ice ($\pi$-O-QSI) \cite{patri2020theory, benton2020ground, li2017symmetry, chen2017spectral, benton2018quantum, taillefumier2017competing}, where the ground state supports a static $\pi$-flux of the emergent gauge field in hexagonal loops of the pyrochlore lattice. That prediction is well established in the perturbative regime (small transverse couplings) of the Ising point where the usual mapping to a compact $U(1)$ lattice gauge theory holds \cite{hermele2004pyrochlore, ross2011quantum,savary2012coulombic, savary2021quantum, lee2012generic, gingras2014quantum}. However, according to recent data analyses, Ce$_2$Zr$_2$O$_7$ may reside far from such a perturbative regime, which makes theoretical analysis rather difficult as, for example, quantum Monte Carlo (QMC) has a sign problem for frustrated transverse couplings. Interestingly, the exact diagonalization (ED) of small system sizes does not find any phase transition \cite{patri2020theory, benton2018quantum, benton2020ground} between the Ising point and the $SU(2)$ symmetric antiferromagnetic Heisenberg point \cite{iqbal2019quantum, astrakhantsev2021broken, huang2016spin, tsunetsugu2001spin, canals1998pyrochlore, canals2000quantum}. Given that the quantum spin ice near the Ising point is described as a coherent superposition of dominant 2-in-2-out spin configurations, it is not obvious how the ground state far away from the Ising limit may be connected to such a perturbative Ising regime. Hence, it is natural to ask whether there exist QSLs beyond the canonical QSI such that spin correlations other than the 2-in-2-out configurations participate significantly, which may offer competing or alternative QSLs. Another motivation to consider competing QSLs is that a recent quantum Monte-Carlo study of the $XYZ$ model with unfrustrated transverse couplings reported the presence of a $\mathbb{Z}_2$ QSL \cite{huang2020extended} in a region of the phase diagram, which was not discovered in the earlier gauge mean-field theory (gMFT) approaches \cite{lee2012generic, huang2014quantum, li2017symmetry, chen2017spectral}. For frustrated transverse couplings, the possibility of a $\mathbb{Z}_2$ QSL has not been explored yet.

In this work, we investigate competing $U(1)$ and $\mathbb{Z}_2$ QSLs for the dipolar-octupolar pyrochlore systems using the Schwinger boson parton construction and the corresponding mean-field theory. We first classify all symmetric $U(1)$ and $\mathbb{Z}_2$ dipolar-octupolar QSLs with bosonic spinons on the pyrochlore lattice using the projective symmetry group (PSG) framework (by symmetric, we refer to QSLs that do not break any space group or time-reversal symmetry, in opposition to chiral QSLs \cite{messio2013time, schneider2021projective}). On top of the 16 symmetric $\mathbb{Z}_2$ bosonic QSLs that Liu, Halàsz, and Balents previously classified \cite{liu2019competing}, we find 4 $U(1)$ symmetric QSLs that are parents of 8 of the $\mathbb{Z}_2$ QSLs. We obtain a mean-field phase diagram for all $\mathbb{Z}_2$ spin liquids where translational symmetries are realized trivially. Our approach captures many aspects of the established theoretical and experimental results. The phase diagram shows that quantum fluctuations are much stronger in the frustrated region of the phase diagram. Hence, it is more likely to stabilize QSLs there than in the unfrustrated region. The resulting phase diagram is remarkably similar to the result of recent ED calculations \cite{patri2020theory}. For example, deviations in phase boundaries (due to quantum fluctuations) from the classical phase diagram are clearly seen, and all the magnetically ordered phases are correctly represented. 

Most importantly, we identify two competing $\mathbb{Z}_2$ QSLs in the frustrated parameter regime that is relevant to Ce$_2$Zr$_2$O$_7$. We compute the equal-time and dynamical spin structure factors for these phases and compare them with experimental results. Interestingly, these gapped $\mathbb{Z}_2$ QSLs show emergent $SU(2)$ symmetry, which signals active participation of spin configurations beyond the 2-in-2-out spin-ice correlations. We obtain the rod-shaped equal-time structure factor as observed in the experiment, which is similar to what one would obtain if all the classical spin configurations, except for the all-in-all-out, are equally summed over a single tetrahedron. This also suggests that the spin correlation length in the gapped $\mathbb{Z}_2$ QSLs is quite short (different tetrahedra are not much correlated) in contrast to the algebraic correlation in canonical QSI. Moreover, we show that the intensity distributions in the equal-time structure factors of competing $\mathbb{Z}_2$ QSLs in the frustrated regime are opposite to those of possible QSLs in the unfrustrated regime, supporting the idea that Ce$_2$Zr$_2$O$_7$ resides in the frustrated regime of the phase diagram. 
Remarkably, the dynamical spin structure factor of one of the two competing $\mathbb{Z}_2$ states in the frustrated regime has high intensities at the X and K points, which is similar to what is observed in neutron scattering experiments \cite{gao2019experimental}. The correspondence between theoretical results for the $\mathbb{Z}_2$ QSL and experiments suggests that it is one of the closely competing QSL states (in addition to the $\pi$-flux QSI) in the frustrated regime of the dipolar-octupolar pyrochlore systems that likely share similar physical properties. 

The rest of the paper is organized as follows: in Sec. \ref{sec: Model} the space group of the pyrochlore lattice and the microscopic model used are discussed in detail before classifying all PSG classes in Sec. \ref{sec: Projective symmetry group}. In Sec. \ref{sec: Analysis of the mean-field ansatze}, the Hamiltonians for each class are built and diagonalized to evaluate the MF phase diagram, the equal-time static spin structure factor, and the dynamical spin structure factor. We finally discuss implications of our work and future directions in Sec. \ref{sec: Discussion}.

\section{\label{sec: Model} Model}

\subsection{\label{subsec: Space group} Pyrochlore lattice - Conventions and space group}

\begin{figure}
\includegraphics[width=0.99\linewidth]{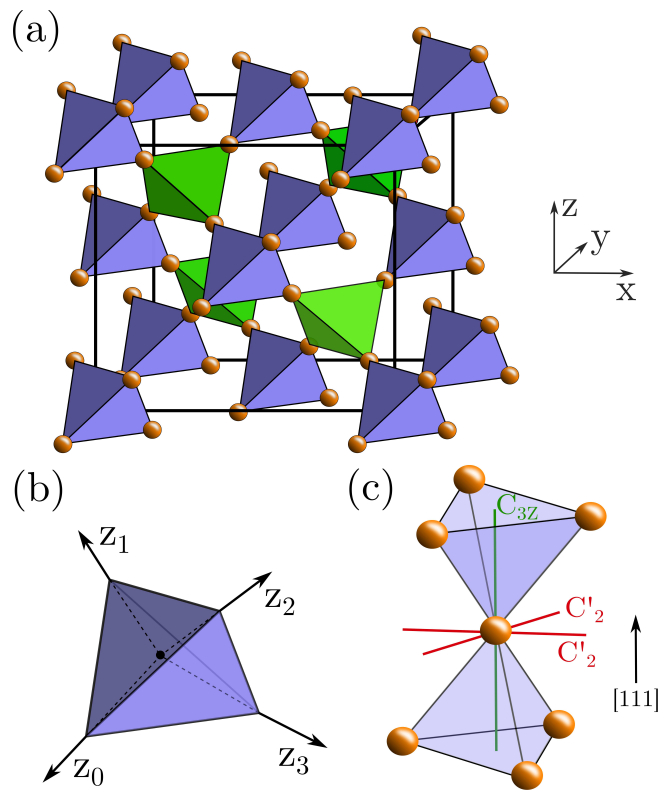}
\caption{ (a) The sites of the pyrochlore lattice form a three-dimensional network of corner-sharing tetrahedra. The up (down) tetrahedra are colored in purple (green). The underlying Bravais lattice is the face-centered cubic (fcc) lattice with four sites (sublattices) per unit cell. (b) Labeling and local $z$ axis for the four sublattices. (c)  The local environment of a rare-earth ion where the $C_{3z}$ and two out of the three $C_{2}'$ rotation axis of the $D_{3d}$ group are shown. \label{fig:pyrochlore lattice}}
\end{figure}

As shown in fig. \ref{fig:pyrochlore lattice} (a), the pyrochlore lattice consists of four FCC sublattices forming a network of corner-sharing tetrahedra. We label these sublattices with the index $\mu=0,1,2,3$ (see fig. \ref{fig:pyrochlore lattice} (b)). In the rest of the article, we use three different coordinate systems. The first two identify the sites on the pyrochlore lattice. The \emph{global cartesian coordinates} (GCC) are the standard frame coordinates of the FCC cube with edge length set to unity. The \emph{sublattice-indexed  pyrochlore coordinates} (SIPC) explicitly identify the unit cell and the sublattice of every site. The position of the unit cell is determined using three basis vectors (as expressed in the GCC)
    \begin{subequations} \label{eq: basis for SIPC expressed in GCC}
        \begin{align}
            &\hat{e}_1 = \frac{1}{2}\left( 0,1,1 \right)\\
            &\hat{e}_2 = \frac{1}{2}\left( 1,0,1 \right)\\
            &\hat{e}_3 = \frac{1}{2}\left( 1,1,0 \right).
        \end{align}
    \end{subequations}
The position of  the site within the unit cell is expressed by defining $\hat{\epsilon}_i = \frac{1}{2}\hat{e}_i$ ($i=1,2,3$) to be the displacement of the $\mu=1,2,3$ sublattices from the $\mu=0$ sublattice respectively (where $\hat{\epsilon}_0=\hat{e}_0=0$). The two coordinate systems for the position of the magnetic ions are related by
\begin{align*}
\left( r_1, r_2, r_3 \right)_{\mu} &= \vec{r}_{\mu} = r_1 \hat{e}_1 + r_2 \hat{e}_2 + r_3 \hat{e}_3 + \hat{\epsilon}_{\mu} \hspace{10mm} \text{(SIPC)} \nonumber\\
&= \frac{1}{2} \left( r_2 + r_3, r_3 + r_1, r_1 + r_2 \right) + \frac{1}{2} \hat{e}_{\mu}. \hspace{1.8mm} \text{(GCC)}  \nonumber
\end{align*}

Finally, the spins on each site $\vec{r}_{\mu}$ are written in a sublattice dependant \emph{local frame} (LF). The local basis on each sublattice $\mu$ is defined in appendix \ref{appendix subsec: Local coordinates and generic model}. As indicated in fig. \ref{fig:pyrochlore lattice} (b) each local quantization axis $\hat{z}_\mu$ is pointing out of the local tetrahedron from its center.

The magnetic ions are located at the 16d Wyckoff position of the cubic space group $F d \overline{3} m$ (No. 227) \cite{curnoe2018exchange}. This space group (SG) is minimally generated by five operators:
\begin{itemize}
    \item \emph{$T_{i}$:} Translation by $\hat{e}_{i}$  ($i=1,2,3$)
    \item \emph{$\overline{C}_6$:} A sixfold rotoreflection around the $\left[111\right]$ axis  ($\hat{e}_{1}+\hat{e}_{2}+\hat{e}_{3}$ ). It can also be written as $\overline{C}_6 = I C_3$, where $C_3$ is a threefold rotation around the [111] axis and $I$ is the inversion.
    \item \emph{$S$:} A nonsymmorphic screw operation. It is the composition of a twofold rotation around $\hat{e}_{3}$ and a fractional translation by $\hat{\epsilon}_{3}$. 
\end{itemize}
These space-group generators transform positions written in the SIPCs according to
\begin{subequations}
\begin{align}
T_{i}:& \vec{r}_{\mu} \to \left(r_{1}+\delta_{i, 1}, r_{2}+\delta_{i, 2}, r_{3}+\delta_{i, 3}\right)_{\mu} \\
\overline{C}_{6}:& \vec{r}_{\mu} \to \left(-r_{3}-\delta_{\mu, 3},-r_{1}-\delta_{\mu, 1},-r_{2}-\delta_{\mu, 2}\right)_{\pi_{123}(\mu)}\\
S:& \vec{r}_{\mu} \rightarrow \nonumber  \\
&\left(-r_{1}-\delta_{\mu, 1},-r_{2}-\delta_{\mu, 2}, r_{1}+r_{2}+r_{3}+1-\delta_{\mu, 0}\right)_{\pi_{03}(\mu)},
\end{align}
\end{subequations}
where $\pi_{123}(\mu)$ and $\pi_{03}(\mu)$ are cyclic permutations of the sublattices 1,2,3 and 0,3 respectively. The point group of this lattice is the cubic group $O_h$.

\subsection{\label{subsec: Microscopic model} Microscopic model}

In rare-earth pyrochlores, every magnetic ion sits at the shared vertex of two nearby tetrahedra. It is thus subjected to a crystalline electric field (CEF) that is symmetric under every transformation of the site symmetry group $D_{3d}$ (see fig. \ref{fig:pyrochlore lattice} (c)). This CEF splits the degenerate spin-orbit coupled $J$ manifold resulting in low-lying doublets. For magnetic ions with an odd number of electrons, Kramers' theorem applies, and all states must be at least two-fold degenerate by time-reversal symmetry (TRS). For a local $D_{3h}$ environment, there are two possible Kramers doublets \cite{rau2019frustrated}: (i) the first commonly referred to as the \emph{effective spin-1/2} where the doublet transforms just like a spinor, and (ii) the so-called \emph{dipolar-octupolar} (DO) doublet that is built from two one-dimensional irreducible representations that only mix under time reversal but not under space group transformations. The single-ion ground state of Ce$_2$Zr$_2$O$_7$ is a dipolar-octupolar doublet. It is found experimentally to be well separated from other CEF states and properly described as a mix of $\ket{J=5/2, m_{J}=\pm 3/2}$ states \cite{gaudet2019quantum, sibille2020quantum}, where the $J_z$ components are defined with respect to the $\hat{z}_\mu$ axis in the local frame. This dipolar-octupolar doublet accordingly determines the low-energy description of the system. It is convenient to represent these single ion degrees of freedom by pseudospins 1/2 operators $\mathbf{S}_{\vec{r}_\mu}$ where two components ($\mathbf{S}^{x}_{\vec{r}_{\mu}}$ and $\mathbf{S}^{z}_{\vec{r}_{\mu}}$) transform as magnetic dipole, and one of them ($\mathbf{S}^{y}_{\vec{r}_{\mu}}$) as a magnetic octupole (see appendix \ref{appendix: Microscopic model} for a detailed discussion). Under an arbitrary space group generator $\mathcal{O}$ and time-reversal $\mathcal{T}$, the pseudospins transform as
\begin{subequations} \label{eq: generic transformaiton of the pseudospin}
  \begin{align}
    &\mathcal{O}: \mathbf{S}_{\vec{r}_{\mu}}^{a} \rightarrow U_{\mathcal{O}} \mathbf{S}_{\mathcal{O}\left(\vec{r}_{\mu}\right)}^{a} U_{\mathcal{O}}^{\dagger}\\
    &\mathcal{T}: \mathbf{S}_{\vec{r}_{\mu}}^{a} \rightarrow \mathcal{K} U_{\mathcal{T}} \mathbf{S}_{\vec{r}_{\mu}}^{a} U_{\mathcal{T}}^{\dagger} \mathcal{K},
  \end{align}
\end{subequations}
where $a\in\left\{x,y,z\right\}$, $\mathcal{K}$ is the complex conjugation operator, and (see appendix \ref{appendix: pseudospin transformation})
\begin{align} \label{eq: matrix form pseudospin trfs}
&U_{T_{i}}=\mathds{1}_{2\times 2}, \hspace{1.0mm}
U_{\mathcal{T}}=i\sigma^{y}, \hspace{1.0mm}  U_{\overline{C}_{6}}=\mathds{1}_{2\times 2}, \hspace{1.0mm} U_{S}=-i\sigma^{y}.
\end{align} 

The most general symmetry-allowed pseudospin Hamiltonian with only nearest-neighbor (NN) coupling can be deduced from these transformations and is given by
\begin{align}  \label{eq: Hamiltonian nn pyrochlore}
    H=\sum_{\langle \vec{r}_{\mu}  \vec{r}_{\nu}' \rangle}&\left[J_{x x} \mathbf{S}_{\vec{r}_{\mu}}^{x} \mathbf{S}_{\vec{r}_{\nu}'}^{x}+J_{y y} \mathbf{S}_{\vec{r}_{\mu}}^{y} \mathbf{S}_{\vec{r}_{\nu}'}^{y}+J_{z z} \mathbf{S}_{\vec{r}_{\mu}}^{z} \mathbf{S}_{\vec{r}_{\nu}'}^{z}\right. \nonumber \\
    \hspace{2mm} &\left.+J_{x z}\left(\mathbf{S}_{\vec{r}_{\mu}}^{x} \mathbf{S}_{\vec{r}_{\nu}'}^{z}+\mathbf{S}_{\vec{r}_{\mu}}^{z} \mathbf{S}_{\vec{r}_{\nu}'}^{x}\right)\right].
\end{align}
The $J_{xz}$ coupling can be eliminated by a uniform rotation along the local $\hat{y}_{\mu}$ axis (i.e. $\tau^{y}=\mathcal{S}^{y}$, $\tau^{x}=\cos\theta\mathcal{S}^{x}-\sin\theta\mathcal{S}^{z}$ and $\tau^{z}=\sin\theta\mathcal{S}^{x}+\cos\theta\mathcal{S}^{z}$, where $\theta$ is defined in appendix \ref{appendix subsec: Local coordinates and generic model}). This yields the simple $XYZ$ pseudospin model
\begin{align} \label{eq: XYZ model}
H = \sum_{a\in\{ x,y,z\}}  \sum_{\langle \vec{r}_{\mu}  \vec{r}_{\nu}' \rangle} \mathcal{J}_{a} \tau_{\vec{r}_{\mu} }^{a} \tau_{\vec{r}_{\nu}'}^{a}.
\end{align}
The values of these coupling constants were estimated  for Ce$_2$Zr$_2$O$_7$ in two recent studies  from a combination of thermodynamic, magnetization and neutron scattering measurements \cite{bhardwaj2021sleuthing, smith2021case}. Both found $J_{xz}\approx0$ (i.e., $J_{aa}\approx\mathcal{J}^{a}$ and $S^{a}\approx\tau^{a}$ for $a=x,y,z$), all couplings to be antiferromagnetic (i.e., $J_{xx}>0$, $J_{yy}>0$, and $J_{zz}>0$), and the octupolar coupling $J_{yy}$ to be the largest or slightly smaller than $J_{xx}$.

\section{\label{sec: Projective symmetry group} Parton Mean-Field Theory and classification of spin liquids}

We now discuss possible dipolar-octupolar symmetric QSLs on the pyrochlore lattice. To do so, we rewrite Eq. \eqref{eq: XYZ model} as a sum of bilinear Schwinger bosons terms using standard mean-field (MF) approximation and then classify all possible $\mathbb{Z}_2$ and $U(1)$ MF solutions using a PSG analysis.

\subsection{\label{subsec: Schwinger bosons and spinon decouping} Schwinger bosons mean-field theory}

To explore possible quantum spin liquid phases, we represent the pseudospin operators in terms of Schwinger bosons \cite{wang2006spin}  
\begin{equation}\label{eq: Schwinger boson rep. for spins}
    \tau^{a}_{\vec{r}_\mu} = \frac{1}{2}\sum_{\alpha,\beta} b_{\vec{r}_\mu,\alpha}^\dag \left[\sigma^{a}\right]_{\alpha\beta} b_{\vec{r}_\mu,\beta},
\end{equation}
where $\alpha,\beta\in\left\{\uparrow,\downarrow\right\}$ and the $b_{\vec{r}_\mu,\alpha}$ operators satify usual bosonic commutation relations. The Schwinger bosons  physically represent deconfined spinons (fractionalized excitations) in the QSL \cite{savary2016quantum}. One should note that such a representation enlarges the initial Hilbert space as the number of Schwinger boson per site is unconstrained. Therefore, by imposing the single-occupancy restriction
\begin{equation} \label{eq: single-occupancy constraint}
    \sum_{\alpha} b_{\vec{r}_\mu,\alpha}^\dag  b_{\vec{r}_\mu,\alpha}= n_{\vec{r}_\mu} = \kappa
\end{equation}
with $\kappa=1$, we recover a faithful representation of the initial pseudospin $1/2$ Hilbert space. In the following, we shall consider $\kappa$ to be a continuous positive parameter \cite{wang2006spin, messio2010schwinger}. At the MF level, this hard local constraint is relaxed by requiring that it is only respected on average over the lattice \cite{wang2006spin}
\begin{equation} \label{eq: relaxed single-occupancy constraint}
    \sum_{\alpha} \expval{b_{\vec{r}_\mu,\alpha}^\dag  b_{\vec{r}_\mu,\alpha}} = \kappa.
\end{equation}

Making the substitution \eqref{eq: Schwinger boson rep. for spins} in Eq. \eqref{eq: Hamiltonian nn pyrochlore} leads to a quartic Hamiltonian in bosonic operators. Applying standard MF decoupling procedure \cite{chern2018magnetic}, every quartic term in the Hamiltonian is first rewritten in the form
\begin{equation} \label{eq: general form Hamiltonian pre-MF decoupling}
H=- \sum_{\langle \vec{r}_\mu \vec{r}_\nu' \rangle} \sum_{A} \left|c^A\right| \hat{A}_{\vec{r}_\mu,\vec{r}_\nu'}^{\dagger} \hat{A}_{\vec{r}_\mu,\vec{r}_\nu'},
\end{equation}
where $\hat{A}_{\vec{r}_\mu,\vec{r}_\nu'}$ represents the auxiliary bond operators \cite{shindou2009s}
\begin{subequations} \label{eq: definitions bond operators}
  \begin{align}
    \hat{\chi}_{\vec{r}_\mu,\vec{r}_\nu'} &=\sum_\alpha b_{\vec{r}_\mu,\alpha}^\dag b_{\vec{r}_\nu',\alpha}, \label{eq: definitions bond operators -> Singlet hopping}\\
    \hat{\Delta}_{\vec{r}_\mu,\vec{r}_\nu'} &= \sum_{\alpha,\beta} b_{\vec{r}_\mu,\alpha} \left[ i\sigma^{y} \right]_{\alpha\beta} b_{\vec{r}_\nu',\beta}, \label{eq: definitions bond operators -> Singlet pairing}\\
    \hat{E}_{\vec{r}_\mu,\vec{r}_\nu'}^a &=\sum_{\alpha,\beta} b_{\vec{r}_\mu,\alpha}^\dag \left[ \sigma^a \right]_{\alpha\beta} b_{\vec{r}_\nu',\beta}, \label{eq: definitions bond operators -> Triplet hopping}\\
    \hat{D}_{\vec{r}_\mu,\vec{r}_\nu'}^a &= \sum_{\alpha,\beta} b_{\vec{r}_\mu,\alpha} \left[i \sigma^y \sigma^a \right]_{\alpha\beta} b_{\vec{r}_\nu',\beta} \label{eq: definitions bond operators -> Triplet pairing}
  \end{align}
\end{subequations}
that correspond to singlet hopping, singlet pairing, triplet hopping, and triplet pairing on sites $\vec{r}_\mu$ and $\vec{r}_\nu'$ respectively. From the above rewriting, a MF decoupling naturally leads to
\begin{align} \label{eq: general form Hamiltonian post-MF decoupling}
H_{MF}=&- \sum_{\langle \vec{r}_\mu \vec{r}_\nu' \rangle}\sum_{A} \left|c^A\right| \left(A_{\vec{r}_\mu,\vec{r}_\nu'}^{*} \hat{A}_{\vec{r}_\mu,\vec{r}_\nu'}+A_{\vec{r}_\mu,\vec{r}_\nu'} \hat{A}_{\vec{r}_\mu,\vec{r}_\nu'}^{\dagger}\right.\nonumber \\
&\left.\hspace{2.8cm}-\left|A_{\vec{r}_\mu,\vec{r}_\nu'}\right|^{2}\right).
\end{align}
Using this form, the MF energy is bounded from below (i.e., the stability requirement is satisfied). It is important to note that $\hat{A}_{\vec{r}_\mu,\vec{r}_\nu'}$ are the bond operators, whereas $A_{\vec{r}_\mu,\vec{r}_\nu'}$ (without the hat) are variational parameters to minimize the MF energy. This set of MF parameters $\left\{ A \right\}\equiv\left\{ \chi,\Delta,E^x,E^y,E^z,D^x,D^y,D^z \right\}$ is referred to as an ansätz.

In our case, a possible choice of rewriting for the different contributions in Eq. \eqref{eq: Hamiltonian nn pyrochlore} in terms of bond operators is (see appendix \ref{appendix subsec: mean-field decoupling})
\begin{widetext}
\begin{equation} \label{eq: MF decoupling for all terms in H pyrochlore nn}
    \mathcal{J}_{a} \tau_{i}^{a} \tau_{j}^{a}= \left\{\begin{array}{l}
-\frac{|\mathcal{J}_a|}{4}\left( 
    \hat{\chi}_{i,j}^\dag \hat{\chi}_{i,j} + \hat{\Delta}_{i,j}^\dag \hat{\Delta}_{i,j} + \hat{E}^{b \dag}_{i,j} \hat{E}^b_{i,j} + \hat{E}^{c \dag}_{i,j} \hat{E}^c_{i,j} + 2\hat{D}^{b \dag}_{i,j} \hat{D}^b_{i,j} + 2\hat{D}^{c \dag}_{i,j} \hat{D}^c_{i,j}
    - 3 n_{j} - 4 n_{i} n_{j}
    \right), \text { for } \mathcal{J}_a<0 \\
-\frac{|\mathcal{J}_{a}|}{4}\left( 
     \hat{\chi}_{i,j}^\dag \hat{\chi}_{i,j} + \hat{\Delta}_{i,j}^\dag \hat{\Delta}_{i,j} + \hat{E}^{a \dag}_{i,j} \hat{E}^a_{i,j} + \hat{E}^{b \dag}_{i,j} \hat{E}^b_{i,j} + \hat{E}^{c \dag}_{i,j} \hat{E}^c_{i,j} + \hat{D}^{a \dag}_{i,j} \hat{D}^a_{i,j}
    - 4 n_{j} - 3 n_{i} n_{j}
    \right), \text { for } \mathcal{J}_a>0
\end{array}\right.
\end{equation}
\end{widetext}
where $a\in\{x,y,z\}$, and $b$ and $c$ are the other two elements in the set. Assuming that \eqref{eq: single-occupancy constraint} is strictly respected, $n_{\vec{r}_\nu'}$ and $n_{\vec{r}_\mu} n_{\vec{r}_\nu'}$ are constants that we will simply drop.

After the decoupling, the MF Hamiltonian \eqref{eq: general form Hamiltonian post-MF decoupling} is of the general form
\begin{align} \label{eq: MF Hamiltonian generic form}
    H_{MF} =& -\sum_{\langle \vec{r}_\mu, \vec{r}_\nu' \rangle}  \left( \vec{b}_{\vec{r}_{\mu}}^\dag \left[ u^h_{\vec{r}_\mu,\vec{r}_\nu'} \right] \vec{b}_{\vec{r}_{\nu}'} + \vec{b}_{\vec{r}_{\mu}}^T\left[ u^p_{\vec{r}_\mu,\vec{r}_\nu'} \right] \vec{b}_{\vec{r}_{\nu}'} +h.c.    \nonumber \right)\\
    &+ \lambda \sum_{\vec{r}_\mu} \left( n_{\vec{r}_\mu} - \kappa \right) + E_0\left(\left\{A\right\}\right),
\end{align}
 where $\vec{b}_{\vec{r}_{\mu}}=\begin{pmatrix}
b_{\vec{r}_\mu,\uparrow} \\ b_{\vec{r}_\mu,\downarrow}
\end{pmatrix}$, $E_0\left(\left\{A\right\}\right) = \sum_{A} |c^A| |A_{\vec{r}_\mu,\vec{r}_\nu'}|^2$ stands for all terms quadratic in the variationnal parameters that make the MF energy bounded from below (see the last term in Eq. \eqref{eq: general form Hamiltonian post-MF decoupling}),
$\lambda$ is a site-independent Lagrange multiplier introduced to enforce the average single occupancy constraint of Eq. \eqref{eq: relaxed single-occupancy constraint}, and $u^h_{\vec{r}_\mu,\vec{r}_\nu'}$ and $u^p_{\vec{r}_\mu,\vec{r}_\nu'}$ are $2\times 2$ matrices that act on the spin degrees of freedom. The labels $p$ an $h$ stand for pairing and hopping respectively. This Hamiltonian is used to generate a variationnal wave function $\ket{\Psi\left(\left\{ A_{\vec{r}_\mu,\vec{r}_\nu} \right\}, \lambda\right)}$ that depends on all variationnal parameters $A_{\vec{r}_\mu,\vec{r}_\nu}$ and on the Lagrange multiplier $\lambda$. After tuning $\lambda$ to respect,
\begin{align} \label{eq: self-consistency equations for average boson number}
    \expval{n_{\vec{r}_\mu}} = \kappa \Longleftrightarrow   \pdv{\expval{H_{MF}}}{\lambda}=0
\end{align}
the variational MF energy can then be minimized with respect to all variational parameters to yield the self-consistency equations
\begin{align} \label{eq: self-consistency equations for MF parameters}
    \pdv{\expval{H_{MF}}}{A_{\vec{r}_{\mu},\vec{r}_{\nu}'}}=0 \Longleftrightarrow A_{\vec{r}_\mu,\vec{r}_{\nu}'}= \expval{\hat{A}_{\vec{r}_\mu,\vec{r}_{\nu}'}}
\end{align}
for $A\in\{\chi,\Delta,E^x,E^y,E^z,D^x,D^y,D^z\}$.

\subsection{\label{subsec: PSG classification} PSG classification}

It is important to emphasize that the Schwinger boson parton representation introduces a $U(1)$ gauge redundancy as the pseudospin operators $\tau^a_{\vec{r}_\mu}$ are invariant under the transformation
\begin{equation} \label{eq: U(1) gauge transformation schwinger boson}
G: b_{\vec{r}_{\mu},\alpha} \rightarrow e^{i \phi\left(\vec{r}_{\mu}\right)} b_{\vec{r}_{\mu},\alpha}.
\end{equation}
As a consequence, the symmetry operations are only realized projectivey at the MF level. That is, the MF ansätz only need to be invariant under the gauge-enriched operations 
$\widetilde{\mathcal{O}}=G_{\mathcal{O}} \circ \mathcal{O}$,
$\widetilde{\mathcal{T}}=G_{\mathcal{T}} \circ \mathcal{T}$ that transform the Schwinger bosons as
\begin{subequations}
\begin{align}
    &\widetilde{\mathcal{O}}: b_{\vec{r}_{\mu},\alpha} \rightarrow e^{i \phi_{\mathcal{O}}\left[\mathcal{O}\left(\vec{r}_{\mu}\right)\right]} \sum_{\alpha} \left[U_{\mathcal{O}}^{\dagger}\right]_{\beta\alpha} b_{\mathcal{O}\left(\vec{r}_{\mu}\right),\alpha} \label{eq: spinon transfromations under SG} \\
    &\widetilde{\mathcal{T}}: b_{\vec{r}_{\mu},\alpha} \rightarrow e^{i \phi_{\mathcal{T}}\left(\vec{r}_{\mu}\right)} \mathcal{K} \sum_{\alpha} \left[U_{\mathcal{T}}^\dag\right]_{\beta\alpha} b_{\vec{r}_{\mu},\alpha}, \label{eq: spinon transfromations under TRS}
\end{align}
\end{subequations}
where $\circ$ denotes composition and $\comm{\mathcal{K}}{U_{\mathcal{T}}}=0$ because $U_{\mathcal{T}}$ is real. The symmetry group of the MF Hamiltonian generated by the gauge-enriched operations $\widetilde{\mathcal{O}}$ and $\widetilde{\mathcal{T}}$ is referred to as the projective symmetry group. As first pointed out by Wen in his seminal work for fermionic spinons on the square lattice \cite{wen2002quantum}, the requirement that symmetric QSLs do not break any space group or time-reversal symmetry constrains the number of physically allowed QSL (PSG classes). Different PSG classes correspond to separate realizations of the gauge-enriched symmetry operations that are not related by any gauge transformation. All these different classes can be enumerated through a group theory analysis. 

Pure gauge transformations associated with the identity operator $G_{\mathds{1}}$ that leave the MF Hamiltonian invariant form a normal subgroup of the PSG called the invariant gauge group (IGG) (i.e., SG=PSG/IGG). The IGG cannot be identified with symmetries but corresponds physically to the emergence gauge structure that interacts with the non-local excitations of the QSL. When both hopping and pairing terms are present in the MF Hamiltonian, the IGG is $\mathbb{Z}_2=\left\{e^{i \pi n} \mid n\in \{0,1\} \right\}$. Alternatively, if the Hamiltonian only contains hopping terms, the IGG is $U(1)=\left\{e^{i \psi} \mid 0 \le \psi < 2\pi \right\}$. If the IGG  is $\mathbb{Z}_2$ ($U(1)$), the resulting QSL is a $\mathbb{Z}_2$ ($U(1)$) spin liquid. 

From the above discussion, it should be noted that the PSG classification within the Schwinger boson formulation is the same in the dipolar-octupolar and the effective spin-1/2 (usual spinor) case. Even though the PSG classes are identical, new behaviors are still to be expected as the ansätz transform differently (because the $U_{\mathcal{O}}$ matrices associated with the pseudospin transformations are different) and will thus lead to different constraints and relations among the MF parameters.

\subsubsection{\label{subsubsec: Z2 QSL} \texorpdfstring{$\mathbb{Z}_2$}{Z2}  symmetric QSL}

The symmetric bosonic $\mathbb{Z}_2$ QSLs were already classified by Liu, Hal\'asz, and Balents \cite{liu2019competing}. They found 16 symmetric QSLs represented by
\begin{subequations} \label{eq: Z2 PSG classification}
\begin{align}
\phi_{T_{1}}\left(\vec{r}_{\mu}\right)=& 0 \\
\phi_{T_{2}}\left(\vec{r}_{\mu}\right)=& n_{1} \pi r_{1} \\
\phi_{T_{3}}\left(\vec{r}_{\mu}\right)=& n_{1} \pi\left(r_{1}+r_{2}\right) \\
\phi_{\mathcal{T}}\left(\vec{r}_{\mu}\right)=& 0 \\
\phi_{\bar{C}_{6}}\left(\vec{r}_{\mu}\right)=&\left[\frac{n_{\bar{C}_{6}}}{2}+\left(n_{1}+n_{S T_{1}}\right) \delta_{\mu=1,2,3}\right] \pi \nonumber \\
&+n_{1}  \pi r_{1} \delta_{\mu=2,3}+n_{1} \pi r_{3}  \delta_{\mu=2} \nonumber \\
&+n_{1} \pi r_{1} \left( r_{2} + r_{3}\right), \\
\phi_{S}\left(\vec{r}_{\mu}\right)=&\left[(-)^{\delta_{\mu=1,2,3}} \frac{n_{1}+n_{S T_{1}}}{2}+\delta_{\mu=2} n_{\bar{C}_{6} S}\right] \pi \nonumber \\
&+\left(n_{1} \delta_{\mu=1,2} + n_{S T_{1}}\right) \pi r_{1} \nonumber\\
&+\left(n_{1} \delta_{\mu=2} + n_{S T_{1}}\right) \pi r_{2}+n_{1} \pi r_{3} \delta_{\mu=1,2}  \nonumber \\
&+\frac{1}{2} n_{1} \pi\left(r_{1}+r_{2}\right)\left(r_{1}+r_{2}+1\right),
\end{align}
\end{subequations}
where $n_1$, $n_{\overline{C}_6 S}$, $n_{ST_1}$ and $n_{\overline{C}_6}$ are all either 0 or 1. We will label these equivalent classes by  $\mathbb{Z}_2 \text{-} n_1 \text{-} n_{\overline{C}_6 S} n_{ST_1} n_{\overline{C}_6}$. Physically, $n_1$, $n_{\overline{C}_6 S}$, $n_{ST_1}$ and $n_{\overline{C}_6}$ describe the phase acquired by a bosonic spinon after being subjected to the transformations $T_i T_{i+1} T_{i}^{-1}T_{i+1}^{-1}$, $\overline{C}_6^6$, $S T_1 S^{-1} T_3^{-1} T_1$ and $(\overline{C}_6 S)^4$ respectively \cite{liu2019competing, liu2021symmetric}. 

It is important to note that in the $n_{1}=1$ case, the unit cell is enlarged as translations by $\hat{e}_2$ and $\hat{e}_3$ act projectively. Furthermore, for $\mathbb{Z}_2$ PSG classes with $n_{ST_{1}}=1$, it is beneficial to shift the entire Brillouin zone by $(\pi,\pi,\pi)$ as the PSG phase factors for the $S$ operation depends on spatial coordinates. This shift can be viewed as a spinon gauge transformation \cite{liu2019competing} and is assumed in the rest of the paper.

\subsubsection{\label{subsubsec: U(1) QSL} \texorpdfstring{$U(1)$}{U(1)} symmetric QSL}

For IGG=$U(1)$, we give the detailed derivation of the PSG solution in appendix \ref{appendix: Classification of symmetric $U(1)$ spin liquids}. We find four $U(1)$ PSG classes that correspond to the phases
\begin{subequations} \label{eq: U(1) PSG classification}
\begin{align}
\phi_{T_{1}}\left(\vec{r}_{\mu}\right)=& 0 \\
\phi_{T_{2}}\left(\vec{r}_{\mu}\right)=& n_{1} \pi r_{1} \\
\phi_{T_{3}}\left(\vec{r}_{\mu}\right)=& n_{1} \pi\left(r_{1}+r_{2}\right) \\
\phi_{\mathcal{T}}\left(\vec{r}_{\mu}\right) =& 0 \\
\phi_{\bar{C}_{6}}\left(\vec{r}_{\mu}\right) =& n_1 \pi \delta_{\mu=1,2,3}  + n_1 \pi r_1 (r_2 + r_3) \nonumber \\
& + n_1 \pi r_1 \delta_{\mu=2,3} + n_1 \pi r_3 \delta_{\mu=2},\\
\phi_{S}\left(\vec{r}_{\mu}\right) =& \left((-)^{\delta_{\mu=1,2,3}} \frac{n_1 }{2} + n_{\overline{C}_6 S} \delta_{\mu=2}\right) \pi \nonumber \\
& + n_1 \pi r_1 \delta_{\mu=1,2} + n_1 \pi r_2 \delta_{\mu=2} + n_1 \pi r_3 \delta_{\mu=1,2} \nonumber \\
& +\frac{1}{2} n_{1} \pi\left(r_{1}+r_{2}\right)\left(r_{1}+r_{2}+1\right),
\end{align}
\end{subequations}
where $n_1$ and $n_{\overline{C}_6 S}$ are also $\mathbb{Z}_2$ parameters that can be either 0 or 1. We label these 4 symmetric QSLs by  $U(1) \text{-} n_1 \text{-} n_{\overline{C}_6 S}$.

A very suggestive gauge fixing was used to make the relation between $U(1)$ and $\mathbb{Z}_2$ symmetric QSL obvious. The $U(1)\text{-}n_1\text{-}n_{\overline{C}_6 S}$ PSG classes phase factors are directly equal to the ones of the $\mathbb{Z}_2 \text{-} n_1 \text{-} (n_{\overline{C}_6 S},0,0)$ classes. Furthermore, the $n_{\overline{C}_6}/2$ factor of the $\phi_{\overline{C}_6}$ phase factor in the $\mathbb{Z}_2$ classification can simply be eliminated by an IGG transformation in the $U(1)$ case (i.e., $\phi_{\overline{C}_6} \to \phi_{\overline{C}_6} - n_{\overline{C}_6}/2$). The $\mathbb{Z}_2 \text{-} n_1 \text{-} (n_{\overline{C}_6 S},0,n_{\overline{C}_6})$ classes are therefore also connected to $U(1)\text{-}n_1\text{-}n_{\overline{C}_6 S}$. In summary, we find the following PSG hierarchy
\begin{equation*}
     U(1)\text{-}n_{1}\text{-}n_{\overline{C}_6S} \longrightarrow \mathbb{Z}_2\text{-}n_{1}\text{-}n_{\overline{C}_6S} 0 n_{\overline{C}_6},
\end{equation*}
where the arrow corresponds to a charge 2 Higgs transition. The parent $U(1)$ state can be obtained by setting all pairing terms of the related $\mathbb{Z}_2$ QSL to zero.

\begin{figure*}[ht!]
\includegraphics[width=0.70\linewidth]{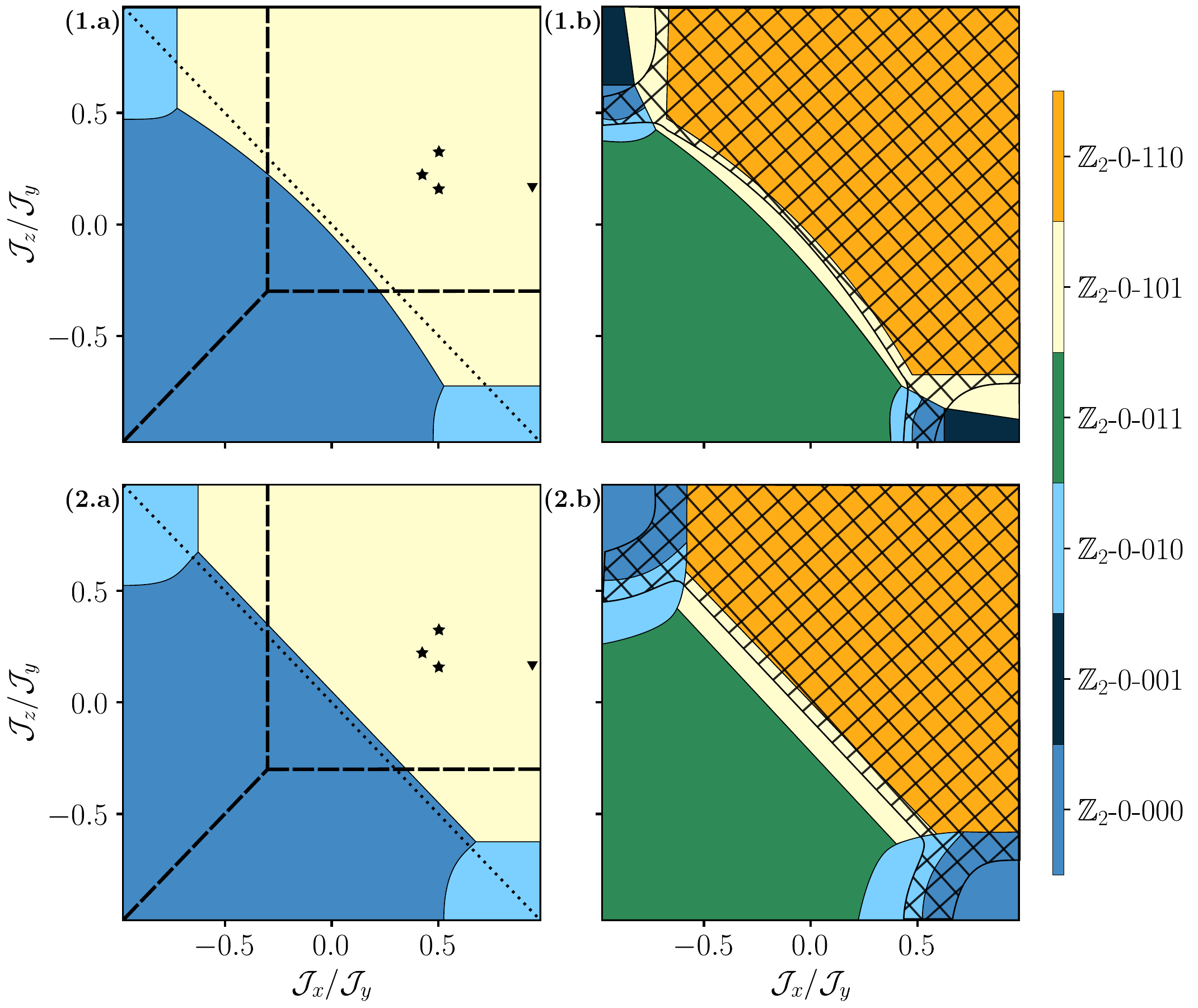}
\caption{\label{fig: phase diagram Z2 0-flux states} Phase diagram in the octupolar quadrant (i.e., $\mathcal{J}_y>0$ and $|\mathcal{J}_{y}|>|\mathcal{J}_{x}|,|\mathcal{J}_{z}|$) for the (a) lowest and (b) second lowest energy states (first and second column respectively) for values of (1) $\kappa=0.1$ and (2) $\kappa=0.14$ (first and second row respectively). The dashed lines denote classical phase boundaries while the $\mathcal{J}_{x} + \mathcal{J}_{z}=0$ dotted line indicates the boundary between the unfrustrated and frustrated regions of exchange couplings. Black stars and the downward triangle represent experimentally determined coupling constants from \cite{bhardwaj2021sleuthing} and \cite{smith2021case} respectively. The $\mathbb{Z}_2\text{-}0\text{-}000$ and $\mathbb{Z}_2\text{-}0\text{-}010$ are both condensed at the $\Gamma$ point (see fig. \ref{fig: dispersion relation 0-flux Z2 states}) and correspond to $X$ and $Z$ all-in-all-out order (see appendix \ref{subsubsec: 0-flux Z2 PSG classes -> condensation pattern}) for all values of the coupling constants and of $\kappa$ displayed. The hatched region corresponds to the area where the difference between the ground state and second-lowest energy state is less than 2.5 percent of the ground state energy (i.e., $|(E_{1}-E_{GS})/E_{GS}|<0.025$)}
\end{figure*}

\section{\label{sec: Analysis of the mean-field ansatze} Analysis of the mean-field ans\"{a}tze}

\subsection{\label{subsubsec: 0-flux Z2 PSG classes -> MF phase diagram} Mean-field phase diagram}

\begin{figure*}[ht!]
\includegraphics[width=1.00\linewidth]{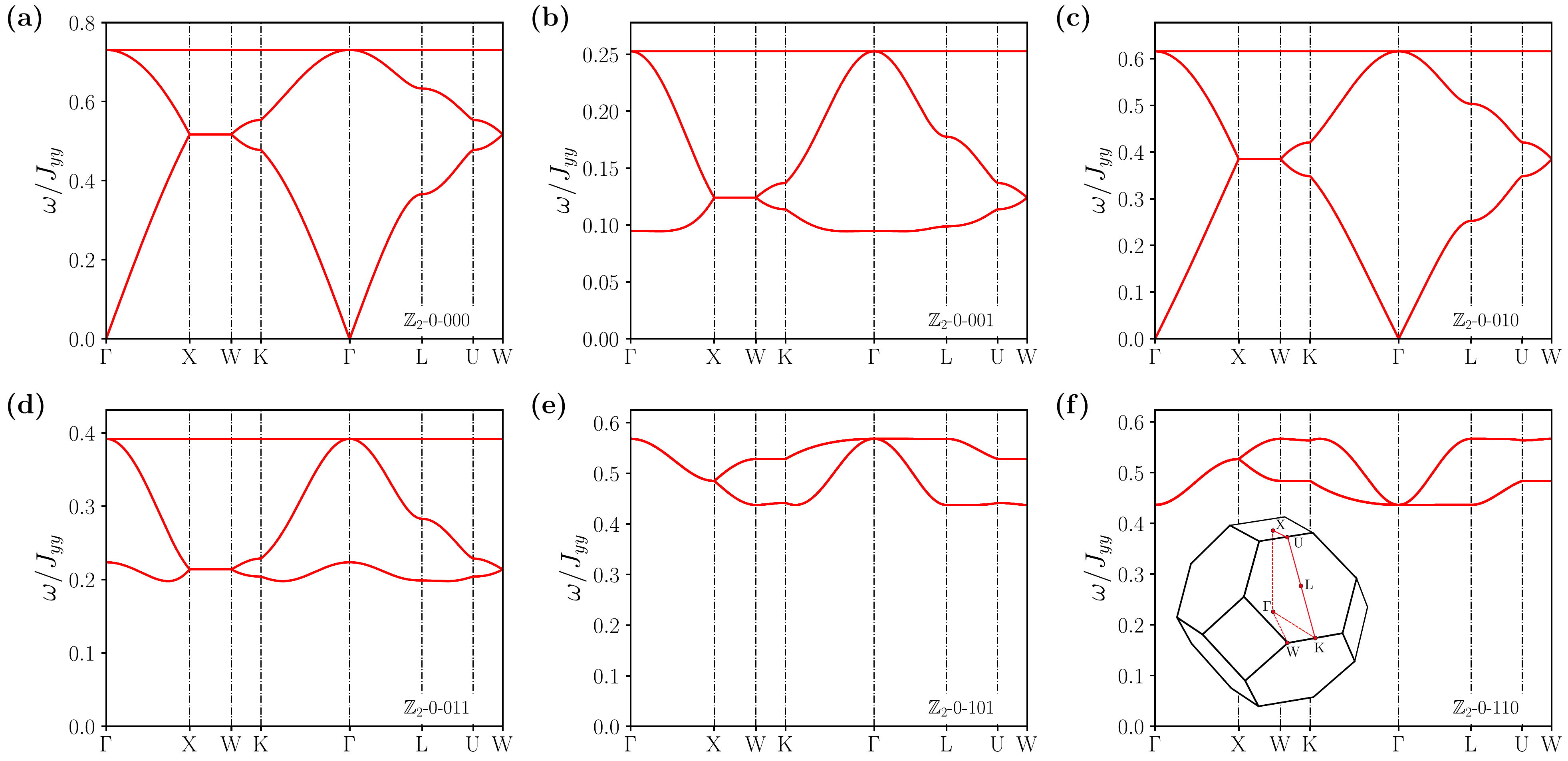}
\caption{\label{fig: dispersion relation 0-flux Z2 states} Spinon dispersion relation with $\kappa=0.14$, $\mathcal{J}_{x}/\mathcal{J}_{y}=0.50$,  $\mathcal{J}_{z}/\mathcal{J}_{y}=0.25$ and $\mathcal{J}_y>0$ along high symmetry lines for the PSG classes (a) $\mathbb{Z}_2 \text{-} 0 \text{-}000$, (b) $\mathbb{Z}_2 \text{-} 0 \text{-} 001$, (c) $\mathbb{Z}_2 \text{-} 0 \text{-}010$, (d) $\mathbb{Z}_2 \text{-} 0 \text{-} 011$, (e) $\mathbb{Z}_2 \text{-} 0 \text{-} 101$ and (f) $\mathbb{Z}_2 \text{-} 0 \text{-} 110$. The inset shows the first Brillouin zone for the face-centered cubic lattice and its high symmetry points. The flat bands are accidental degeneracies that would be lifted upon inclusion of next-nearest-neighbor couplings.}
\end{figure*}

Now that all symmetric $\mathbb{Z}_2$ and $U(1)$ QSLs with bosonic spinons have been classified, it is possible to build the corresponding MF Hamiltonian of each PSG class (see appendix \ref{appendix: Diagonalization of the mean-field Hamiltonian}), and subsequently diagonalize it using Bogoliubov transformations (see appendix \ref{appendix: From PSG to mean-field Hamiltonian}). The MF energy can then be minimized by finding the ansätz that solves the self-consistency equations. We numerically solve the self-consistency equations for all $\mathbb{Z}_2$ symmetric QSLs with $n_1=0$ that do not have an emergent $U(1)$ symmetry at the nearest-neighbor level (i.e., at least one symmetry allowed pairing term is present, see table \ref{tab: independant non-zero MF parameters}) and compare their energy to obtain a MF phase diagram. The lowest and second-lowest energy states for different values of $\kappa$ in the experimentally relevant octupolar quadrant (i.e., $\mathcal{J}_y>0$ and $|\mathcal{J}_{y}|>|\mathcal{J}_{x}|,|\mathcal{J}_{z}|$) are presented in fig. \ref{fig: phase diagram Z2 0-flux states}. Information about the excitations of these states can be retrieved from the spinon dispersion relations. These are presented for coupling constants close to the experimentally determined ones in fig. \ref{fig: dispersion relation 0-flux Z2 states}.

The $U(1)$ PSG classes are not considered separately as they are all related to two $\mathbb{Z}_2$ PSG classes by setting the pairing terms to zero and the self-consistency equations are solved for the parameters that minimize the MF energy. At the MF level, the energy of the $U(1)$ QSL is always greater or equal to their descendant $\mathbb{Z}_2$ QSLs. The only way for a $U(1)$ PSG class to appear in the MF phase diagram would be if all pairing terms of one of its $\mathbb{Z}_2$ descendant state vanish. However, it is found that all of the studied $\mathbb{Z}_2$ PSG classes have at least one non-vanishing pairing term after solving the self-consistency equations for all coupling constants in the octupolar quadrant.

It is well-known from previous studies using the Schwinger boson MF formalism that deconfined phases are usually only present for values of $\kappa$ much smaller than $\kappa=1$ \cite{wang2006spin, wang2010schwinger, yang2016schwinger, sachdev1992kagome}. This is because the MF theory does not properly capture the full quantum fluctuations \cite{messio2010schwinger, auerbach1988spin, arovas1988functional}. Thus, a smaller value of $\kappa$ may represent more physical situations. As there are no clear ways to estimate the value of $\kappa$ that should reproduce physical results at the MF level, we examine the MF phase diagram for different values of $\kappa$ and check the consistency with other numerical computations. For example, a phase diagram that closely resembles ED results is obtained for values around $\kappa=0.14$ (see the second row of fig. \ref{fig: phase diagram Z2 0-flux states}).

The part of the phase diagram where classical ordering is expected (to the left and below the dashed lines in fig. \ref{fig: phase diagram Z2 0-flux states} (1.a) and (2.a)) is occupied by the $\mathbb{Z}_{2}\text{-}0\text{-}000$ and $\mathbb{Z}_{2}\text{-}0\text{-}010$ PSG classes. These two QSLs are gapless at the $\Gamma$ point (see fig. \ref{fig: dispersion relation 0-flux Z2 states} (a) and (c)). This leads to a condensation of the Schwinger boson that corresponds to magnetic ordering with ordering wavevector $\vec{k}_c=\vec{0}$. By considering the eigenspace at the critical momentum, it can be further showed that the spins are in either a $X$ or $Z$ all-in-all-out ordered phase as is expected classically for $|\mathcal{J}_x|>|\mathcal{J}_z|$ and $|\mathcal{J}_x|<|\mathcal{J}_z|$ respectively (see appendix \ref{subsubsec: 0-flux Z2 PSG classes -> condensation pattern} for details).

An octupolar ($\tau^{y}$) two-in-two-out state is present classically in the small triangular region illustrated in fig. \ref{fig: phase diagram Z2 0-flux states} close to the classical octupolar spin-ice point ($\mathcal{J}_{x}=\mathcal{J}_{z}=0$ and $\mathcal{J}_{y}>0$). In gauge mean-field theory and ED calculations, this region is replaced by a 0-flux $U(1)$ octupolar quantum spin ice (0-O-QSI) \cite{huang2014quantum, patri2020theory}. Our approach does not capture such a phase; instead, a classically ordered state occupies that region of parameter space. This is most likely due to the fact that 0-O-QSI requires a different parton construction (gMFT) and is not faithfully represented by the Schwinger boson representation.

Most interestingly, in the frustrated region  (i.e., $\mathcal{J}_{x}+\mathcal{J}_{y}>0$), the gaped $\mathbb{Z}_2\text{-}0\text{-}101$ QSL is stabilized. The region of parameter space it occupies expands beyond the classical phase boundaries for large values of $\mathcal{J}_x/\mathcal{J}_y$ and $\mathcal{J}_z/\mathcal{J}_y$. Previous ED simulations also show this non-trivial behavior of the phase boundaries \cite{patri2020theory}. When the value of $\kappa$ is decreased, this PSG class expands further in the phase diagram, as can be inferred by comparing the first and second row in fig. \ref{fig: phase diagram Z2 0-flux states}. This state has an extremely narrow spinon dispersion and a large energy gap (see fig. \ref{fig: dispersion relation 0-flux Z2 states}.(e)) which makes it stable against magnetic ordering. We indeed find that the state remains gaped even for large values of $\kappa$ greater than unity. It is important to note that there is a strong competition with the other gapped $\mathbb{Z}_2\text{-}0\text{-}110$ state for that region of parameter space. To depict this competition, we plot the second lowest energy states for both values of $\kappa$ in the second column of fig. \ref{fig: phase diagram Z2 0-flux states}. The relative energy difference between the ground state and second-lowest state $|(E_{GS}-E_{1})/E_{GS}|$ is lower than 0.12 for the whole phase diagram and both values of $\kappa$ presented in fig. \ref{fig: phase diagram Z2 0-flux states}. However, as is depicted by the hatched region, this energetic competition is strongest at the phase boundaries and for the frustrated region. This means that the relative energy of these two states can easily be changed if the fluctuations beyond MF theory are taken into account. Therefore, these two states should be considered on an equal footing as possible candidates for a stable bosonic $\mathbb{Z}_2$ bosonic QSL in that region of parameter space. For both $\mathbb{Z}_2\text{-}0\text{-}101$ and $\mathbb{Z}_2\text{-}0\text{-}110$, all hopping terms become vanishingly small upon solving the self-consistency equations. The only non-zero MF parameter is the singlet pairing $\Delta_{\vec{r}_{\mu},\vec{r}_{\nu}'}$. These two states thus have an emergent $SU(2)$ symmetry.

\subsection{\label{subsubsec: 0-flux Z2 PSG classes -> Correlations} Spin structure factor}

\subsubsection{\label{subsubsubsec: 0-flux Z2 PSG classes -> static spin structure factor} Equal-time static spin structure factor}

\begin{figure*}[ht!]
\includegraphics[width=0.85\linewidth]{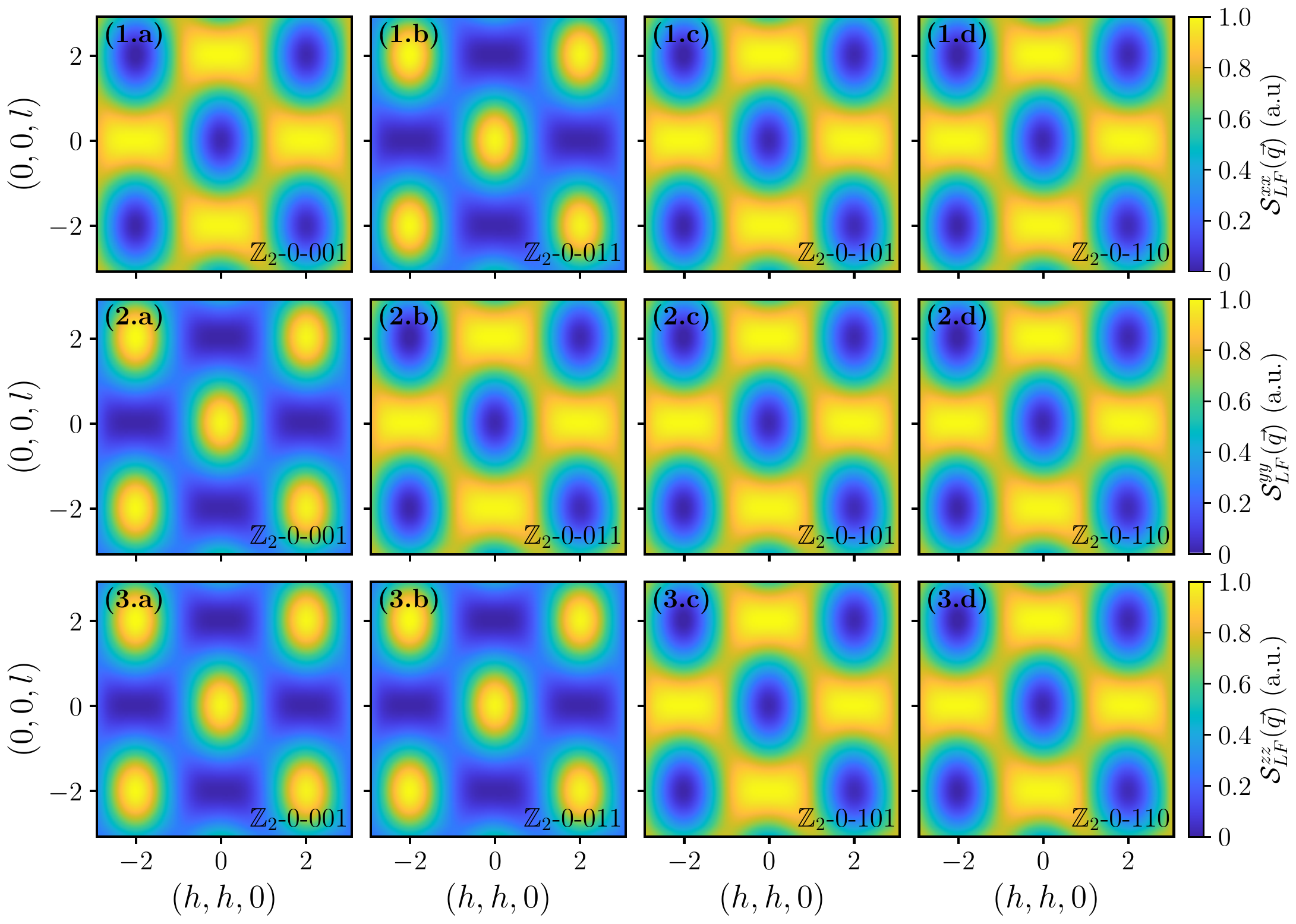}
\caption{\label{fig: SSSF in LF} Normalized equal-time spin structure factor in the local frame with $\kappa=0.14$, $\mathcal{J}_{x}/\mathcal{J}_{y}=0.50$,  $\mathcal{J}_{z}/\mathcal{J}_{y}=0.25$ and $\mathcal{J}_y>0$ in the [hhl] plane for the (1) $\mathcal{S}^{xx}_{LF}$, (2) $\mathcal{S}^{yy}_{LF}$ and (3) $\mathcal{S}^{zz}_{LF}$ component of the (a) $\mathbb{Z}_2\text{-}0\text{-}001$, (b) $\mathbb{Z}_2\text{-}0\text{-}011$, (c) $\mathbb{Z}_2\text{-}0\text{-}101$, and (d) $\mathbb{Z}_2\text{-}0\text{-}110$ PSG classes. The momenta are measured in reciprocal lattice unit, (i.e., 1 r.l.u.=$2 \pi$ because we take the lattice constant to be unity).}
\end{figure*}

To compare the signatures of our $\mathbb{Z}_2$ QSLs with experiments and other theoretical studies, we examine spin-spin correlations. We first evaluate the equal-time spin structure factor in the local frame 
\begin{align}
\mathcal{S}^{ab}_{LF}(\vec{q}) &= \frac{1}{N_{u.c.}} \sum_{\vec{r}_{\mu}, \vec{r}_{\nu}' } e^{i \vec{q}\cdot \left(\vec{r}_{\mu}-\vec{r}_{\nu}' \right)}\left\langle \tau^{a}_{\vec{r}_{\mu}}  \tau^{b}_{\vec{r}_{\nu} '} \right\rangle. 
\end{align}
This expression can be rewritten in terms of Bogoliubov transformations \cite{schneider2021projective, liu2019competing}. The resulting $xx$, $yy$, and $zz$ components of the normalized equal-time spin-spin correlation in the local frame in the $[h,h,l]$ plane are presented in fig. \ref{fig: SSSF in LF} for the four gaped $\mathbb{Z}_2$ QSLs. On account of the emergent spin rotational $SU(2)$ symmetry that appears after solving the self-consistency equations, we find that $\mathcal{S}^{xx}_{LF}=\mathcal{S}^{yy}_{LF}=\mathcal{S}^{zz}_{LF}$ for both $\mathbb{Z}_2\text{-}0\text{-}101$ and $\mathbb{Z}_2\text{-}0\text{-}110$. Both also have almost identical spin structure factor. They have a minimum at the zone center and maxima at $[002]$ and symmetry-related points. These local maxima are located where the pinch points of classical spin ice would be present \cite{isakov2004dipolar, moessner1998low, moessner1998properties}. $\mathbb{Z}_2\text{-}0\text{-}011$ ($\mathbb{Z}_2\text{-}0\text{-}001$) has the same general behavior but with the inverted pattern for the intensity of the $xx$ and $zz$ ($yy$ and $zz$) components. It is interesting to note that $\mathbb{Z}_2\text{-}0\text{-}011$ is the second-lowest energy state in the part of the phase diagram where the transverse couplings are unfrustrated. This seems to indicate that frustrated couplings favor spin correlations in the local frame with a minimum at the zone center and maxima at $[002]$ and symmetry-related points, whereas unfrustrated couplings favor the opposite.

\begin{figure}
\includegraphics[width=0.99\linewidth]{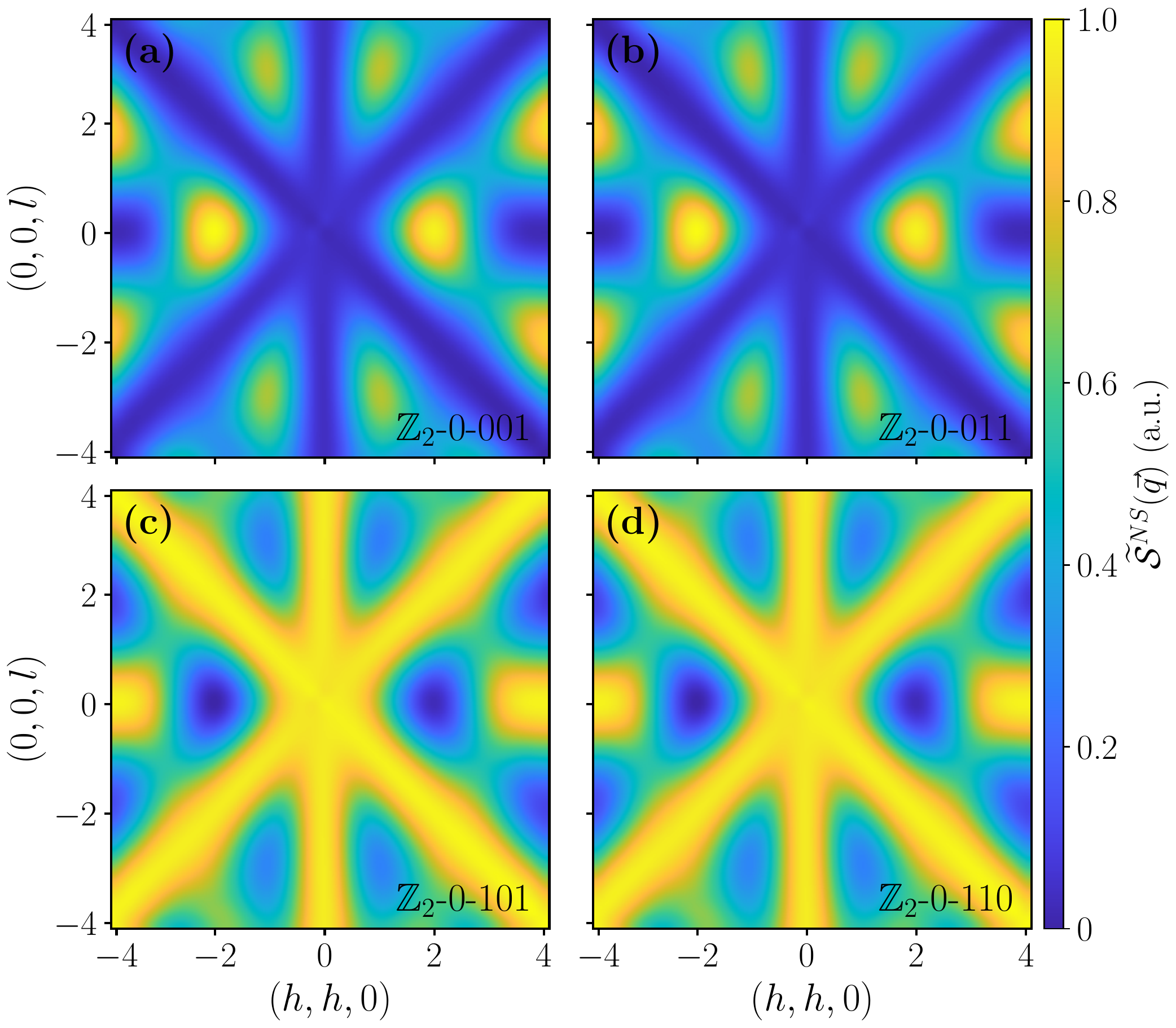}
\caption{\label{fig: SSSF INS} Normalized total equal-time static neutron scattering amplitude with $\kappa=0.14$, $\mathcal{J}_{x}/\mathcal{J}_{y}=0.50$,  $\mathcal{J}_{z}/\mathcal{J}_{y}=0.25$ and $\mathcal{J}_y>0$ in the [hhl] plane for the (a) $\mathbb{Z}_2\text{-}0\text{-}001$, (b) $\mathbb{Z}_2\text{-}0\text{-}011$, (c) $\mathbb{Z}_2\text{-}0\text{-}101$, and (d) $\mathbb{Z}_2\text{-}0\text{-}110$ PSG classes.
}
\end{figure}

On the other hand, to make direct contact with what is measured in experiments, one needs to look at the total equal-time static neutron scattering (NS) amplitude 
\begin{subequations}
\begin{align}
\widetilde{\mathcal{S}}^{NS}(\vec{q}) &= \sum_{ab} \left( \delta_{ab} - \frac{\vec{q}_{a} \vec{q}_{b}}{|\vec{q}|^2} \right) \left\langle m^{a}(\vec{q},0) m^{b}(-\vec{q},0) \right\rangle\\
m^{a}(\vec{q},t)&=\frac{1}{\sqrt{N_{u.c.}}} \sum_{\vec{r}_\mu} e^{i\vec{q}\cdot\vec{r}_\mu} \sum_{b} g_{\mu}^{ab}\widetilde{\mathbf{S}}^b_{\vec{r}_\mu}(t)
\end{align}
\end{subequations}
where we have introduced the magnetic moments $m^a(\vec{q},t)$, the pseudospin operators in the global frame $\widetilde{\mathbf{S}}_{\vec{r}_\mu}$ and the corresponding g-tensor $g_{\mu}^{ab}$ (see appendix \ref{appendix subsec: Local to global frame mapping}). The total scattering amplitude for the $\mathbb{Z}_2\text{-}0\text{-}001$, $\mathbb{Z}_2\text{-}0\text{-}011$, $\mathbb{Z}_2\text{-}0\text{-}101$, and $\mathbb{Z}_2\text{-}0\text{-}110$ PSG classes are presented in fig. \ref{fig: SSSF INS}. The $\mathbf{S}^{y}$ component of the pseudospin does not appreciably contribute to the amplitude since the octupolar moments do not linearly couple to neutrons, and the $\mathbf{S}^x$ component also has a vanishingly small g-tensor \cite{bhardwaj2021sleuthing}. Therefore, the scattering amplitude is dominated by $\mathbf{S}^z$ for Ce$_2$Zr$_2$O$_7$. This is reflected in our calculations: the two pairs $\mathbb{Z}_2\text{-}0\text{-}001$ \& $\mathbb{Z}_2\text{-}0\text{-}011$, and $\mathbb{Z}_2\text{-}0\text{-}001$ \& $\mathbb{Z}_2\text{-}0\text{-}011$ respectively have identical inelastic neutron scattering amplitude since their correlations $\mathcal{S}^{zz}_{LF}$ are also identical (see last row of fig. \ref{fig: SSSF in LF}). Moreover, the total equal-time static NS amplitude of $\mathbb{Z}_2\text{-}0\text{-}001$ and $\mathbb{Z}_2\text{-}0\text{-}011$ has inverted intensity compared to $\mathbb{Z}_2\text{-}0\text{-}001$ and $\mathbb{Z}_2\text{-}0\text{-}011$ since the same trend is observed for the $\mathcal{S}^{zz}_{LF}$ correlation of these states. This observation combined with our previous conclusion that ferromagnetic (antiferromagnetic) coupling favor a maximum (minimum) in the spin correlation at the zone center, provides a simple way to experimentally determine if the system is in the region of parameter space with frustrated or unfrustrated transverse coupling. The total equal-time NS amplitude has the rod motifs that were obtained from molecular dynamics simulations in the frustrated region of the phase diagram \cite{smith2021case, bhardwaj2021sleuthing}. While this feature is roughly consistent with neutron scattering experiments, the higher intensity at (001) and (003) reported in empirical studies does not show up in the theoretical results for these $\mathbb{Z}_2$ QSLs states. Such a difference was attributed to the presence of next-nearest-neighbor exchange interactions in \cite{bhardwaj2021sleuthing}.

\begin{figure*}[ht!]
\includegraphics[width=0.99\linewidth]{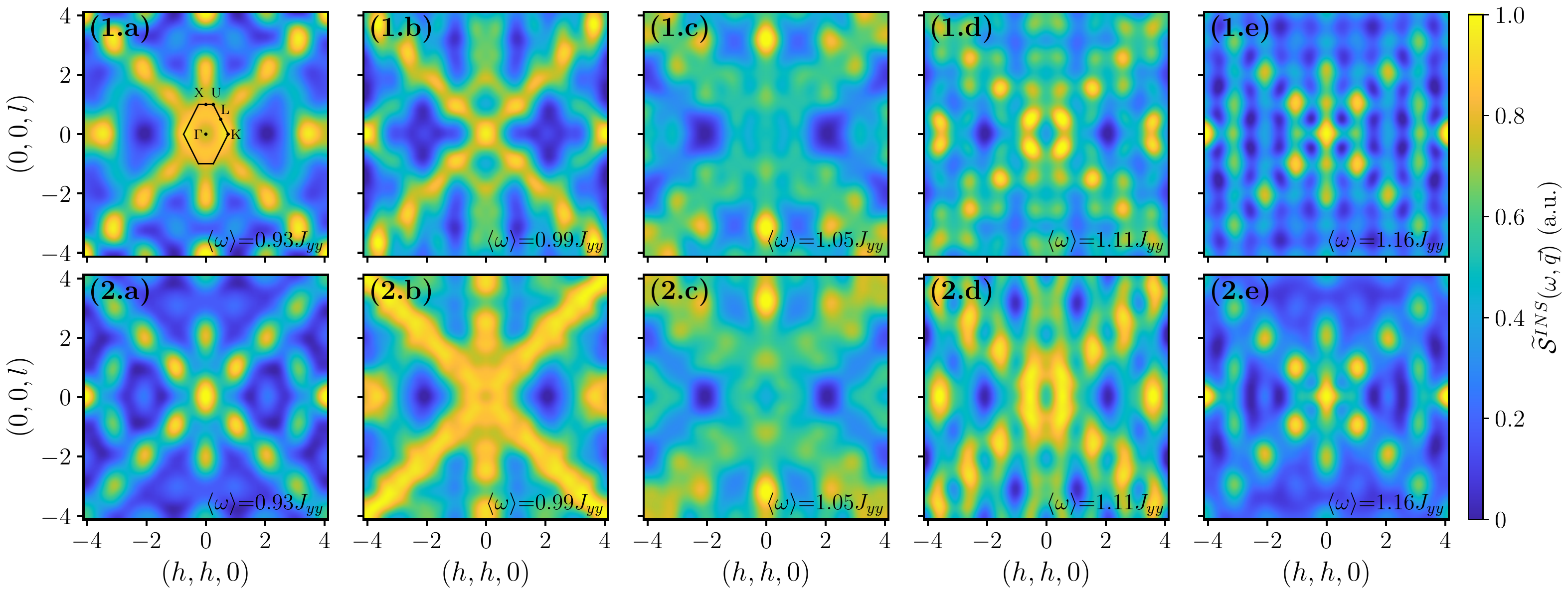}
\caption{\label{fig: DSSF in [hhl]} Dynamical spin structure factors of the two candidate spin liquid states (1) $\mathbb{Z}_2$-0-101 and (2) $\mathbb{Z}_2$-0-110 in the [hhl] plane integrated over the energy range $\left(\expval{\omega}-\Delta\omega, \expval{\omega}+\Delta\omega \right)$ for (a) $\expval{\omega}=0.93 J_{yy}$, (b) $\expval{\omega}=0.99 J_{yy}$, (c) $\expval{\omega}=1.05 J_{yy}$, (d) $\expval{\omega}=1.11 J_{yy}$, and (e) $\expval{\omega}=1.16 J_{yy}$ with $\Delta \omega=0.3 J_{yy}$. All plots are normalized independently. Figure (1.a) shows the first Brillouin zone and corresponding high symmetry points.}
\end{figure*}

For polarized neutrons, the total equal-time static NS amplitude can be further divided into spin-flip (SF)
\begin{equation}
    \widetilde{\mathcal{S}}^{SF} = \sum_{ab}\frac{(\vec{P}\cross \vec{q})_a (\vec{P}\cross \vec{q})_b}{|\vec{q}|^2} \left\langle m^{a}(\vec{q},0) m^{b}(-\vec{q},0) \right\rangle,
\end{equation}
and non spin slip (NSF) channels 
\begin{equation}
    \widetilde{\mathcal{S}}^{NSF} = \sum_{ab} \vec{P}_a \vec{P}_b \left\langle m^{a}(\vec{q},0) m^{b}(-\vec{q},0) \right\rangle 
\end{equation}
where $\vec{P}=(1,-1,0)/\sqrt{2}$ is the neutron polarization vector and the simple relation $\widetilde{\mathcal{S}}^{NS}(\vec{q}) = \widetilde{\mathcal{S}}^{SF}(\vec{q}) + \widetilde{\mathcal{S}}^{NSF}(\vec{q})$ holds in the $[hhl]$ plane. In our calculations, the NSF channel is always empty. The amplitude in the SF channel is accordingly identical to fig. \ref{fig: SSSF INS}. 

Finally, it is intriguing to note the correspondence between the NS amplitude of our most stable states in the part of phase diagram with frustrated transverse couplings ($\mathbb{Z}_2$-0-101 and $\mathbb{Z}_2$-0-110) and the one obtained for a single tetrahedron summed uniformly over all classical spin configurations but the all-in-all-out ones, which was discussed by Castelnovo and Moessner \cite{castelnovo2019rod}. Further comparisons with this single tetrahedron calculation presented in appendix \ref{appendix: comparison with single tetrahedron calculation} support the conclusion that the total equal-time static NS amplitudes are equivalent. The agreement between the full lattice average and this single tetrahedron calculation signals that $\mathbf{S}^z$ spin correlations are short-ranged and neighboring tetrahedra are essentially decoupled. Because of the emergent $SU(2)$ symmetry of $\mathbb{Z}_2\text{-}0\text{-}101$ and $\mathbb{Z}_2\text{-}0\text{-}110$, the same statement applies to the $\mathbf{S}^x$ and $\mathbf{S}^y$ components. This is different from QSI. The QSI wave function is a coherent superposition of 2-in-2-out configurations where spin-flip excitations (i.e., monopoles or spinons depending on the convention) are only allowed provided that they form nearby dipoles \cite{savary2021quantum}. This leads to algebraic correlations that are cut off only if single monopoles (spinons) proliferate \cite{castelnovo2019rod}.

\subsubsection{\label{subsubsubsec: 0-flux Z2 PSG classes -> dynamical spin structure factor} Dynamical spin structure factor}

Our two most likely stable MF states in the frustrated region of the octupolar quadrant have identical total equal-time static NS amplitude. Having said that, both $\mathbb{Z}_2$-0-101 and $\mathbb{Z}_2$-0-110 should have distinct signature in their dynamical spin structure factor (DSSF)
\begin{align}
&\widetilde{\mathcal{S}}^{INS}(\omega, \vec{q}) =\nonumber \\
&\quad \sum_{ab} \int_{-\infty}^{\infty} \mathrm{d} t e^{i \omega t} \left( \delta_{ab} - \frac{\vec{q}_{a} \vec{q}_{b}}{|\vec{q}|^2} \right) \left\langle m^{a}(\vec{q},t) m^{b}(-\vec{q},0) \right\rangle
\end{align}
because their dispersion relations differ (see fig. \ref{fig: dispersion relation 0-flux Z2 states}). Several energy cuts of $\widetilde{\mathcal{S}}^{INS}(\vec{q},\omega)$ integrated over small windows $\left(\expval{\omega}-\Delta \omega, \expval{\omega}+\Delta \omega \right)$ (i.e., $\int_{\expval{\omega}-\Delta\omega}^{\expval{\omega}+\Delta\omega} \mathrm{~d} \omega^{\prime} \mathcal{S}\left(\mathbf{q}, \omega^{\prime}\right)$) in the [hhl] plane for our two candidate $\mathbb{Z}_2$ QSLs are presented in fig. \ref{fig: DSSF in [hhl]}. The low energy part of the DSSF for the two QSL presents different traits and could potentially be used to differentiate the two. The contributions to the dynamical spin structure factor come from processes where two Bogoliubov quasiparticles are created at time $0$ and then annihilating at a later time $t$. It follows that the DSSF is zero for energies smaller than twice the minimum and larger than twice the maximum of the spinon dispersion relation. We only consider energies within that range.

\begin{figure}
\includegraphics[width=0.99\linewidth]{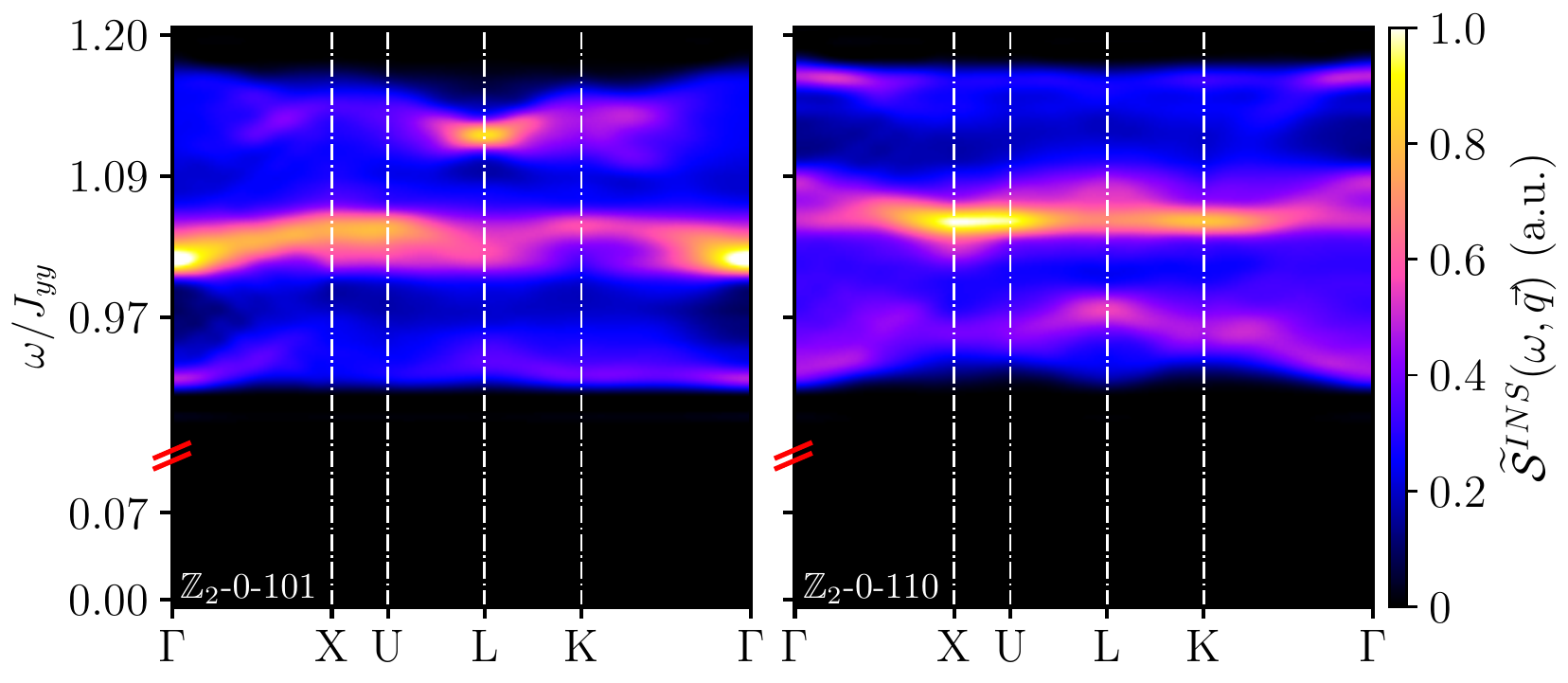}
\caption{\label{fig: DSSF along path} Normalized dynamical spin structure factor with $\kappa=0.14$, $\mathcal{J}_{x}/\mathcal{J}_{y}=0.50$,  $\mathcal{J}_{z}/\mathcal{J}_{y}=0.25$ and $\mathcal{J}_y>0$ along high symmetry lines in the first Brillouin zone for the  $\mathbb{Z}_2\text{-}0\text{-}101$ (left), and $\mathbb{Z}_2\text{-}0\text{-}110$ (right) PSG classes. A line cut indicated by the two red lines was introduced along the energy axis.
}
\end{figure}

For a given momentum transfer, the total equal-time static NS amplitude should be the same as the DSSF integrated over all energies. This is illustrated in fig. \ref{fig: DSSF in [hhl]} where the contribution of every energy cut to the overall rod motifs can be identified. It is important to note that every energy cut is normalized independently for visibility. To get the relative intensity at different energies, one can look at the DSSF along high symmetry lines presented in fig. \ref{fig: DSSF along path}. For both $\mathbb{Z}_2$ QSL, the DSSF has its highest intensity around $\expval{\omega}\approx1.05 J_{yy}$. Interestingly, $\mathbb{Z}_2$-0-110 has maxima in its DSSF at the X and K points as measured in experiments \cite{gao2019experimental}. Taking  $J_{yy}$ to be in its experimentally determined range of $\left(0.06,0.09\right)$ meV \cite{smith2021case, bhardwaj2021sleuthing}, the position of the maxima is also consistent with measurements. In comparison, molecular dynamics simulation of the classical model only produces DSSF that is constant along the $\Gamma\to$X direction or has a marginally increased intensity at X upon inclusion of next-nearest-neighbor interaction \cite{bhardwaj2021sleuthing}. Hence, it is possible that high intensity at X and K arises from purely quantum effects.

\section{\label{sec: Discussion} Discussion}

In this paper, we used a PSG approach together with a Schwinger boson representation of the pseudospin operators to give a complete classification of symmetric $\mathbb{Z}_2$ and $U(1)$ dipolar-octupolar QSLs on the pyrochlore lattice. For each PSG class, a Schwinger boson MF theory for the most general nearest-neighbor symmetry-allowed Hamiltonian was carried out. We have compared the energy of these different classes after solving the self-consistency equations to evaluate the MF phase diagram at different values of $\kappa$, which controls the amount of quantum fluctuations. The phase diagram for $\kappa=0.14$ captures many characteristics of the previous ED calculations \cite{patri2020theory}: we observed the effect of quantum fluctuations via the bending of the classical phase boundaries and the correct magnetic ordering (i.e., $X$ and $Z$ all-in-all-out) in the relevant parameter regime, although the 0-flux QSI in the vicinity of the Ising limit in the unfrustrated regime cannot be identified because the QSI is represented in a different parton construction. Most importantly, we found competing QSL phases beyond the QSI limit in the frustrated region of the phase diagram, where quantum fluctuations are significantly stronger. The two competing gapped $\mathbb{Z}_2$ QSLs at the experimentally determined coupling constants have a very narrow spinon band dispersion. The equal-time structure factors of these QSLs show a rod-like pattern, similar to what is observed in neutron scattering experiments on Ce$_2$Zr$_2$O$_7$. Moreover, one of these two states (the $\mathbb{Z}_2$-0-110 PSG class) has a high intensity in its dynamical spin structure factor at the X and K point as reported in inelastic neutron scattering experiments.

More work is still required to determine if the $\mathbb{Z}_2$-0-110 state is truly a ground state candidate that may be competing with $\pi$-O-QSI. In neutron scattering experiments on Ce$_2$Zr$_2$O$_7$, the transverse dipolar component $S^z$ dominates the coupling to the neutron spin. Hence, if two quantum spin liquid states have similar $S^z$ correlations, the resulting neutron scattering spectra will turn out to be similar as well. This means we need new theoretical predictions involving the octupolar $S^y$ degree of freedom to possibly distinguish the two QSLs. This is particularly so because the $\mathbb{Z}_2$-0-110 state has an emergent SU(2) symmetry. It is known that the low-temperature behavior of heat capacity is different for $\mathbb{Z}_2$ and $U(1)$ QSLs: it scales as $T^3$ in the $U(1)$ case due to the linearly dispersing emergent photon, and has a thermally activated exponential form in the $\mathbb{Z}_2$ case. While recent measurements may be consistent with an exponential extrapolation and an energy gap of approximately $0.035$K \cite{smith2021case}, the energy scale associated with the emergent photon is very small, and hence the detection of photons may require much lower temperatures. It will also be useful to investigate the temperature dependence of the thermal conductivity at low temperatures to check consistency with the heat capacity. 

Another plausible scenario is that $\mathbb{Z}_2$-0-110 and $\mathbb{Z}_2$-0-101 may not be the true ground state of the pristine $XYZ$ model, but rather a part of the manifold of closely competing QSL states in the frustrated region of the phase diagram with the Ising(transverse) octupolar(dipolar) degree of freedom. These competing states may share similar physical properties and are therefore hard to differentiate. In this case, it is conceivable that small perturbations may favor one of the competing ground states in real materials. On a separate note, we demonstrated that putative QSL states in the unfrustrated region of the phase diagram have very different physical properties, which show completely opposite intensity distributions in equal-time structure factor in comparison to the experiments on Ce$_2$Zr$_2$O$_7$. Hence it strongly supports the picture that the QSL ground state of Ce$_2$Zr$_2$O$_7$ lies in the frustrated region of the phase diagram. 

On the theoretical front, it will be helpful to develop a framework that can fairly compare different parton constructions such as the Schwinger boson, Abrikosov fermions, or gMFT representations in three-dimensional QSLs. Just like the well-known correspondence between $\mathbb{Z}_2$ bosonic and fermionic QSLs in two dimensions, some of the QSLs in different parton constructions may turn out to be smoothly connected to each other.

\begin{acknowledgments}
We thank Chunxiao Liu and Adarsh Patri for helpful discussions. We acknowledge support from the Natural Sciences and Engineering Research Council of Canada (NSERC) and the Centre of Quantum Materials at the University of Toronto. The computations were performed on the Niagara cluster, which SciNet host in partnership with Compute Canada.
\end{acknowledgments}


%

\appendix

\section{\label{appendix: Microscopic model} Microscopic model}

\subsection{\label{appendix subsec: Local coordinates and generic model} Local coordinates}

The basis in the local frame at each sublattice is defined in table \ref{tab: Local basis}.

\begin{table}[h!]
\caption{\label{tab: Local basis}%
Local sublattice basis vectors
}
\begin{ruledtabular}
\begin{tabular}{ccccc}
$\mu$ & 0 & 1  & 2  & 3 \\
\hline
$\hat{\mathbf{z}}_{\mu}$ & $\frac{-1}{\sqrt{3}}\left(1,1,1\right)$ & $\frac{1}{\sqrt{3}}\left(-1,1,1\right)$  & $\frac{1}{\sqrt{3}}\left(1,-1,1\right)$  & $\frac{1}{\sqrt{3}}\left(1,1,-1\right)$   \\[2mm]
$\hat{\mathbf{y}}_{\mu}$ & $\frac{1}{\sqrt{2}}\left(0,-1,1\right)$  & $\frac{1}{\sqrt{2}}\left(1,0,1\right)$  & $\frac{1}{\sqrt{2}}\left(1,1,0\right)$ & $\frac{1}{\sqrt{2}}\left(1,-1,0\right)$  \\[2mm]
$\hat{\mathbf{x}}_{\mu}$ & $\frac{-1}{\sqrt{6}}\left(-2,1,1\right)$ & $\frac{-1}{\sqrt{6}}\left(1,2,-1\right)$  & $\frac{-1}{\sqrt{6}}\left(-1,1,2\right)$  & $\frac{1}{\sqrt{6}}\left(1,1,2\right)$    \\
\end{tabular}
\end{ruledtabular}
\end{table}

\subsection{pseudospin}

The site symmetry group  $D_{3d}$ contains a $C_3$ rotation, three $C_2$ rotations and their composition with inversion (see fig. \ref{fig:pyrochlore lattice} (c)). The CEF splits the degenerate spin-orbit coupled $J$ manifold resulting in low-lying doublets that can be further divided into irreducible representations (irreps) of the double group of $D_{3d}$. The "effective spin-1/2" corresponds to the $\Gamma_{10}$ irrep, whereas the dipolar-octupolar (DO) doublet transforms as the  $\Gamma_5^+\oplus\Gamma_6^+$ irrep.

The ground state doublet of Ce$_2$Zr$_2$O$_7$ is a mix of $\ket{J=5/2, m_{J}=\pm 3/2}$ states and can thus be written generically as \cite{gaudet2019quantum},
\begin{equation}
    \ket{\Gamma_5^+\oplus\Gamma_6^+}_{\pm} = \alpha \ket{\pm 3/2} \pm \beta \ket{\mp 3/2}.
\end{equation}
The peculiar nature of this ground state doublet can be made evident by studying the active multipole moments in the ground state manifold. These are one dipole ($J_{z}$), and two octupoles ($J_{x}^3-\overline{J_x J_y J_y}$ and $J_{y}^3-\overline{J_y J_x J_x}$) \cite{patri2020theory}. The dipole and the first octupole transform as the $\Gamma_2$ irrep, whereas the second octupole transforms as $\Gamma_1$ under the point group symmetries. These multipolar moments can be represented as pseudospin operators
\begin{subequations}
  \begin{align}
    \mathbf{S}_{\vec{r}_\mu}^x &=  \mathcal{P} \left( \mathcal{C}_{0}\left( (J^{x}_{\vec{r}_\mu})^{3}-\overline{J^{x}_{\vec{r}_\mu} J^{y}_{\vec{r}_\mu} J^{y}_{\vec{r}_\mu}}\right)+\mathcal{C}_{1} J^{z}_{\vec{r}_\mu} \right) \mathcal{P}  \\
    \mathbf{S}_{\vec{r}_\mu}^y &= \mathcal{C}_{2} \mathcal{P} \left( (J^{y}_{\vec{r}_\mu})^{3}-\overline{J^{y}_{\vec{r}_\mu} J^{x}_{\vec{r}_\mu} J^{x}_{\vec{r}_\mu}} \right) \mathcal{P}   \\
    \mathbf{S}_{\vec{r}_\mu}^z &= \mathcal{C}_3 \mathcal{P}J_{\vec{r}_\mu}^{z}\mathcal{P},
  \end{align}
\end{subequations}
where $\mathcal{P}$ is a projection operator into the ground state manifold, the overline symbolizes symmetrized products \cite{onoda2011quantum} and we introduced normalization constants $\mathcal{C}_0$, $\mathcal{C}_1$, $\mathcal{C}_2$ and $\mathcal{C}_3$ determined by CEF parameters so that these operators can be identified with the Pauli matrices $\sigma^a/2$ for $a\in \{x,y,z\}$ in the $\{\ket{\Gamma_5^+\oplus\Gamma_6^+}_{+},\ket{\Gamma_5^+\oplus\Gamma_6^+}_{-}\}$ basis and thus form a new local $\mathfrak{su}(2)$ algebra (i.e., $\comm{\mathbf{S}_{\vec{r}_\mu}^m}{\mathbf{S}_{\vec{r}_\mu}^n}= i\sum_l \varepsilon^{mnl}\mathbf{S}_{\vec{r}_\mu}^{l}$) in the ground state manifold. It is important to note that even though the $x$ component of the pseudospin operators is composed of both dipolar and octupolar moments, it transforms identically to a simple dipole ($\Gamma_2$) and is thus referred to as a dipolar moment.

\subsection{XYZ model}

To map the initial Hamiltonian in the familiar form of pseudospin on the pyrochlore lattice Eq. \eqref{eq: Hamiltonian nn pyrochlore} into the much simpler $XYZ$ form of Eq. \eqref{eq: XYZ model}, we perform a uniform rotation in the local pseudospin frame along the $\hat{y}_{\mu}$ axis
\begin{align} \label{eq: Trf: nn pyrochlore -> XYZ}
\tau^{x} &= \cos (\theta) \mathbf{S}^{x}-\sin (\theta) \mathbf{S}^{z} \\
\tau^{y} &= \mathbf{S}^{y},\\
\tau^{z} &= \sin (\theta) \mathbf{S}^{x}+\cos (\theta) \mathbf{S}^{z}
\end{align}
where $\tan (2 \theta)=\frac{2 J_{x z}}{J_{z z}-J_{x x}}$. The coupling constants are accordingly transformed to
\begin{align} \label{eq: Relation couplings nn pyrochlore -> XYZ}
&\mathcal{J}_{x}=\frac{J_{z z}+J_{x x}}{2}-\frac{\sqrt{\left(J_{z z}-J_{x x}\right)^{2}+4 J_{x z}^{2}}}{2}, \\
&\mathcal{J}_{y}=J_{y y}, \\
&\mathcal{J}_{z}=\frac{J_{z z}+J_{x x}}{2}+\frac{\sqrt{\left(J_{z z}-J_{x x}\right)^{2}+4 J_{x z}^{2}}}{2}.
\end{align}

\section{\label{appendix: pseudospin transformation} pseudospin transformation}
\subsection{Space group operations}

The transformation of the pseudospin under a space group generator can be deduced by first performing the active transformation on the pseudospin and then changing the local coordinates due to sublattice mixing. For the first step, we note that the inversion in the $\overline{C}_6$ operation does not affect the pseudospins as we are working within the ground state manifold where the total angular momentum $J=5/2$. So the effects of $C_3$ and $\overline{C}_6=C_3\circ I$ on the angular momentum $\vec{J}$ are the same in this case. Accordingly, the transformation of the angular momentum $\vec{J}$ can be evaluated by acting on $\ket{\Gamma_5^+\oplus \Gamma_6^+}_{\pm}$ with
\begin{equation}
    R_{\overline{C}_6,\mu} = e^{-i \frac{2 \pi}{3} \hat{n}_{\overline{C}_{6},\mu}\cdot \vec{J}}, \hspace{2mm} R_{S,\mu} = e^{-i \pi \hat{n}_{S,\mu} \cdot \vec{J}}, \label{eq: rotation matrices for J}
\end{equation}
where $\hat{n}_{\mathcal{O},\mu}$ is the rotation axis of the generator $\mathcal{O}$ (i.e., in the GCC these are $\frac{1}{\sqrt{3}}(1,1,1)$ for $\overline{C}_6$ and $\frac{1}{\sqrt{2}}(1,1,0)$ for $S$) written in the local frame on the $\mu$ sublattice.

Next, for the sublattice mixing, we can define orthogonal matrices $O_{\mu\to \nu}$ that map a local frame $\mu$ to the new local frame $\nu$. These $SO(3)$ matrices can be deduced from the inner product of basis vectors in table \ref{tab: Local basis}  
\begin{align}
    [O_{\mu\to\nu}]_{i,j} = \vec{x}_{\nu,i} \cdot \vec{x}_{\mu,j},
\end{align}
where $i,j\in\left\{1,2,3 \right\}$ and $\vec{x}_{\mu}= (\hat{\mathbf{x}}_{\mu},\hat{\mathbf{y}}_{\mu},\hat{\mathbf{z}}_{\mu})$. We note that obviously $O_{\mu\to\nu}=(O_{\nu\to\mu})^{-1}$ and $O_{\mu\to\mu}=\mathds{1}_{3\times 3}$. Once these matrices have been found, the change of basis for the pseudospin can be performed by acting on the doublet $\ket{\Gamma_5^+\oplus \Gamma_6^+}_{\pm}$ with $R_{\mu\to\nu}=e^{-i(\theta_{\alpha}J_z)} e^{-i(\theta_{\beta}J_y)} e^{-i(\theta_{\gamma}J_z)}$, where $\theta_{\alpha}$, $\theta_{\beta}$ and $\theta_{\gamma}$ are the Euler angles associated with the SO(3) matrix $O_{\mu\to\nu}$ \cite{tinkham2003group}. Putting all of this together, the final transformations at sublattice $\mu$ are
\begin{equation}
    U_{\mathcal{O},\mu} = \mathcal{P}  R_{\mu\to\mathcal{O}(\mu)} R_{\mathcal{O},\mu}  \mathcal{P}
\end{equation}
Evaluating everything explicitly, the transformation does not have any sublattice dependence and are given in Eq. \eqref{eq: matrix form pseudospin trfs}. Acting with these matrices explicitly, the pseudospin transformations under the SG elements and TRS are
\begin{subequations} \label{eq: pseudospin transformation under SG generators}
  \begin{align}
    \mathcal{T} : \qty\Big{ \mathbf{S}^{x}_{\vec{r}_\mu}, \mathbf{S}^{y}_{\vec{r}_\mu}, \mathbf{S}^{z}_{\vec{r}_\mu} } \longrightarrow &  \qty\Big{  -\mathbf{S}^{x}_{\vec{r}_\mu},  -\mathbf{S}^{y}_{\vec{r}_\mu}, -\mathbf{S}^{z}_{\vec{r}_\mu} } \\
     T_i :  \qty\Big{ \mathbf{S}^{x}_{\vec{r}_\mu}, \mathbf{S}^{y}_{\vec{r}_\mu}, \mathbf{S}^{z}_{\vec{r}_\mu} } \longrightarrow &  \qty\Big{  \mathbf{S}^{x}_{T_i(\vec{r}_\mu)}, \mathbf{S}^{y}_{T_i(\vec{r}_\mu)}, \mathbf{S}^{z}_{T_i(\vec{r}_\mu)} }\\
    \overline{C}_6 :  \qty\Big{  \mathbf{S}^{x}_{\vec{r}_\mu}, \mathbf{S}^{y}_{\vec{r}_\mu}, \mathbf{S}^{z}_{\vec{r}_\mu} } \longrightarrow &  \qty\Big{  \mathbf{S}^x_{\overline{C}_6(\vec{r}_\mu)} ,   \mathbf{S}^y_{\overline{C}_6(\vec{r}_\mu)}, \mathbf{S}^z_{\overline{C}_6(\vec{r}_\mu)} } \\
    S :  \qty\Big{ \mathbf{S}^{x}_{\vec{r}_\mu}, \mathbf{S}^{y}_{\vec{r}_\mu}, \mathbf{S}^{z}_{\vec{r}_\mu} } \longrightarrow &  \qty\Big{   -\mathbf{S}^{x}_{S(\vec{r}_\mu)}  ,   \mathbf{S}^{y}_{S(\vec{r}_\mu)} , -\mathbf{S}^{z}_{S(\vec{r}_\mu)} }
  \end{align}
\end{subequations}

\subsection{Local to global frame mapping \label{appendix subsec: Local to global frame mapping}}

In a similar spirit, one can rotate the spins in the local frame $\mathbf{S}^{a}_{\vec{r}_{\mu}}$ to the  global frame $\widetilde{\mathbf{S}}^{a}_{\vec{r}_{\mu}}$ (i.e.,frame with basis vectors  $\hat{\mathbf{x}}_{\text{GF}}=(1,0,0), \hat{\mathbf{y}}_{\text{GF}}=(0,1,0)$ and $\hat{\mathbf{z}}_{\text{GF}}=(0,0,1)$). To do so we adapt the presentation of appendix A in \cite{bhardwaj2021sleuthing} to our convention for the local basis. The local to global frame map is 
\begin{equation}
\left(\begin{array}{c}
\mathbf{S}^{x}_{\vec{r}_{\mu}} \\
\mathbf{S}^{y}_{\vec{r}_{\mu}} \\
\mathbf{S}^{z}_{\vec{r}_{\mu}} 
\end{array}\right)
=
R_{\mu}
\left(\begin{array}{c}
\widetilde{\mathbf{S}}^{x}_{\vec{r}_{\mu}} \\
\widetilde{\mathbf{S}}^{y}_{\vec{r}_{\mu}} \\
\widetilde{\mathbf{S}}^{z}_{\vec{r}_{\mu}}
\end{array}\right)
\end{equation}
where the sublattice dependant $SO(3)$ transformations $R_{\mu}$ are 
\begin{align*}
    R_0 &= 
    \left(\begin{array}{ccc}
        \sqrt{\frac{2}{3}} & -\frac{1}{\sqrt{6}} & -\frac{1}{\sqrt{6}} \\
        0 & -\frac{1}{\sqrt{2}} & \frac{1}{\sqrt{2}} \\
        -\frac{1}{\sqrt{3}} & -\frac{1}{\sqrt{3}} & -\frac{1}{\sqrt{3}}
    \end{array}\right),
    \hspace{1mm} R_1 = 
    \left(\begin{array}{ccc}
        -\frac{1}{\sqrt{6}} & -\sqrt{\frac{2}{3}} & \frac{1}{\sqrt{6}} \\
        \frac{1}{\sqrt{2}} & 0 & \frac{1}{\sqrt{2}} \\
        -\frac{1}{\sqrt{3}} & \frac{1}{\sqrt{3}} & \frac{1}{\sqrt{3}}
    \end{array}\right)
     \\
    R_2 &= 
    \left(\begin{array}{ccc}
        \frac{1}{\sqrt{6}} & -\frac{1}{\sqrt{6}} & -\sqrt{\frac{2}{3}} \\
        \frac{1}{\sqrt{2}} & \frac{1}{\sqrt{2}} & 0 \\
        \frac{1}{\sqrt{3}} & -\frac{1}{\sqrt{3}} & \frac{1}{\sqrt{3}}
    \end{array}\right)
    ,
    \hspace{1mm} R_3 = 
    \left(\begin{array}{ccc}
        \frac{1}{\sqrt{6}} & \frac{1}{\sqrt{6}} & \sqrt{\frac{2}{3}} \\
        \frac{1}{\sqrt{2}} & -\frac{1}{\sqrt{2}} & 0 \\
        \frac{1}{\sqrt{3}} & \frac{1}{\sqrt{3}} & -\frac{1}{\sqrt{3}}
    \end{array}\right).
\end{align*}
Starting with the initial Hamiltonian in the local frame with an applied magnetic field $\vec{h} = \mu_B \mu_0 \vec{H}$
\begin{align} 
    H=&\sum_{\langle \vec{r}_{\mu}  \vec{r}_{\nu}' \rangle}\left[J_{x x} \mathbf{S}_{\vec{r}_{\mu}}^{x} \mathbf{S}_{\vec{r}_{\nu}'}^{x}+J_{y y} \mathbf{S}_{\vec{r}_{\mu}}^{y} \mathbf{S}_{\vec{r}_{\nu}'}^{y}+J_{z z} \mathbf{S}_{\vec{r}_{\mu}}^{z} \mathbf{S}_{\vec{r}_{\nu}'}^{z}\right. \nonumber \\
    & \hspace{1.0cm} \left.+J_{x z}\left(\mathbf{S}_{\vec{r}_{\mu}}^{x} \mathbf{S}_{\vec{r}_{\nu}'}^{z}+\mathbf{S}_{\vec{r}_{\mu}}^{z} \mathbf{S}_{\vec{r}_{\nu}'}^{x}\right)\right] \nonumber \\
    & - \sum_{\vec{r}_\mu} (\hat{z}_{\mu}\cdot \vec{h}) (g_{x} \mathbf{S}_{\vec{r}_{\mu}}^{x} + g_{z} \mathbf{S}_{\vec{r}_{\mu}}^{z}),
\end{align}
where $g_x$ and $g_{z}$ are the coupling strength of the dipolar pseudospin components to the local $z$ component of $\vec{h}$, we can use these transformations to rewrite it in the global frame
\begin{align}
    H=&\sum_{\langle \vec{r}_{\mu}, \vec{r}_{\nu}' \rangle } \sum_{ab} J_{\mu \nu}^{ab} \widetilde{\mathbf{S}}_{\vec{r}_\mu}^{a} \widetilde{\mathbf{S}}_{\vec{r}_\nu'}^{b} - \sum_{\vec{r}_\mu} \sum_{ab} h^{a} g^{ab}_{\mu} \widetilde{\mathbf{S}}_{\vec{r}_{\mu}}^b.
\end{align}
with $a,b\in\{x,y,z\}$. The new exchange coupling constants found by performing the transformations are
\begin{align}
    &J_{01} = 
    \left(\begin{array}{ccc}
        J_{1} & J_{2} & -J_{1} \\
        J_{3} & -J_{1} & J_{4} \\
        -J_{4} & -J_{1} & -J_{3}
    \end{array}\right), 
    J_{02} = 
    \left(\begin{array}{ccc}
        -J_{1} & J_{1} & J_{2} \\
        J_{4} & J_{3} & -J_{1} \\
        -J_{3} & -J_{4} & -J_{1}
    \end{array}\right) \nonumber \\
    &J_{03} = 
    \left(\begin{array}{ccc}
        -J_{1} & -J_{1} & -J_{2} \\
        J_{4} & -J_{3} & J_{1} \\
        -J_{3} & J_{4} & J_{1}
    \end{array}\right), 
    J_{12} = 
    \left(\begin{array}{ccc}
        -J_{3} & -J_{4} & -J_{1} \\
        J_{1} & -J_{1} & -J_{2} \\
        -J_{4} & -J_{3} & J_{1}
    \end{array}\right) \nonumber \\
    &J_{13} = 
    \left(\begin{array}{ccc}
        -J_{3} & J_{4} & J_{1} \\
        J_{1} & J_{1} & J_{2} \\
        -J_{4} & J_{3} & -J_{1}
    \end{array}\right), 
    J_{23} = 
    \left(\begin{array}{ccc}
        -J_{4} & J_{3} & -J_{1} \\
        -J_{3} & J_{4} & J_{1} \\
        J_{1} & J_{1} & J_{2}
    \end{array}\right) 
\end{align}
where we have introduced
\begin{subequations}
\begin{align}
    J_{1} &= \frac{1}{6} \left( -2 J_{xx} + 2 J_{zz} - \sqrt{2} J_{xz} \right) \\
    J_{2} &= \frac{1}{3} \left( -2 J_{xx} -  J_{zz} + 2 \sqrt{2} J_{xz}  \right) \\
    J_{3} &= \frac{1}{6} \left( J_{xx} - 3 J_{yy} + 2 J_{zz} + 2 \sqrt{2} J_{xz} \right)\\
    J_{4} &= \frac{1}{6} \left( -J_{xx} - 3 J_{yy} - 2J_{zz} - 2 \sqrt{2} J_{xz} \right).
\end{align}
\end{subequations}
Equivalently, the new g-matrices are
\begin{align}
    g_0 &= 
    \left(\begin{array}{lll}
        -g_{-} & g_{+} & g_{+} \\
        -g_{-} & g_{+} & g_{+} \\
        -g_{-} & g_{+} & g_{+}
    \end{array}\right), 
    g_1 = 
    \left(\begin{array}{ccc}
        g_{+} & g_{-} & -g_{+} \\
        -g_{+} & -g_{-} & g_{+} \\
        -g_{+} & -g_{-} & g_{+}
    \end{array}\right)  \nonumber \\
    g_2 &= 
    \left(\begin{array}{ccc}
        g_{+} & -g_{+} & -g_{-} \\
        -g_{+} & g_{+} & g_{-} \\
        g_{+} & -g_{+} & -g_{-}
    \end{array}\right), 
    g_3 = 
    \left(\begin{array}{ccc}
        g_{+} & g_{+} & g_{-} \\
        g_{+} & g_{+} & g_{-} \\
        -g_{+} & -g_{+} & -g_{-}
    \end{array}\right). 
\end{align}
with $g_{+}=\frac{1}{3}\left(g_{x}/\sqrt{2} + g_{z}\right)$ and $g_{-}=\frac{1}{3}\left(\sqrt{2} g_{x} - g_{z} \right)$. The coupling strengths $g_{x}=0$ and $g_{y}=2.4$, or $g_{x}=-0.2324$ and $g_{y}=2.35$ were determined to be in good agreement with experiments \cite{bhardwaj2021sleuthing}. We consider this first set of data and $J_{xz}=0$ when computing the spin structure factors.

\section{\label{appendix subsec: mean-field decoupling} Schwinger boson decoupling}

\subsection{General decoupling}

To rewrite the Hamiltonian in terms of bond operators, a meaningful comparison of the different terms must be carried out. Thus, we introduce a canonical ordering of the bosonic operators, where (in priority order): creation operators are to the left of annihilation operators, operators at $\vec{r}_\mu$ are to the left of those at $\vec{r}_\nu'$, and operators for spin up ($\uparrow$) are to the left of spin-down ($\downarrow$) operators. Using this convention, the product of bond operators are 
\begin{widetext}
\begin{subequations} \label{eq: appendix product bond operators as schwinger bosons}
  \begin{align}
    \hat{\chi}_{\vec{r}_\mu,\vec{r}_\nu'}^\dag \hat{\chi} _{\vec{r}_\mu,\vec{r}_\nu'} &= n_{\vec{r}_\nu'}+b_{\vec{r}_\mu \downarrow}^{\dagger}  b_{\vec{r}_\nu' \downarrow}^{\dagger}  b_{\vec{r}_\mu \downarrow}  b_{\vec{r}_\nu' \downarrow}+b_{\vec{r}_\mu \downarrow}^{\dagger}  b_{\vec{r}_\nu' \uparrow}^{\dagger}  b_{\vec{r}_\mu \uparrow}  b_{\vec{r}_\nu'\downarrow} + b_{\vec{r}_\mu \uparrow}^{\dagger} b_{\vec{r}_\nu' \downarrow}^{\dagger}  b_{\vec{r}_\mu \downarrow}  b_{\vec{r}_\nu' \uparrow}+b_{\vec{r}_\nu' \uparrow}^{\dagger}  b_{\vec{r}_\nu' \uparrow}^{\dagger}  b_{\vec{r}_\nu \uparrow}  b_{\vec{r}_\nu' \uparrow} \\
    \hat{\Delta}_{\vec{r}_\mu,\vec{r}_\nu'}^\dag  \hat{\Delta}_{\vec{r}_\mu,\vec{r}_\nu'} &= b_{\vec{r}_\mu \downarrow}^{\dagger}   b_{\vec{r}_\nu' \uparrow}^{\dagger}  b_{\vec{r}_\mu \downarrow}  b_{\vec{r}_\nu' \uparrow}-b_{\vec{r}_\mu \downarrow}^{\dagger} b_{\vec{r}_\nu' \uparrow}^{\dagger}  b_{\vec{r}_\mu \uparrow}  b_{\vec{r}_\nu' \downarrow}-b_{\vec{r}_\mu \uparrow}^{\dagger}  b_{\vec{r}_\nu' \downarrow}^{\dagger}  b_{\vec{r}_\mu \downarrow}  b_{\vec{r}_\nu' \uparrow}+b_{\vec{r}_\mu \uparrow}^{\dagger}  b_{\vec{r}_\nu' \downarrow}^{\dagger}  b_{\vec{r}_\mu \uparrow}  b_{\vec{r}_\nu' \downarrow} \\
    \hat{E}^{x \dag}_{\vec{r}_\mu,\vec{r}_\nu'} \hat{E}^x_{\vec{r}_\mu,\vec{r}_\nu'} &=n_{\vec{r}_\nu'}+b_{\vec{r}_\mu \downarrow}^{\dagger}  b_{\vec{r}_\nu' \downarrow}^{\dagger}  b_{\vec{r}_\mu \uparrow}  b_{\vec{r}_\nu' \uparrow}+b_{\vec{r}_\mu \downarrow}^{\dagger}  b_{\vec{r}_\nu' \uparrow}^{\dagger}  b_{\vec{r}_\mu \downarrow}  b_{\vec{r}_\nu' \uparrow}+b_{\vec{r}_\nu' \uparrow}^{\dagger}  b_{\vec{r}_\nu' \downarrow}^{\dagger}  b_{\vec{r}_\mu \uparrow}  b_{\vec{r}_\nu' \downarrow}+b_{\vec{r}_\nu' \uparrow}^{\dagger}  b_{\vec{r}_\nu' \uparrow}^{\dagger}  b_{\vec{r}_\nu \downarrow}  b_{\vec{r}_\nu' \downarrow} \\
    \hat{E}^{y \dag}_{\vec{r}_\mu,\vec{r}_\nu'} \hat{E}^y_{\vec{r}_\mu,\vec{r}_\nu'} &= n_{\vec{r}_\nu'}-b_{\vec{r}_\mu \downarrow}^{\dagger}  b_{\vec{r}_\nu' \downarrow}^{\dagger} b_{\vec{r}_\mu \uparrow} b_{\vec{r}_\nu' \uparrow}+b_{\vec{r}_\mu \downarrow}^{\dagger}  b_{\vec{r}_\nu' \uparrow}^{\dagger}  b_{\vec{r}_\mu \downarrow}  b_{\vec{r}_\nu' \uparrow}+b_{\vec{r}_\nu' \uparrow}^{\dagger}  b_{\vec{r}_\nu' \downarrow}^{\dagger}  b_{\vec{r}_\mu \uparrow}  b_{\vec{r}_\nu' \downarrow}-b_{\vec{r}_\mu \uparrow}^{\dagger}  b_{\vec{r}_\nu' \uparrow}^{\dagger}  b_{\vec{r}_\mu \downarrow}  b_{\vec{r}_\nu' \downarrow} \\
    \hat{E}^{z \dag}_{\vec{r}_\mu,\vec{r}_\nu'} \hat{E}^z_{\vec{r}_\mu,\vec{r}_\nu'} &= n_{\vec{r}_\nu'}+
    b_{\vec{r}_\mu \downarrow}^{\dagger}  b_{\vec{r}_\nu' \downarrow}^{\dagger}  b_{\vec{r}_\mu \downarrow}   b_{\vec{r}_\nu' \downarrow} -b_{\vec{r}_\mu \downarrow}^{\dagger}  b_{\vec{r}_\nu' \uparrow}^{\dagger}  b_{\vec{r}_\mu \uparrow}  b_{\vec{r}_\nu' \downarrow}-b_{\vec{r}_\mu \uparrow}^{\dagger}  b_{\vec{r}_\nu' \downarrow}^{\dagger}  b_{\vec{r}_\mu \downarrow}  b_{\vec{r}_\nu' \uparrow}+b_{\vec{r}_\mu \uparrow}^{\dagger}  b_{\vec{r}_\nu' \uparrow}^{\dagger}  b_{\vec{r}_\mu \uparrow}  b_{\vec{r}_\nu' \uparrow}\\
    \hat{D}^{x \dag}_{\vec{r}_\mu,\vec{r}_\nu'} D^x_{\vec{r}_\mu,\vec{r}_\nu'} &= b_{\vec{r}_\mu \downarrow}^{\dagger} b_{\vec{r}_\nu' \downarrow}^{\dagger}  b_{\vec{r}_\mu \downarrow}  b_{\vec{r}_\nu' \downarrow}-b_{\vec{r}_\mu \downarrow}^{\dagger} b_{\vec{r}_\nu' \downarrow}^{\dagger}  b_{\vec{r}_\mu \uparrow}  b_{\vec{r}_\nu' \uparrow}-b_{\vec{r}_\mu \uparrow}^{\dagger}  b_{\vec{r}_\nu' \uparrow}^{\dagger}  b_{\vec{r}_\mu \downarrow}  b_{\vec{r}_\nu' \downarrow}+b_{\vec{r}_\mu \uparrow}^{\dagger} b_{\vec{r}_\nu' \uparrow}^{\dagger} b_{\vec{r}_\mu \uparrow}  b_{\vec{r}_\nu' \uparrow} \\
    \hat{D}^{y \dag}_{\vec{r}_\mu,\vec{r}_\nu'} \hat{D}^y_{\vec{r}_\mu,\vec{r}_\nu'} &= b_{\vec{r}_\mu \downarrow}^{\dagger}  b_{\vec{r}_\nu' \downarrow}^{\dagger}  b_{\vec{r}_\mu \downarrow}  b_{\vec{r}_\nu' \downarrow} +b_{\vec{r}_\mu \downarrow}^{\dagger}  b_{\vec{r}_\nu' \downarrow}^{\dagger}  b_{\vec{r}_\mu \uparrow}  b_{\vec{r}_\nu' \uparrow}+b_{\vec{r}_\mu \uparrow}^{\dagger}  b_{\vec{r}_\nu' \uparrow}^{\dagger}  b_{\vec{r}_\mu \downarrow}  b_{\vec{r}_\nu' \downarrow} +b_{\vec{r}_\mu \uparrow}^{\dagger}  b_{\vec{r}_\nu' \uparrow}^{\dagger}  b_{\vec{r}_\mu \uparrow}  b_{\vec{r}_\nu' \uparrow} \\
    \hat{D}^{z \dag}_{\vec{r}_\mu,\vec{r}_\nu'} \hat{D}^z_{\vec{r}_\mu,\vec{r}_\nu'} &= b_{\vec{r}_\mu \downarrow}^{\dagger}  b_{\vec{r}_\nu' \uparrow}^{\dagger}  b_{\vec{r}_\mu \downarrow}  b_{\vec{r}_\nu' \uparrow}+b_{\vec{r}_\mu \downarrow}^{\dagger} b_{\vec{r}_\nu' \uparrow}^{\dagger}  b_{\vec{r}_\mu \uparrow}  b_{\vec{r}_\nu' \downarrow}+b_{\vec{r}_\mu \uparrow}^{\dagger}  b_{\vec{r}_\nu' \downarrow}^{\dagger} b_{\vec{r}_\mu \downarrow}  b_{\vec{r}_\nu' \uparrow}+b_{\vec{r}_\mu \uparrow}^{\dagger}  b_{\vec{r}_\nu' \downarrow}^{\dagger}  b_{\vec{r}_\mu \uparrow}  b_{\vec{r}_\nu' \downarrow},
  \end{align}
\end{subequations}
\end{widetext}
where $n_{\vec{r}_\nu'}= b^\dag_{\vec{r}_\nu'\downarrow}b_{\vec{r}_\nu'\downarrow} + b^\dag_{\vec{r}_\nu'\uparrow}b_{\vec{r}_\nu'\uparrow}$. Similarly, the term proportional to $\mathcal{J}_{z}$ in \eqref{eq: Hamiltonian nn pyrochlore} are
\begin{align} \label{eq: appendix Hamiltonian terms as schwinger bosons}
    \mathcal{J}_{z} \tau_{\vec{r}_\mu }^{z} \tau_{\vec{r}_\nu'}^{z} =& \frac{\mathcal{J}_{z}}{4} \left( b_{\vec{r}_\mu\uparrow}^\dag b_{\vec{r}_\nu'\uparrow}^\dag b_{\vec{r}_\mu\uparrow} b_{\vec{r}_\nu'\uparrow} -b_{\vec{r}_\mu\downarrow}^\dag b_{\vec{r}_\nu'\uparrow}^\dag b_{\vec{r}_\mu\downarrow} b_{\vec{r}_\nu'\uparrow} \right. \nonumber \\
    &\left. -b_{\vec{r}_\mu\uparrow}^\dag b_{\vec{r}_\nu'\downarrow}^\dag b_{\vec{r}_\mu\uparrow} b_{\vec{r}_\nu'\downarrow} + b_{\vec{r}_\mu\downarrow}^\dag b_{\vec{r}_\nu'\downarrow}^\dag b_{\vec{r}_\mu\downarrow} b_{\vec{r}_\nu'\downarrow}\right)
\end{align}

As a consequence of our canonical ordering, we see that there are only nine different inequivalent terms. Thus, a system of linear equations can be solved to rewrite each term in the Hamiltonian as a product of bond operators. By doing so, one finds the general solution
\begin{subequations} \label{eq: appendix Hamiltonian terms as bonds operators General Form}
  \begin{align}
    \tau_{\vec{r}_\mu }^{z} \tau_{\vec{r}_\nu'}^{z} &= \frac{1}{4}\left[ A \hat{\chi}_{\vec{r}_\mu,\vec{r}_\nu'}^\dag \hat{\chi}_{\vec{r}_\mu,\vec{r}_\nu'} + B \hat{\Delta}_{\vec{r}_\mu,\vec{r}_\nu'}^\dag  \hat{\Delta}_{\vec{r}_\mu,\vec{r}_\nu'} \right. \nonumber \\
    &+ C^x  \hat{E}^{x \dag}_{\vec{r}_\mu,\vec{r}_\nu'} \hat{E}^x_{\vec{r}_\mu,\vec{r}_\nu'} + C^y  \hat{E}^{y \dag}_{\vec{r}_\mu,\vec{r}_\nu'} \hat{E}^y_{\vec{r}_\mu,\vec{r}_\nu'}\nonumber \\
    &+C^z  \hat{E}^{z \dag}_{\vec{r}_\mu,\vec{r}_\nu'} \hat{E}^z_{\vec{r}_\mu,\vec{r}_\nu'} + \left( 1 - A + B + C^x\right)  \hat{D}^{x \dag}_{\vec{r}_\mu,\vec{r}_\nu'} \hat{D}^x_{\vec{r}_\mu,\vec{r}_\nu'} \nonumber\\
    &+\left( 1 - A + B + C^y \right) \hat{D}^{y \dag}_{\vec{r}_\mu,\vec{r}_\nu'} \hat{D}^y_{\vec{r}_\mu,\vec{r}_\nu'} \nonumber\\
    &+ \left( -A + B + C^z \right) \hat{D}^{z \dag}_{\vec{r}_\mu,\vec{r}_\nu'} \hat{D}^z_{\vec{r}_\mu,\vec{r}_\nu'} \nonumber\\
    & + \left( -A - C^x - C^y - C^z \right)n_{\vec{r}_\nu'} \nonumber\\
    & \left. + \left( -1 + A - 2 B - C^x - C^y - C^z \right) n_{\vec{r}_\mu} n_{\vec{r}_\nu'} \right]
    \end{align}
\end{subequations}
where $A$, $B$, $C^x$, $C^y$, and $C^z$ are arbitrary real constants to be fixed.

\subsection{Choice of decoupling}

For this work, we made the specific choice of decouplings presented in Eq. \eqref{eq: MF decoupling for all terms in H pyrochlore nn}. This choice is far from unique. It is well documented that, once a MF approximation has been carried out, the choice of decoupling has a significant impact on the results \cite{mezio2011test}. Therefore, one usually chooses a decoupling that reproduces some analytical or reliable numerical results in a certain limit of the regime of interest. As such a limit does not exist in our case, our MF decoupling follows from two main considerations: (i) ensuring that the MF energy is bounded from below (stable decoupling), and (ii) making sure that every term in our Hamiltonian, i.e., $\mathcal{J}_{x}$, $\mathcal{J}_{y}$ and $\mathcal{J}_{z}$, contributes to the energy of each PSG class (see table \ref{tab: independant non-zero MF parameters}). Even though the stability requirement is not an absolute necessity per se, we would like to argue that it is of great practical advantage as one can solve the self-consistency equations robustly with minimization algorithms (e.g., simulated annealing, gradient descent) without explicitly considering the bounds on the MF parameters (see appendix \ref{appendix: Bounds of the MF parameters}). Moreover, by using different MF decouplings over parameter space, it is easy to spot any artifact caused by a specific choice of decoupling as these will yield inconsistent values of the observables (e.g., MF parameters, MF energy) at the boundary between regions where different decouplings are used.

\section{\label{appendix: Bounds of the MF parameters} Bounds of the MF parameters}

We prove in this appendix that the modulus of the MF parameters $|\Delta|$, $|D^a|$, $|\chi|$ and $|E^a|$ that solve the self-consistency equations have an upper bound. From the general inequality for arbitrary operators $\hat{u}$ and $\hat{v}$
\begin{equation}
    \left| \expval{\hat{u} \hat{v}} \right| \le \frac{\expval{\hat{u} \hat{u}^\dag} + \expval{\hat{v}^\dag \hat{v}}}{2},
\end{equation}
we have from our definitions in Eqs. \eqref{eq: definitions bond operators} that 
\begin{subequations}
\begin{align}
\left|\left\langle\widehat{\Delta}_{\vec{r}_\mu,\vec{r}_\nu'}\right\rangle\right| &\le \frac{\left\langle\widehat{n}_{\vec{r}_{\mu}}+\widehat{n}_{\vec{r}_{\nu}'}+2\right\rangle}{2}\\
\left|\left\langle\widehat{D}^a_{\vec{r}_\mu,\vec{r}_\nu'}\right\rangle\right| &\le \frac{\left\langle\widehat{n}_{\vec{r}_{\mu}}+\widehat{n}_{\vec{r}_{\nu}'}+2\right\rangle}{2}\\
\quad\left|\left\langle\widehat{\chi}_{\vec{r}_\mu,\vec{r}_\nu'} \right\rangle\right| &\le \frac{\left\langle\widehat{n}_{\vec{r}_\mu}+\widehat{n}_{\vec{r}_\nu'}\right\rangle}{2} \\
\quad\left|\left\langle\widehat{E}^a_{\vec{r}_\mu,\vec{r}_\nu'} \right\rangle\right| &\le \frac{\left\langle\widehat{n}_{\vec{r}_\mu}+\widehat{n}_{\vec{r}_\nu'}\right\rangle}{2} .
\end{align}
\end{subequations}
Then using Eq. \eqref{eq: single-occupancy constraint}, this leads to the constraints
\begin{align}
\left| \expval{\hat{\Delta}_{\vec{r}_\mu,\vec{r}_\nu'}} \right| &\le \left( \kappa + 1 \right)  \\
\left| \expval{\hat{D}^a_{\vec{r}_\mu,\vec{r}_\nu'}} \right| &\le \left( \kappa + 1 \right)  \\
\left| \expval{\hat{\chi}_{\vec{r}_\mu,\vec{r}_\nu'}} \right| &\le \kappa\\
\left| \expval{\hat{E}^a_{\vec{r}_\mu,\vec{r}_\nu'}} \right| &\le \kappa
\end{align}
after solving the self-consistency equations.

\section{\label{subsec: Transformation of the Mean-field ansätz} Transformation of the mean-field ansätz}

The transformation properties of the MF ansätz are necessary to determine the algebraic PSG constraints from the SG transformations, and the MF properties of a given PSG solution. The form of the ansätz after such operations can be deduced by first finding how the spinon operators transform and then using the fact that, by definition, the Hamiltonian must be invariant under these symmetry operations: the gauge-enriched operators $\widetilde{\mathcal{O}}$ and $\widetilde{\mathcal{T}}$ are symmetries of the MF Hamiltonian ($\widetilde{\mathcal{O}}:H_{MF}\to H_{MF}$ and $\widetilde{\mathcal{T}}:H_{MF}\to H_{MF}$). From the general form of the MF Hamiltonian in Eq. \eqref{eq: MF Hamiltonian generic form}, and the spinon operators transformations \eqref{eq: spinon transfromations under SG} and \eqref{eq: spinon transfromations under TRS}, the pairing and hopping matrices must transform as
\begin{subequations} \label{eq: transformation pairing and hopping matrices under SG}
  \begin{align}
   u_{\mathcal{O}(\vec{r}_\mu),\mathcal{O}(\vec{r}_\nu)'}^p &= U_{\mathcal{O}}^*  u_{\vec{r}_\mu,\vec{r}_\nu'}^p   U_\mathcal{O}^{\dag} e^{i\left( \phi_{\mathcal{O}}(\mathcal{O}(\vec{r}_\mu)) + \phi_{\mathcal{O}}(\mathcal{O}(\vec{r}_\nu')) \right)}  \label{eq: transformation pairing matrix under SG}\\  u_{\mathcal{O}(\vec{r}_\mu),\mathcal{O}(\vec{r}_\nu)'}^h &= U_{\mathcal{O}} u_{\vec{r}_\mu,\vec{r}_\nu'}^h U_\mathcal{O}^{\dag} e^{i\left(- \phi_{\mathcal{O}}(\mathcal{O}(\vec{r}_\mu)) + \phi_{\mathcal{O}}(\mathcal{O}(\vec{r}_\nu')) \right)}\label{eq: transformation hopping matrix under SG}
  \end{align}
\end{subequations}
under space group operations $\widetilde{\mathcal{O}}$ and as
\begin{subequations}  \label{eq: transformation pairing and hopping matrices under TR}
    \begin{align}
     u_{\vec{r}_\mu,\vec{r}_\nu'}^p  &= U_{\mathcal{T}}^* \left(u_{\vec{r}_\mu,\vec{r}_\nu'}^p\right)^*   U_\mathcal{T}^\dag e^{i\left( \phi_{\mathcal{T}}(\vec{r}_\mu) + \phi_{\mathcal{T}}(\vec{r}_\nu') \right)} \label{eq: transformation pairing matrix under TR} \\
     u_{\vec{r}_\mu,\vec{r}_\nu'}^h &=  U_{\mathcal{T}} \left(u_{\vec{r}_\mu,\vec{r}_\nu'}^h\right)^* U_\mathcal{T}^\dag e^{i\left( -\phi_{\mathcal{T}}(\vec{r}_\mu) + \phi_{\mathcal{T}}(\vec{r}_\nu') \right)} \label{eq: transformation hopping matrix under TR}
    \end{align}
\end{subequations}
under time-reversal symmetry $\widetilde{\mathcal{T}}$ for the MF Hamiltonian to be invariant. 

If we parametrize the $u^h$ and $u^p$ matrices in Eq. \eqref{eq: MF Hamiltonian generic form} by
\begin{subequations} \label{eq: u matrices parametrization}
    \begin{align}
    u_{\vec{r}_\mu,\vec{r}_\nu'}^p =& a^{p}_{\vec{r}_\mu,\vec{r}_\nu'} i \sigma^y + b^{p}_{\vec{r}_\mu,\vec{r}_\nu'} i\sigma^y\sigma^x \nonumber\\
    &+ c^{p}_{\vec{r}_\mu,\vec{r}_\nu'} i\sigma^y\sigma^y + d^{p}_{\vec{r}_\mu,\vec{r}_\nu'} i\sigma^y\sigma^z  \\
    u_{\vec{r}_\mu,\vec{r}_\nu'}^h =& a^{h}_{\vec{r}_\mu,\vec{r}_\nu'} \mathds{1}_{2\times 2} +  b^{h}_{\vec{r}_\mu,\vec{r}_\nu'} \sigma^x \nonumber \\
    &+ c^{h}_{\vec{r}_\mu,\vec{r}_\nu'} \sigma^y + d^{h}_{\vec{r}_\mu,\vec{r}_\nu'} \sigma^z  
    \end{align}
\end{subequations}
where $a^p_{\vec{r}_\mu,\vec{r}_\nu'}$, $b^p_{\vec{r}_\mu,\vec{r}_\nu'}$, $c^p_{\vec{r}_\mu,\vec{r}_\nu'}$, $d^p_{\vec{r}_\mu,\vec{r}_\nu'}$, $a^h_{\vec{r}_\mu,\vec{r}_\nu'}$, $b^h_{\vec{r}_\mu,\vec{r}_\nu'}$, $c^h_{\vec{r}_\mu,\vec{r}_\nu'}$ and $d^h_{\vec{r}_\mu,\vec{r}_\nu'}$ are real linear combinations of the MF parameters ($\chi,\Delta,E^x,E^y,E^z,D^x,D^y,D^z$) with the coupling constants ($\mathcal{J}_{x}$, $\mathcal{J}_{y}$ and $\mathcal{J}_{z}$) as prefactors, the space group and time-reversal operations (without the gauge transformations $G_{\mathcal{O}}$ and $G_{\mathcal{T}}$) then correspond to $O(4)$ matrices acting on $(a^{p}_{\vec{r}_\mu,\vec{r}_\nu'}, b^{p}_{\vec{r}_\mu,\vec{r}_\nu'}, c^{p}_{\vec{r}_\mu,\vec{r}_\nu'}, d^{p}_{\vec{r}_\mu,\vec{r}_\nu'})^T$ and $(a^{h}_{\vec{r}_\mu,\vec{r}_\nu'},b^{h}_{\vec{r}_\mu,\vec{r}_\nu'},c^{h}_{\vec{r}_\mu,\vec{r}_\nu'},d^{h}_{\vec{r}_\mu,\vec{r}_\nu'})^T$ as
\begin{align} \label{eq: O(4) transformations matrix}
    &\mathcal{R}^{\overline{C}_6}= \mathds{1}_{4\times 4}  ,\hspace{2mm} 
    \mathcal{R}^{S}= 
    \begin{pmatrix}
    1 & 0 & 0 & 0 \\
    0 & -1 & 0  & 0 \\
    0 & 0 & 1 & 0 \\
    0 & 0 & 0 & -1 
    \end{pmatrix}, \nonumber \\ &\mathcal{R}^{T_i}=\mathds{1}_{4\times 4}, \hspace{3mm}  \mathcal{R}^{\mathcal{T}}=
    \begin{pmatrix}
    1 & \\
     & -\mathds{1}_{3 \times 3} 
    \end{pmatrix} \mathcal{K}
\end{align}

\section{\label{appendix: Classification of symmetric $U(1)$ spin liquids} Classification of symmetric \texorpdfstring{$U(1)$}{U(1)} spin liquids}

\subsection{\label{appendix subsec: PSG solution - Generalities} Generalities}

To classify all PSG classes, one starts from all space group algebraic constraints
\begin{equation} \label{eq: algebraic SG relations}
\mathcal{O}_{1} \circ \mathcal{O}_{2} \circ \cdots=1
\end{equation}
which translate directly to the gauge-enriched relations
\begin{equation} \label{eq: algebraic SG relations gauge-enriched}
\widetilde{\mathcal{O}}_{1} \circ \widetilde{\mathcal{O}}_{2} \circ \cdots=\left(G_{\mathcal{O}_{1}} \circ \mathcal{O}_{1}\right) \circ\left(G_{\mathcal{O}_{2}} \circ \mathcal{O}_{2}\right) \circ \cdots=\mathcal{G},
\end{equation}
where $\mathcal{G}$ is a pure gauge transformation. All these gauge-enriched constraints correspond to phase relations of the form
\begin{equation}
\begin{aligned}\label{eq: phase equations for PSG}
&\phi_{\mathcal{O}_{1}}\left(\vec{r}_{\mu}\right) +\phi_{\mathcal{O}_{2}}\left[\mathcal{O}_{1}^{-1}\left(\vec{r}_{\mu}\right)\right] \\
&+\phi_{\mathcal{O}_{3}}\left[\mathcal{O}_{2}^{-1}\circ\mathcal{O}_{1}^{-1}\left(\vec{r}_{\mu}\right)\right]+\cdots= \psi\hspace{3mm} \text{mod}2\pi,
\end{aligned}
\end{equation}
with $\psi\in\text{IGG}$, where we have made use of the conjugation relation
\begin{equation}
\mathcal{O}_{i} \circ G_{\mathcal{O}_{j}} \circ \mathcal{O}_{i}^{-1}: b_{\vec{r}_{\mu}} \rightarrow e^{i \phi_{\mathcal{O}_{j}}\left[\mathcal{O}_{i}^{-1}\left(\vec{r}_{\mu}\right)\right]} b_{\vec{r}_{\mu}}.
\end{equation}
To get the PSG constraints for algebraic relations that include time-reversal, it is possible to proceed in the exact same way by using the conjugation relation
\begin{equation}
\begin{aligned}
\mathcal{T} \circ & G_{\mathcal{O}} \circ \mathcal{T}^{-1} = \mathcal{T} \circ G_{\mathcal{O}} \circ \mathcal{T}: \\
& b_{\vec{r}_{\mu}} \rightarrow \mathcal{K} U_{\mathcal{T}} e^{i \phi_{\mathcal{O}}\left(\vec{r}_{\mu}\right)} U_{\mathcal{T}} \mathcal{K} b_{\vec{r}_{\mu}}=e^{-i \phi_{\mathcal{O}}\left(\vec{r}_{\mu}\right)} b_{\vec{r}_{\mu}}.
\end{aligned}
\end{equation}

The PSG classes for a given IGG are then obtained by listing the gauge inequivalent solutions of all phases equations of the form \eqref{eq: phase equations for PSG}. That is, it must be impossible to relate two distinct PSG classes by a general gauge transformation $G$ that transforms the phase factor as
\begin{align}
    \phi_{\mathcal{O}}(\vec{r}_\mu) \to \phi_{\mathcal{O}}(\vec{r}_\mu) + \phi_{G}(\vec{r}_\mu) - \phi_{G}(\mathcal{O}^{-1}(\vec{r}_\mu))
\end{align}
All gauge degrees of freedom must be fixed in the process of solving the algebraic equations. In our case, there are four distinct gauge transformations for each sublattice ($\mu\in\{0,1,2,3\}$ ) in every direction ($r_1$, $r_2$ and $r_3$)
\begin{align}
    G_{i,\mu}: \phi_{G_i}(\vec{r}_\mu) = \psi_{G_i,\mu} r_i,
\end{align}
one constant gauge transformations for every sublattice 
\begin{align}
    G_{\nu}^{cst}: \phi_{G_i}(\vec{r}_\mu) = \psi_{\nu} \delta_{\mu,\nu},
\end{align}
where $\psi_{G_{i},\mu}$ and $\psi_{\nu}$ are elements of the IGG. We are also free to add a site independent phase factor that is an element of the IGG to any of our five SG phases. Therefore, a total of 16 local gauge and 5 IGG phase factors have to be fixed to get unambiguously inequivalent results \cite{schneider2021projective}.

\subsection{\label{appendix subsec: PSG solution - Algebraic constraints} Algebraic constraints}

For the pyrochlore lattice, the algebraic constraints are
\begin{subequations} \label{eq: SG generators constraints}
  \begin{align}
    T_{i} T_{i+1} T_{i}^{-1} T_{i+1}^{-1} &=1,  i=1,2,3 \\
    \bar{C}_{6}^{6} &=1, \\
    S^{2} T_{3}^{-1} &=1,  \\
    \bar{C}_{6} T_{i} \bar{C}_{6}^{-1} T_{i+1} &=1, i=1,2,3 \\
    S T_{i} S^{-1} T_{3}^{-1} T_{i} &=1,  i=1,2, \\
    S T_{3} S^{-1} T_{3}^{-1} &=1  \\
    \left(\bar{C}_{6} S\right)^{4} &=1 \\
    \left(\bar{C}_{6}^{3} S\right)^{2} &=1,  \\
    \mathcal{T}^{2} &=-1  \\
    \mathcal{T O T}^{-1} \inv{\mathcal{O}} &=1, \forall \mathcal{O} \in\text{SG} .
  \end{align}
\end{subequations}
These commutation relations are also respected for the pseudospin transformations as can be verified from Eq. \eqref{eq: matrix form pseudospin trfs}. The corresponding PSG equations are
\begin{subequations}
\begin{align*}
\left(G_{T_{i}} T_{i}\right)\left(G_{T_{i+1}} T_{i+1}\right)\left(G_{T_{i}} T_{i}\right)^{-1}\left(G_{T_{i+1}} T_{i+1}\right)^{-1}&\in IGG, \\
\left(G_{\bar{C}_{\theta}} \bar{C}_{6}\right)^{6} &\in IGG, \\
\left(G_{S} S\right)^{2}\left(G_{T_{3}} T_{3}\right)^{-1} &\in IGG,\\
\left(G_{\bar{C}_{6}} \bar{C}_{6}\right)\left(G_{T_{i}} T_{i}\right)\left(G_{\bar{C}_{6}} \bar{C}_{6}\right)^{-1}\left(G_{T_{i+1}} T_{i+1}\right) &\in IGG, \\
\left(G_{S} S\right)\left(G_{T_{i}} T_{i}\right)\left(G_{S} S\right)^{-1}\left(G_{T_{3}} T_{3}\right)^{-1}\left(G_{T_{i}} T_{i}\right) &\in IGG, \\
\left(G_{S} S\right)\left(G_{T_{3}} T_{3}\right)\left(G_{S} S\right)^{-1}\left(G_{T_{3}} T_{3}\right)^{-1} &\in IGG, \\
\left[\left(G_{\bar{C}_{6}} \bar{C}_{6}\right)\left(G_{S} S\right)\right]^{4} &\in IGG,\\
\left[\left(G_{\bar{C}_{6}} \bar{C}_{6}\right)^{3}\left(G_{S} S\right)\right]^{2} &\in IGG .
\end{align*}
\end{subequations}
In the case where $\text{IGG}=U(1)$, these constraints are explicitly
\begin{widetext}
\begin{subequations} 
\begin{empheq}[]{align}
\phi_{T_{i}}\left(\vec{r}_{\mu}\right)+\phi_{T_{i+1}}\left[T_{i}^{-1}\left(\vec{r}_{\mu}\right)\right]-\phi_{T_{i}}\left[T_{i+1}^{-1}\left(\vec{r}_{\mu}\right)\right]-\phi_{T_{i+1}}\left(\vec{r}_{\mu}\right) &=\psi_{T_i},\label{eq: psg classification T_i}\\
\phi_{\bar{C}_{6}}\left(\vec{r}_{\mu}\right)+\phi_{\bar{C}_{6}}\left[\bar{C}_{6}^{-1}\left(\vec{r}_{\mu}\right)\right]+\phi_{\bar{C}_{6}}\left[\bar{C}_{6}^{-2}\left(\vec{r}_{\mu}\right)\right]+\phi_{\bar{C}_{6}}\left[\bar{C}_{6}^{-3}\left(\vec{r}_{\mu}\right)\right]+\phi_{\bar{C}_{6}}\left[\bar{C}_{6}^{-4}\left(\vec{r}_{\mu}\right)\right]+\phi_{\bar{C}_{6}}\left[\bar{C}_{6}^{-5}\left(\vec{r}_{\mu}\right)\right] &=\psi_{\bar{C}_{6}} \label{eq: psg classification C} \\
\phi_{S}\left(\vec{r}_{\mu}\right)+\phi_{S}\left[S^{-1}\left(\vec{r}_{\mu}\right)\right]-\phi_{T_{3}}\left(\vec{r}_{\mu}\right) &=\psi_{S} \label{eq: psg classification S}  \\
\phi_{\bar{C}_{6}}\left(\vec{r}_{\mu}\right)+\phi_{T_{i}}\left[\bar{C}_{6}^{-1}\left(\vec{r}_{\mu}\right)\right]-\phi_{\bar{C}_{6}}\left[T_{i+1}\left(\vec{r}_{\mu}\right)\right]+\phi_{T_{i+1}}\left[T_{i+1}\left(\vec{r}_{\mu}\right)\right] &=\psi_{\bar{C}_{6} T_{i}} \label{eq: psg classification CT_i}\\
\phi_{S}\left(\vec{r}_{\mu}\right)+\phi_{T_{i}}\left[S^{-1}\left(\vec{r}_{\mu}\right)\right]-\phi_{S}\left[T_{3}^{-1} T_{i}\left(\vec{r}_{\mu}\right)\right]-\phi_{T_{3}}\left[T_{i}\left(\vec{r}_{\mu}\right)\right]+\phi_{T_{i}}\left[T_{i}\left(\vec{r}_{\mu}\right)\right]&=\psi_{S T_{i}} \label{eq: psg classification S T_i}  \\
\phi_{S}\left(\vec{r}_{\mu}\right)+\phi_{T_{3}}\left[S^{-1}\left(\vec{r}_{\mu}\right)\right]-\phi_{S}\left[T_{3}^{-1}\left(\vec{r}_{\mu}\right)\right]-\phi_{T_{3}}\left(\vec{r}_{\mu}\right) &=\psi_{S T_{3}} \label{eq: psg classification S T_3} \\
\phi_{S_{6}}\left(\vec{r}_{\mu}\right)+\phi_{S}\left[\bar{C}_{6}^{-1}\left(\vec{r}_{\mu}\right)\right]+\phi_{\bar{C}_{6}}\left[\left(\bar{C}_{6} S\right)^{-1}\left(\vec{r}_{\mu}\right)\right]+\phi_{S}\left[\left(\bar{C}_{6} S \bar{C}_{6}\right)^{-1}\left(\vec{r}_{\mu}\right)\right]+\phi_{\bar{C}_{6}}\left[\left(\bar{C}_{6} S \bar{C}_{6} S\right)^{-1}\left(\vec{r}_{\mu}\right)\right] & \nonumber\\
+\phi_{S}\left[\left(\bar{C}_{6} S \bar{C}_{6} S \bar{C}_{6}\right)^{-1}\left(\vec{r}_{\mu}\right)\right]+\phi_{\bar{C}_{6}}\left[\left(\bar{C}_{6} S \bar{C}_{6} S \bar{C}_{6} S\right)^{-1}\left(\vec{r}_{\mu}\right)\right]+\phi_{S}\left[\left(\bar{C}_{6} S \bar{C}_{6} S \bar{C}_{6} S \bar{C}_{6}\right)^{-1}\left(\vec{r}_{\mu}\right)\right] &=\psi_{\bar{C}_{6} S} \label{eq: psg classification CS} \\
\phi_{\bar{C}_{6}}\left(\vec{r}_{\mu}\right)+\phi_{\bar{C}_{6}}\left[\bar{C}_{6}^{-1}\left(\vec{r}_{\mu}\right)\right]+\phi_{\bar{C}_{6}}\left[\bar{C}_{6}^{-2}\left(\vec{r}_{\mu}\right)\right]+\phi_{S}\left[\bar{C}_{6}^{-3}\left(\vec{r}_{\mu}\right)\right] & \nonumber\\
+\phi_{\bar{C}_{6}}\left[\left(\bar{C}_{6}^{3} S\right)^{-1}\left(\vec{r}_{\mu}\right)\right]+\phi_{\bar{C}_{6}}\left[\left(\bar{C}_{6}^{3} S \bar{C}_{6}\right)^{-1}\left(\vec{r}_{\mu}\right)\right]+\phi_{\bar{C}_{6}}\left[\left(\bar{C}_{6}^{3} S \bar{C}_{6}^{2}\right)^{-1}\left(\vec{r}_{\mu}\right)\right]+\phi_{S}\left[S\left(\vec{r}_{\mu}\right)\right] &=\psi_{S \bar{C}_{6}}  \label{eq: psg classification SC}
\end{empheq}
\end{subequations}
\end{widetext}
where all $\psi$ are $U(1)$ factors, Eq. \eqref{eq: psg classification T_i} and \eqref{eq: psg classification CT_i} represent three equations ($i=1,2,3$) and Eq. \eqref{eq: psg classification S T_i} stands for two equations ($i=1,2$). It is important to note that all of the following equations are defined modulo $2\pi$. For simplicity's sake, we will not indicate that subtlety explicitly.

\subsection{\label{appendix subsec: PSG solution - Solution of the PSG constraints: inter-unit cell part} Solution of the PSG constraints: inter-unit cell part}

Let's first solve Eq. \eqref{eq: psg classification T_i}. Using our gauge freedom, we can set $\psi_{T_1}(r_1,r_2,r_3)_\mu=\psi_{T_2}(0,r_2,r_3)_\mu=\psi_{T1}(0,0,r_3)_\mu=0$, which then leads to 
\begin{subequations} \label{eq U(1) classification: T1, T2, T3 first equation}
  \begin{empheq}[left={ }\empheqlbrace]{align}
    \phi_{T_1}(\vec{r}_\mu) &= 0 \\
    \phi_{T_2}(\vec{r}_\mu) &= -\psi_{T_1} r_1\\
    \phi_{T_3}(\vec{r}_\mu) &= \psi_{T_3} r_1 - \psi_{T_2} r_2.
  \end{empheq}
\end{subequations}

Using this form for the translation phase factors, Eq. \eqref{eq: psg classification CT_i} yields 
\begin{align*}
    \phi_{\overline{C}_6}(r_1,r_2,r_3)_\mu -\phi_{\overline{C}_6}(r_1,r_2+1,r_3)_\mu -r_1 \psi_{T_1} &= \psi_{\overline{C}_6 T_1}  \\
    \phi_{\overline{C}_6}(r_1,r_2,r_3)_\mu -\phi_{\overline{C}_6}(r_1,r_2,r_3+1)_\mu & \\
    + \psi_{T_1} (r_2 + \delta_{\mu=2}) - \psi_{T_2} r_2 + \psi_{T_3} r_1 &= \psi_{\overline{C}_6 T_2}  \\
    \phi_{\overline{C}_6}(r_1,r_2,r_3)_\mu -\phi_{\overline{C}_6}(r_1+1,r_2,r_3)_\mu &\\
    + \psi_{T_2}(r_3 + \delta_{\mu=3}) -\psi_{T_3}(r_3 + \delta_{\mu=2}) &= \psi_{\overline{C}_6 T_3}.
\end{align*}
Solving these, we find that $\psi_{T_1}=\psi_{T_2}=\psi_{T_3}$ and
\begin{align} \label{eq U(1) classification: first equation for phi C6}
    \phi_{\overline{C}_6}(\vec{r}_\mu) =& \phi_{\overline{C}_6}(\vec{0}_\mu) - r_2 \psi_{\overline{C}_6 T_1} -r_3 (\psi_{\overline{C}_6 T_2} - \delta_{\mu=2}\psi_{T_1}) \nonumber \\
    &- r_1 (\psi_{\overline{C}_6 T_{3}} + (\delta_{\mu=2}-\delta_{\mu=3})\psi_{T_1}) \nonumber \\
    &- \psi_{T_1}(r_1 r_2 - r_1 r_3). 
\end{align}

For the screw axis phase factor, we can once again plug in the translation phase factors \eqref{eq U(1) classification: T1, T2, T3 first equation} in the constraints \eqref{eq: psg classification S T_i} and \eqref{eq: psg classification S T_3} to find
\begin{align*}
    \phi_{S}(r_1,r_2,r_3)_\mu -\phi_{S}(r_1+1,r_2,r_3-1)_\mu &\\
    + (-1-r_1+r_2)\psi_{T_1} &= \psi_{S T_1}  \\
    \phi_{S}(r_1,r_2,r_3)_\mu -\phi_{S}(r_1,r_2+1,r_3-1)_\mu &\\
    + (1-r_1+r_2+\delta_{\mu=1}) \psi_{T_1}  &= \psi_{S T_2}  \\
    \phi_{S}(r_1,r_2,r_3)_\mu -\phi_{S}(r_1+1,r_2,r_3)_\mu &\\
    +(2r_2-2r_1 - \delta_{\mu=1}+\delta_{\mu=2})\psi_{T_1} &= \psi_{S T_3}
\end{align*}
which can only be satisfied if $\psi_{T_1}=n_1 \pi$ with $n_1 \in\left\{0, 1 \right\}$ and 
\begin{align}\label{eq U(1) classification: first equation for phi S}
    \phi_{S}(\vec{r}_\mu) &= \phi_{S}(\vec{0}_\mu) -r_1 \psi_{ST_1} - r_2 \psi_{S T_2}  \nonumber \\
    &+ \frac{1}{2} n_1 \pi \left( -r_1 + r_2 + 2 r_1 r_2 - r_1^2 + r_2^2 + 2 \delta_{\mu=1} r_2 \right) \nonumber \\
    &+ (r_1 + r_2 + r_3) (n_1 \pi \delta_{\mu=1,2} + \psi_{ST_3} ).
\end{align}

Having found the general form for the five space group phase factors, we can plug these in the remaining equations to find all constraints. The finite order of the rotoreflection $\overline{C}_6$ expressed in \eqref{eq: psg classification C} imposes the two equations
\begin{subequations}
  \begin{align}
    6\phi_{\overline{C}_6}(\vec{0}_0) &= \psi_{\overline{C}_6} \\
    2( \phi_{\overline{C}_6}(\vec{0}_1) + \phi_{\overline{C}_6}(\vec{0}_2) + \phi_{\overline{C}_6}(\vec{0}_3) ) \nonumber \\
    + (\psi_{\overline{C}_6 T_1} + \psi_{\overline{C}_6 T_2} + \psi_{\overline{C}_6 T_3}) &=\psi_{\overline{C}_6}.
    \end{align}
\end{subequations}

Eq. \eqref{eq: psg classification S} gives the equations
\begin{align*}
    (-1+r_1 +r_2+2r_3) \psi_{S T_3} + \phi_S(\vec{0}_0) + \phi_S(\vec{0}_3) &= \psi_S \\
    n_1\pi + \psi_{S T_1} + 2 \phi_{S}(\vec{0}_1) + \psi_{S T_3} (-1+r_1+r_2 + 2r_3) &=\psi_S \\
    n_1\pi + \psi_{S T_2} + 2 \phi_{S}(\vec{0}_2) + \psi_{S T_3} (-1+r_1+r_2 + 2r_3) &=\psi_S 
\end{align*}
that, once simplified, lead to 
\begin{subequations}
  \begin{align}
    \psi_{ST_3} &=0\\
    \phi_{S}(\vec{0}_0)+ \phi_{S}(\vec{0}_3) &=\psi_S\\
    2 \phi_{S}(\vec{0}_1) + \psi_{ST_1} +n_1 \pi &= \psi_S \\
    2 \phi_{S}(\vec{0}_2) + \psi_{ST_2} +n_1 \pi &= \psi_S .
  \end{align}
\end{subequations}

At last, Eq. \eqref{eq: psg classification CS} and  \eqref{eq: psg classification SC} give
\begin{equation}
    \sum_{i=1}^3 \psi_{\overline{C}_6 T_i} + \sum_{j=0}^3 \left( \phi_{\overline{C}_6}(\vec{0}_j) + \phi_{S}(\vec{0}_j) \right) = \psi_{\overline{C}_6 S}
\end{equation}
and
\begin{subequations}
  \begin{align}
    3 \psi_{\overline{C}_6 T_1} - 3 \psi_{\overline{C}_6 T_2} + \psi_{\overline{C}_6 T_3} - 2 \psi_{S T_1} &= 0 \label{eq: psg inter unit cell constraint 1} \\
    3 \psi_{\overline{C}_6 T_1} +  \psi_{\overline{C}_6 T_2} - 3 \psi_{\overline{C}_6 T_3} - 2 \psi_{S T_2} &= 0 \label{eq: psg inter unit cell constraint 2} \\
    (\psi_{\overline{C}_6 T_2} + \psi_{\overline{C}_6 T_3}) + 3 \phi_{\overline{C}_6}(\vec{0}_0)  \hspace{1cm} &\nonumber \\
    + \sum_{i=1}^3 \phi_{\overline{C}_6}(\vec{0}_i)  + \phi_{S}(\vec{0}_0) + \phi_{S}(\vec{0}_3) &=\psi_{S\overline{C_6}} \\
    (2 \psi_{\overline{C}_6 T_1} + 2 \psi_{\overline{C}_6 T_3} )\hspace{2.5cm}&\nonumber \\
    +2 \left( \sum_{i=1}^3 \phi_{\overline{C}_6}(\vec{0}_i)  + \phi_{S}(\vec{0}_1)  \right) &= \psi_{S \overline{C}_6} \\
    (2 \psi_{\overline{C}_6 T_1} + 2 \psi_{\overline{C}_6 T_2})\hspace{2.5cm} &\nonumber \\
    +2 \left( \sum_{i=1}^3 \phi_{\overline{C}_6}(\vec{0}_i)  + \phi_{S}(\vec{0}_2)  \right) &= \psi_{S \overline{C}_6} .
  \end{align}
\end{subequations}

\subsection{\label{appendix subsec: PSG solution - Solution of the PSG constraints: time-reversal} Solution of the PSG constraints: time-reversal}

Turning our attention to the effect of TRS, from the constraint 
\begin{equation}
    (G_{\mathcal{T}}\mathcal{T})(G_{\mathcal{O}}\mathcal{O}) (G_{\mathcal{T}} \mathcal{T} )^{-1} (G_{\mathcal{O}} \mathcal{O})^{-1} \in \text{IGG}
\end{equation}
where $\mathcal{O}$ stands for all space group generators, we get the five equations
\begin{align}
\phi_{\mathcal{T}}\left(\vec{r}_{\mu}\right)-\phi_{\mathcal{T}}\left[T_{i}^{-1}\left(\vec{r}_{\mu}\right)\right]-2 \phi_{T_{i}}\left(\vec{r}_{\mu}\right) &= \psi_{\mathcal{T} T_{i}} \label{eq: psg classification TRS Ti} \\
\phi_{\mathcal{T}}\left(\vec{r}_{\mu}\right)-\phi_{\mathcal{T}}\left[\bar{C}_{6}^{-1}\left(\vec{r}_{\mu}\right)\right]-2 \phi_{\bar{C}_{6}}\left(\vec{r}_{\mu}\right) &= \psi_{\mathcal{T} \bar{C}_{6}}  \label{eq: psg classification TRS C}  \\
\phi_{\mathcal{T}}\left(\vec{r}_{\mu}\right)-\phi_{\mathcal{T}}\left[S^{-1}\left(\vec{r}_{\mu}\right)\right]-2 \phi_{S}\left(\vec{r}_{\mu}\right) &= \psi_{\mathcal{T} S}.   \label{eq: psg classification TRS S} 
\end{align}

Solving the three equations summarized in Eq.  \eqref{eq: psg classification TRS Ti}, we get
\begin{equation} \label{eq U(1) classification: first equation TRS}
    \phi_{\mathcal{T}}(\vec{r}_\mu) = \phi_{\mathcal{T}}(\vec{0}_\mu) + \sum_{i=1}^3 \psi_{\mathcal{T} T_i} r_i.
\end{equation}
Replacing this expression in \eqref{eq: psg classification TRS C} leads to the constraints
\begin{subequations}
\begin{align}
    2 \psi_{\overline{C}_6 T_1} + \psi_{\mathcal{T} T_1} + \psi_{\mathcal{T}T_2} &= 0 \label{eq: psg TRS C condition 1} \\
    2 \psi_{\overline{C}_6 T_3} + \psi_{\mathcal{T}T_1}  + \psi_{\mathcal{T}T_3} &= 0 \label{eq: psg TRS C condition 2} \\
     2 \psi_{\overline{C}_6 T_2} + \psi_{\mathcal{T}T_2} + \psi_{\mathcal{T}T_3} &= 0  \label{eq: psg TRS C condition 3} \\
    -2 \phi_{\overline{C}_6}(\vec{0}_0) &= \psi_{\mathcal{T} \overline{C}_6} \\
    \psi_{\mathcal{T}T_3}  - 2 \phi_{\overline{C}_6}(\vec{0}_1) - \phi_{\mathcal{T}}(\vec{0}_3) + \phi_{\mathcal{T}}(\vec{0}_1) &= \psi_{\mathcal{T} \overline{C}_6} \\
    \psi_{\mathcal{T}T_1} - 2 \phi_{\overline{C}_6}(\vec{0}_2) - \phi_{\mathcal{T}}(\vec{0}_1) + \phi_{\mathcal{T}}(\vec{0}_2)  &=\psi_{\mathcal{T} \overline{C}_6} \\
    \psi_{\mathcal{T}T_2}  - 2 \phi_{\overline{C}_6}(\vec{0}_3) - \phi_\mathcal{T}(\vec{0}_2) + \phi_\mathcal{T}(\vec{0}_3)  &= \psi_{\mathcal{T} \overline{C}_6} .
\end{align}
\end{subequations}
Doing the same with Eq. \eqref{eq: psg classification TRS S}, one finds
\begin{subequations}
\begin{align}
    2 \psi_{S T_1} + 2 \psi_{\mathcal{T} T_1} - \psi_{\mathcal{T} T_3} &= 0 \label{eq: psg TRS S condition 1} \\
    2 \psi_{S T_2} + 2 \psi_{\mathcal{T} T_2} - \psi_{\mathcal{T} T_3} &= 0 \label{eq: psg TRS S condition 2} \\
    \psi_{\mathcal{T} T_3} - 2 \phi_{S}(\vec{0}_0) + \phi_{\mathcal{T}} (\vec{0}_0) - \phi_{\mathcal{T}} (\vec{0}_3) &=\psi_{\mathcal{T} S} \\
    \psi_{\mathcal{T} T_1} - 2 \phi_{S}(\vec{0}_1) &= \psi_{\mathcal{T} S} \\
    \psi_{\mathcal{T} T_2} - 2 \phi_{S}(\vec{0}_2) &= \psi_{\mathcal{T} S} \\
    - 2 \phi_{S}(\vec{0}_3) - \phi_{\mathcal{T}} (\vec{0}_0) + \phi_{\mathcal{T}} (\vec{0}_3) &=\psi_{\mathcal{T} S}.
\end{align}
\end{subequations}

\subsection{\label{appendix subsec: PSG solution - Solution of the PSG constraints: intra-unit cell part} Solution of the PSG constraints: gauge fixing and intra-unit cell part}

Before fixing our gauge and solving all intra-unit cell constraints, let's recollect ourselves and quickly summarize what we found thus far. From the space group constraints we obtained the phase equations \eqref{eq U(1) classification: T1, T2, T3 first equation}, \eqref{eq U(1) classification: first equation for phi C6} and \eqref{eq U(1) classification: first equation for phi S}, and the constraints
\begin{subequations} \label{eq: intra-cell condition} 
\begin{align}
    6\phi_{\overline{C}_6}(\vec{0}_0) &= \psi_{\overline{C}_6} \label{eq: intra-cell condition Eq.1} \\
    \sum_{i=1}^3 \left( 2 \phi_{\overline{C}_6} (\vec{0}_i) +\psi_{\overline{C_6} T_i} \right) &= \psi_{\overline{C}_6}  \label{eq: intra-cell condition Eq.2} \\
    \phi_{S}(\vec{0}_0)+ \phi_{S}(\vec{0}_3) &= \psi_{S}  \label{eq: intra-cell condition Eq.3}\\
    2 \phi_{S}(\vec{0}_1) + \psi_{S T_1} + n_1 \pi  &= \psi_{S}  \label{eq: intra-cell condition Eq.4}\\
    2 \phi_{S}(\vec{0}_2) + \psi_{S T_2} + n_1 \pi &= \psi_{S}   \label{eq: intra-cell condition Eq.5}\\
    \sum_{i=1}^3 \psi_{\overline{C}_6 T_i} + \sum_{j=0}^3 \left( \phi_{\overline{C}_6}(\vec{0}_j) + \phi_{S}(\vec{0}_j) \right)&=\psi_{\overline{C}_6 S}  \label{eq: intra-cell condition Eq.6} \\
    (\psi_{\overline{C}_6 T_2} + \psi_{\overline{C}_6 T_3}) + 3 \phi_{\overline{C}_6}(\vec{0}_0) 
    &\nonumber \\
    + \sum_{i=1}^3 \phi_{\overline{C}_6}(\vec{0}_i) + \phi_{S}(\vec{0}_0) + \phi_{S}(\vec{0}_3) &= \psi_{S \overline{C}_6}  \label{eq: intra-cell condition Eq.7}\\
    2(\psi_{\overline{C}_6 T_1} + \psi_{\overline{C}_6 T_3})\hspace{3cm}&\nonumber  \\
    + 2 \left(\phi_{S}(\vec{0}_1) +  \sum_{i=1}^3 \phi_{\overline{C}_6}(\vec{0}_i)  \right)  &= \psi_{S \overline{C}_6} \label{eq: intra-cell condition Eq.8} \\
    2(\psi_{\overline{C}_6 T_1} + \psi_{\overline{C}_6 T_2})\hspace{3cm}&\nonumber  \\
    2 \left( \phi_{S}(\vec{0}_2) +  \sum_{i=1}^3 \phi_{\overline{C}_6}(\vec{0}_i)  \right)  &= \psi_{S \overline{C}_6}  \label{eq: intra-cell condition Eq.9}\\
    3 \psi_{\overline{C}_6 T_1} -3 \psi_{\overline{C}_6 T_2} + \psi_{\overline{C}_6 T_3} - \psi_{S T_1} &=0 \label{eq: intra-cell condition Eq.10}\\
    3 \psi_{\overline{C}_6 T_1} + \psi_{\overline{C}_6 T_2} - 3 \psi_{\overline{C}_6 T_3} - \psi_{S T_2} &=0. \label{eq: intra-cell condition Eq.11}
\end{align}
\end{subequations}
From TRS, we got the phase equation \eqref{eq U(1) classification: first equation TRS} and the constraints
\begin{subequations}  \label{eq: intra-cell condition Eq. TRS}
\begin{align}
    2 \psi_{\overline{C_6}T_1} + \psi_{\mathcal{T}T_1} + \psi_{\mathcal{T} T_2} &= 0 \label{eq: intra-cell condition Eq.TRS 1} \\
    2 \psi_{\overline{C_6}T_2} + \psi_{\mathcal{T}T_2} + \psi_{\mathcal{T} T_3} &= 0 \label{eq: intra-cell condition Eq.TRS 2} \\
    2 \psi_{\overline{C_6}T_3} + \psi_{\mathcal{T}T_3} + \psi_{\mathcal{T} T_1} &= 0 \label{eq: intra-cell condition Eq.TRS 3}\\
     -2 \phi_{\overline{C}_6}(\vec{0}_0) &= \psi_{\mathcal{T}\overline{C}_6}  \label{eq: intra-cell condition Eq.TRS 4} \\ 
    \psi_{\mathcal{T}T_3} - 2\phi_{\overline{C}_6}\left(0_{1}\right) - \phi_{\mathcal{T}}\left(0_{3}\right) + \phi_{\mathcal{T}}\left(0_{1}\right) &= \psi_{\mathcal{T} \overline{C}_6}  \label{eq: intra-cell condition Eq.TRS 5}\\
    \psi_{\mathcal{T}T_1} - 2\phi_{\overline{C}_6}\left(0_{2}\right) - \phi_{\mathcal{T}}\left(0_{1}\right) + \phi_{\mathcal{T}}\left(0_{2}\right) &= \psi_{\mathcal{T} \overline{C}_6}  \label{eq: intra-cell condition Eq.TRS 6}\\
    \psi_{\mathcal{T}T_2} - 2\phi_{\overline{C}_6}\left(0_{3}\right) - \phi_{\mathcal{T}}\left(0_{2}\right) + \phi_{\mathcal{T}}\left(0_{3}\right) &= \psi_{\mathcal{T} \overline{C}_6}  \label{eq: intra-cell condition Eq.TRS 7}\\
    2 \psi_{ST_1} + 2 \psi_{\mathcal{T} T_1}- \psi_{\mathcal{T} T_3} &= 0 \label{eq: intra-cell condition Eq.TRS 8} \\
    2 \psi_{ST_2} + 2 \psi_{\mathcal{T} T_2}- \psi_{\mathcal{T} T_3} &= 0 \label{eq: intra-cell condition Eq.TRS 9} \\
    \psi_{\mathcal{T}T_3} - 2 \phi_{S}(\vec{0}_0) + \phi_{\mathcal{T}}(0_0) - \phi_{\mathcal{T}}(0_3) &= \psi_{\mathcal{T}S}  \label{eq: intra-cell condition Eq.TRS 10} \\
    \psi_{\mathcal{T}T_1} - 2 \phi_{S}(\vec{0}_1) &= \psi_{\mathcal{T}S}  \label{eq: intra-cell condition Eq.TRS 11} \\
    \psi_{\mathcal{T}T_2} - 2 \phi_{S}(\vec{0}_2) &= \psi_{\mathcal{T}S}  \label{eq: intra-cell condition Eq.TRS 12} \\
    -\phi_{S}(\vec{0}_3) - \phi_{\mathcal{T}}(\vec{0}_0) + \phi_{\mathcal{T}}(\vec{0}_3) &= \psi_{\mathcal{T} S}.  \label{eq: intra-cell condition Eq.TRS 13}
\end{align}
\end{subequations}

These results can be simplified by partially fixing our gauge freedom to remove redundant gauge equivalent solutions. We note that the phase associated with $T_1$, $T_2$ and $T_3$ appear an odd number of times in Eq. \eqref{eq: psg classification CT_i} and \eqref{eq: psg classification S T_i} respectively. Consequently, we can make use of our gauge freedom and IGG structure ($\phi_{\mathcal{O}}\to\phi_{\mathcal{O}}+\chi$, where $\chi \in\text{IGG}$) for $\phi_{T_1}$, $\phi_{T_2}$ and $\phi_{T_3}$ to set $\psi_{\overline{C}_6 T_1}=\psi_{\overline{C}_6 T_2}=\psi_{S T_2}=0$. Next, we can use a constant sublattice dependent gauge transformation of the form
\begin{equation}
    \phi(\vec{r}_\mu) = \phi_{\mu} \hspace{5mm} (\mu=0,1,2,3).
\end{equation}
As the phase factor transform according to $\phi_{\mathcal{O}}(\vec{r}_\mu)\to \phi_{\mathcal{O}}(\vec{r}_\mu) + \phi(\vec{r}_\mu) - \phi\left[ \mathcal{O}^{-1}(\vec{r}_\mu) \right]$ for a general gauge transformation, our initial gauge fixing for $\phi_{T_1}$, $\phi_{T_{2}}$ and $\phi_{T_{3}}$ are unaffected while $\phi_{\overline{C}_6}$ and $\phi_{S}$ transform like
\begin{subequations}
  \begin{empheq}[left={ }\empheqlbrace]{align*}
    \phi_{\overline{C_6}}(\vec{0})_0 &\to \phi_{\overline{C_6}}(\vec{0})_0 \\
    \phi_{\overline{C_6}}(\vec{0})_i &\to \phi_{\overline{C_6}}(\vec{0})_i+\phi_{i}-\phi_{i-1} \hspace{2mm} \text{for $i\in \{ 1,2,3 \}$} \\
    \phi_{S}(\vec{0})_0 &\to \phi_{S}(\vec{0})_0 + \phi_0 - \phi_3 \\
    \phi_{S}(\vec{0})_{1,2} &\to \phi_{S}(\vec{0})_{1,2} \\
    \phi_{S}(\vec{0})_3 &\to \phi_{S}(\vec{0})_0 + \phi_3 - \phi_0 \\
    \phi_{\mathcal{T}}(\vec{0})_{\mu} &\to \phi_{\mathcal{T}}(\vec{0})_{\mu} + 2\phi_{\mu} .
  \end{empheq}
\end{subequations}
Then we can choose $\phi_{\mu}$ to fix
\begin{equation}
    \phi_{\mathcal{T}} (\vec{0}_\mu) = 0. \label{eq U(1) classification: gauge fixing phi TRS}
\end{equation}
Even after these major simplifications, we still have some gauge freedom left: our $U(1)$ gauge freedom for $\phi_{\overline{C}_6}$ and $\phi_{S}$, and a discrete gauge freedom
\begin{equation}
    \phi(\vec{r}_\mu) = \pi \delta_{\mu,\nu} \label{eq U(1) classification: discrete gauge freedom}
\end{equation}
as it does not affect our gauge fixing \eqref{eq U(1) classification: gauge fixing phi TRS}.

Let's now move on to solve all TRS constraints. It is first evident from Eq. \eqref{eq: intra-cell condition Eq.TRS 1} and \eqref{eq: intra-cell condition Eq.TRS 3} that  $\psi_{\mathcal{T}T_3}=-\psi_{\mathcal{T}T_2}=\psi_{\mathcal{T}T_1}$ and from Eq. \eqref{eq: intra-cell condition Eq.TRS 2} that 
\begin{equation}
    \psi_{\mathcal{T}T_1}= -\psi_{\overline{C}_6 T_3} + \pi n_{\mathcal{T}T_1},
\end{equation}
where $n_{\mathcal{T}T_1}\in\{0,1\}$. Putting all of this in eqs. \eqref{eq: intra-cell condition Eq.TRS 8} and \eqref{eq: intra-cell condition Eq.TRS 9} one further finds
\begin{align}
    \psi_{\overline{C}_6 T_3} &= n_{\mathcal{T}T_1} \pi + 2 \psi_{S T_1}\\
    \psi_{S T_1} & = \frac{n_{S T_1} \pi}{3},
\end{align}
where $n_{ST_1}\in\{0,1,3,4,5\}$. 
Using the remaining constraints in \eqref{eq: intra-cell condition Eq. TRS} we get
\begin{subequations} \label{eq U(1) classification: phase factor after solving TRS constraints}
\begin{align}
\phi_S(\vec{0}_0) &= -\frac{n_{ST_1} \pi}{3} - \frac{\psi_{\mathcal{T}S}}{2} + \pi m_{0} \\
\phi_S(\vec{0}_1) &= -\frac{n_{ST_1} \pi}{3} - \frac{\psi_{\mathcal{T}S}}{2} + \pi m_1 \\
\phi_S(\vec{0}_2) &=  \frac{n_{ST_1}\pi}{3} - \frac{\psi_{\mathcal{T}S}}{2} + \pi m_2 \\
\phi_S(\vec{0}_3) &= - \frac{\psi_{\mathcal{T}S}}{2} + \pi m_3 \\
\phi_{\overline{C}_6}(\vec{0}_0) &= -\frac{\psi_{\mathcal{T} \overline{C}_6}}{2}+\pi p_0\\
\phi_{\overline{C}_6}(\vec{0}_1) &= -\frac{n_{ST_1}\pi}{3} -\frac{\psi_{\mathcal{T} \overline{C}_6}}{2}+\pi p_1\\
\phi_{\overline{C}_6}(\vec{0}_2) &= -\frac{n_{ST_1}\pi}{3}-\frac{\psi_{\mathcal{T} \overline{C}_6}}{2}+\pi p_2 \\
\phi_{\overline{C}_6}(\vec{0}_3) &= \frac{n_{ST_1}\pi}{3}-\frac{\psi_{\mathcal{T} \overline{C}_6}}{2}+\pi p_3.
\end{align}
\end{subequations}
Using our discrete gauge freedom \eqref{eq U(1) classification: discrete gauge freedom}, we can fix $p_1=p_2=p_3 \equiv p$ by choosing $\nu=1,2,3$. We are still left with discrete gauge freedom on the zero sublattice $\phi(\vec{r}_\mu) = \pi \delta_{\mu,0}$.

Now, the only thing left is to solve for the intra-cell conditions \eqref{eq: intra-cell condition}. One should first note that equations \eqref{eq: intra-cell condition Eq.1}, \eqref{eq: intra-cell condition Eq.2}, \eqref{eq: intra-cell condition Eq.8}, \eqref{eq: intra-cell condition Eq.9}, \eqref{eq: intra-cell condition Eq.10} and \eqref{eq: intra-cell condition Eq.11} directly impose that $n_{\mathcal{T}T_1}=0 $  and $\psi_{S\overline{C}_6}=\psi_{\overline{C}_6}=-3\psi_{\mathcal{T}\overline{C}_6}$. Then with \eqref{eq: intra-cell condition Eq.7} we get
\begin{equation}
m_3 = m_0 + p_0 + p .
\end{equation}
Using these in \eqref{eq: intra-cell condition Eq.6} naturally leads to
\begin{align}
\psi_{\overline{C}_6 S} &= (m_1 + m_2) \pi  - 2 \psi_{\mathcal{T}\overline{C}_6}  2 \psi_{\mathcal{T}S}
\end{align}
Next, we find from Eq. \eqref{eq: intra-cell condition Eq.5} that 
\begin{equation}
\psi_{S} = n_1 \pi + \frac{2 n_{S T_1} \pi}{3} - \psi_{\mathcal{T}S}
\end{equation}
and from \eqref{eq: intra-cell condition Eq.4} that 
\begin{equation}
n_{ST_1}=0.
\end{equation}
Finally using our last constraint Eq. \eqref{eq: intra-cell condition Eq.3}, we obtain
\begin{equation}
p = n_1 + p_0.
\end{equation}
Therefore, we have 
\begin{align}
    \phi_{\overline{C}_6}(\vec{0}_\mu) &= p_0 \pi - \frac{\psi_{\mathcal{T}\overline{C}_6}}{2} + n_1 \pi \delta_{\mu=1,2,3} \\
    \phi_{S}(\vec{0}_\mu) &= (m_0  \delta_{\mu=0,3} + m_1 \delta_{\mu=1} + m_2 \delta_{\mu=2})\pi - \frac{\psi_{\mathcal{T}S}}{2}.
\end{align}
This solution can be simplified by using the IGG transformations
\begin{align}
    \phi_{\overline{C}_6} &\to \phi_{\overline{C}_6}+ p_0\pi + \frac{\psi_{\mathcal{T}\overline{C}_6}}{2} \\
    \phi_{S} &\to \phi_{S}+ m_1\pi + \frac{\psi_{\mathcal{T}S}}{2} - \frac{n_1 \pi}{2}.
\end{align}
It is now perfectly appropriate to use some of our leftover discrete gauge freedom on the zero sublattices to perform the gauge transformation
\begin{equation}
    \phi(\vec{r}_{\mu}) = (m_1 + n_1) \pi \delta{\mu,0}.
\end{equation}
That yields 
\begin{align}
    \phi_{\overline{C}_6}(\vec{0}_\mu) &=  n_1 \pi \delta_{\mu=1,2,3} \\
    \phi_{S}(\vec{0}_\mu) &= (-)^{\delta_{\mu=1,2,3}} \frac{n_1 \pi}{2} + m_2 \pi \delta_{\mu=2}.
\end{align}

With this gauge fixing, we can also rewrite our PSG classification in such a way that the physical interpretation is more explicit by noting that 
\begin{align}
    \psi_{\overline{C}_6S} &= n_{\overline{C}_6 S} \pi = m_2\pi \\
    \Longrightarrow m_2 &= n_{\overline{C}_6 S}.
\end{align}

In conclusion, with our choice of gauge fixing, we find 
\begin{align}
    \phi_{\overline{C}_6}(\vec{0}_\mu) &= n_1 \pi \delta_{\mu=1,2,3} \\
    \phi_{S}(\vec{0}_\mu) &= (-)^{\delta_{\mu=1,2,3}} \frac{n_1 \pi}{2} + n_{\overline{C}_6 S} \pi \delta_{\mu=2}
\end{align}
and the final solution presented in Eq. \eqref{eq: U(1) PSG classification}.

\section{\label{appendix: From PSG to mean-field Hamiltonian} From PSG to mean-field Hamiltonian}

\subsection{\label{appendix subsec: From PSG to mean-field Hamiltonian -> Generalities} Generalities}

With the transformation of the hopping (Eqs. \eqref{eq: transformation hopping matrix under SG} and \eqref{eq: transformation hopping matrix under TR}) and pairing (Eqs. \eqref{eq: transformation pairing matrix under SG} and \eqref{eq: transformation pairing matrix under TR}) matrices and our PSG classifications (eqs. \eqref{eq: Z2 PSG classification} and \eqref{eq: U(1) PSG classification}), we can build the fully constrained MF Hamiltonian for all PSG classes.

First, using the time-reversal transformations in Eq. \eqref{eq: O(4) transformations matrix}, we have the requirements that 
\begin{align}
    (a^{p}_{\vec{r}_\mu,\vec{r}_\nu'}, b^{p}_{\vec{r}_\mu,\vec{r}_\nu'},& c^{p}_{\vec{r}_\mu,\vec{r}_\nu'}, d^{p}_{\vec{r}_\mu,\vec{r}_\nu'})\nonumber \\
    &= (a^{p*}_{\vec{r}_\mu,\vec{r}_\nu'}, -b^{p*}_{\vec{r}_\mu,\vec{r}_\nu'}, -c^{p*}_{\vec{r}_\mu,\vec{r}_\nu'}, -d^{p*}_{\vec{r}_\mu,\vec{r}_\nu'})  \\
    (a^{h}_{\vec{r}_\mu,\vec{r}_\nu'}, b^{h}_{\vec{r}_\mu,\vec{r}_\nu'},& c^{h}_{\vec{r}_\mu,\vec{r}_\nu'}, d^{h}_{\vec{r}_\mu,\vec{r}_\nu'}) \nonumber\\
    &= (a^{h*}_{\vec{r}_\mu,\vec{r}_\nu'}, -b^{h*}_{\vec{r}_\mu,\vec{r}_\nu'}, -c^{h*}_{\vec{r}_\mu,\vec{r}_\nu'}, -d^{h*}_{\vec{r}_\mu,\vec{r}_\nu'})
\end{align}
for a Kramers doublet. These relations follow from the fact that $\phi_{\mathcal{T}}(\vec{r}_\mu)=0$ for all PSG classes. Thus, all singlet parameters are real and all triplet parameters are imaginary.

Next, the SG operations provide a map between different NN bonds $( \vec{r}_\mu \to \vec{r}_\nu')$ on the pyrochlore lattice. This defines an equivalence relation in the space of all NN bonds. It is thus possible to classify all bonds in equivalence classes where members of the same class are related by a space group operation. As the relation between MF parameters on different bonds that are symmetry-related is known, we only need the MF parameters on a single bond of each equivalence class to build the whole MF Hamiltonian. It turns out that all NN bonds are symmetry-related on the pyrochlore lattice. The MF ansätz on a single representative bond, that we choose to be $\vec{0}_0\to\vec{0}_1$, is required to build the whole MF Hamiltonian
\begin{subequations}
\begin{align} \label{eq: Mf parameters for representation bond 0->1}
    u_{\vec{0}_0,\vec{0}_1}^p =& a^{p} i \sigma^y + b^{p} i\sigma^y\sigma^x + c^{p} i\sigma^y\sigma^y + d^{p} i\sigma^y\sigma^z  \\
    u_{\vec{0}_0,\vec{0}_1}^h =& a^{h} \mathds{1}_{2\times 2} +  b^{h} \sigma^x + c^{h} \sigma^y + d^{h} \sigma^z, 
\end{align}
\end{subequations}
where, for simplicity's sake, we have suppressed the position indices when referring to this representative bond (e.g., $a^p_{\vec{0}_0,\vec{0}_1} \equiv a^p$).

However, one should anticipate that the mapping between two bonds may not be unique. If two non-equivalent space group elements do indeed map between the same bond, this leads to non-trivial constraints on the MF parameters of the representative bond. To determine all non-trivial constraints on the MF parameters, one needs to find all the symmetry operations that leave the representative bonds invariant (the so-called "stabilizers" or "little group"). More formally, this corresponds to determining the kernel of the homomorphism from the SG to the space of undirected NN bond transformations. If we require the MF parameters of the representative bond to be invariant under all transformations in the kernel, any ambiguity previously mentioned is resolved, as should be seen from the fundamental homomorphism theorem \cite{dummit2004abstract}. To determine all non-trivial transformations that map the representative bond to itself, we can restrict our attention to a single unit cell as a translation of the unit cell acts trivially on the MF ansätz.

\subsection{\label{appendix: Point group structure} Point group and stabilizer}

\begin{figure}
\includegraphics[width=0.65\linewidth]{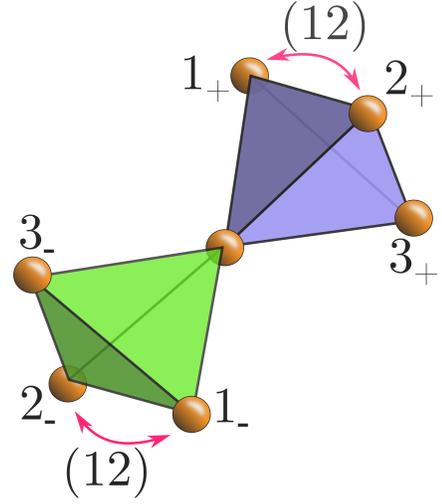}
\caption{Primitve cell in the $n_1=0$ case. All seven vertices are identified using the convention used in the text. The effect of the operation (12) is illustrated. \label{fig: primitve cell 0-flux case}}
\end{figure}

In the $n_1=0$ case, all primitive translations act trivially. Only a simple unit cell that is composed of a pair of corner-sharing tetrahedra, as illustrated in fig. \ref{fig: primitve cell 0-flux case}, needs to be considered as all other translation-related bonds are equivalent to the one in that primitive unit cell. This is equivalent to restricting our attention to the point group of the $Fd\overline{3}m$ space group: $O_h\simeq Fd\overline{3}m/\mathbf{T}_0$ where $\mathbf{T}_0$ is the translation normal subgroup generated by $\{T_1, T_2,T_3\}$. $O_h$ has a direct product structure $O_h \simeq S_4\times \mathbb{Z}_2$. Labeling all seven vertices of the primitive cell by $\mu_\pm$, where $\mu=0,1,2,3$ is the usual sublattice index and $+$ ($-$) identified the upper (lower) tetrahedron. Then $S_4$ identifies operations that permute elements on the same tetrahedron, while $\mathbb{Z}_2$ interchanges the same sublattice sites of the upper and lower tetrahedra. $O_h$ can thus be understood as a permutation over the two sets $\{0,1,2,3\}$ and $\{+,-\}$. The same sublattice permutations group $S_4$ is generated by two operators $C_3=(123)$ and $\Sigma=S \circ I = (03)$. The inversion $I=(+-)$ is the generator of $\mathbb{Z}_2$ part. Note that we use the usual permutation notation where a cycle $(p_1 p_2 \text{...} p_n)$ represents the mapping of $p_i$ to $p_{i+1}$. 

\begin{figure}
\includegraphics[width=1.0\linewidth]{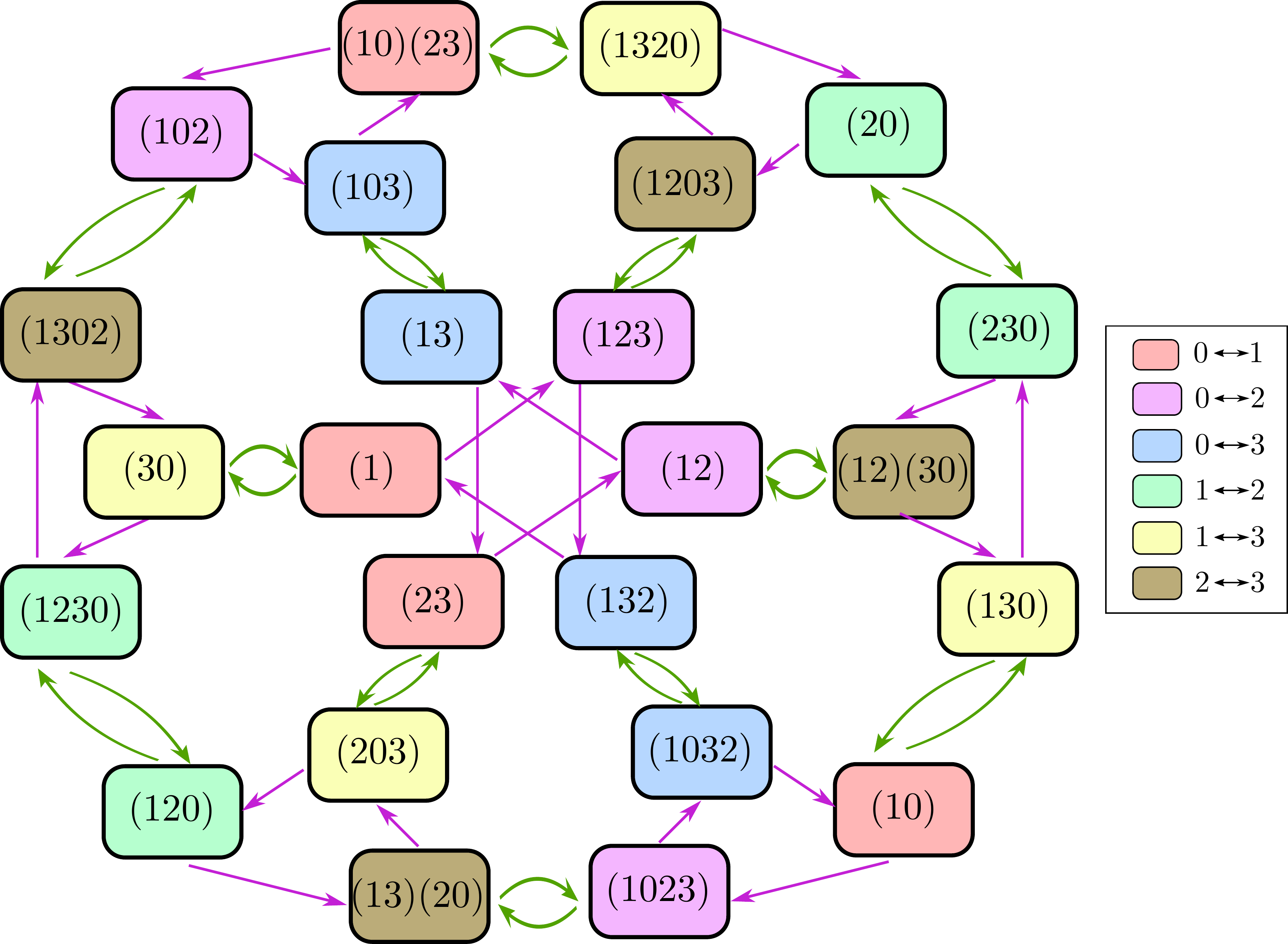}
\caption{Cayley graph for the $S_{4}$ group. Magenta arrows represent the group action by the generator $C_3=(123)$ and green ones by $\Sigma=(03)$. All elements are classified by their action on the undirected bond ($0\leftrightarrow 1$). \label{fig: Cayley graph S4}}
\end{figure}

We note that inversion always maps a bond that relates sites within the same upper tetrahedron to a bond that relates two different upper tetrahedra (and vice versa). For instance $(0_{+}\to 1_{+})$ is mapped to $(0_{+}\to 1_{-})$ by inversion. So no group elements containing the inversion will be in the stabilizers. We can therefore restrict our attention to $S_4 = O_h/\mathbb{Z}_2$. The relation of 24 elements of $S_4$ with the two generators can be directly read off the Cayley graph presented in fig. \ref{fig: Cayley graph S4}. The Cayley graph also displays how the different group elements transform the representative undirected bond $(0\leftrightarrow 1)$ ($(0\leftrightarrow 1)$ stands for both $ (0\to 1)$ and $( 1\to 0)$). Upon inspection of these transformations, we find that the kernel of the homomorphism comprises four elements: (1), (23), (10), and (10)(23). (1) and (23) map the bond $(0_+ \to 1_+)$ directly to itself, while (10) and (10)(34) simply invert its orientation (i.e., they map it to $(1_+\to 0_+)$). In terms of the SG generators, $E$ and $S \circ \overline{C}_{6}\circ S \circ \overline{C}_{6}^{-1}\circ S\circ \overline{C}_{6}^{-1}$ map the representative bond to itself, whereas $\overline{C}_{6}\circ S^{-1}\circ \overline{C}_{6}\circ \overline{C}_{6}$ and $\overline{C}_{6}\circ S^{-1}\circ \overline{C}_{6}\circ \overline{C}_{6}\circ S\circ \overline{C}_{6}\circ S\circ \overline{C}_{6}^{-1}\circ S\circ \overline{C}_{6}^{-1}$ map it to the inverted one. 

In the case $n_1=\pi$, the primitive unit cell is doubled. It now contains four tetrahedra at position (in the pyrochlore coordinates) $(0,0,0)$,  $(0,1,0)$, $(0,0,1)$, and $(0,1,1)$. So the point group is $Fd\overline{3}m/\mathbf{T}_{\pi}$ where $\mathbf{T}_\pi$ is the translation normal subgroup generated by $\{T_1 , T_2\circ T_2, T_3\circ T_3\}$). To get all non-trivial constraints, we consider the group generated by $\{ \overline{C}_6, S, T_2, T_3 \}$, where $T_i^2=E$ for $i=2,3$ because of the periodicity condition. Mapping these group elements to the space of oriented bond transformations, we find that the stabilizer is the same than for $n_1=0$.

\subsection{\label{appendix: constraints on the mean-field parameters} Constraints on the MF parameters}

\begin{table}
\caption{\label{tab: independant non-zero MF parameters}
Independent non-zero nearest-neighbor MF parameters for all $\mathbb{Z}_2$ and $U(1)$ PSG classes in the dipolar-octupolar case
}
\begin{tabular}{c|cc|c}
\hline\hline
\multirow{2}{*}{Class} & \multicolumn{2}{c|}{\underline{Independent non-zero parameters}} &  \multirow{2}{*}{Comment}  \\
      & Pairing & Hopping &  \\
\hline
$\mathbb{Z}_2\text{-}0\text{-}(0,0,0)$ & $\IM{c^{p}}$ & $\RE{a^{h}}$,  & \\
$\mathbb{Z}_2\text{-}0\text{-}(0,0,1)$ & $\IM{b^{p}}$, $\IM{d^{p}}$ & $\RE{a^{h}}$ & \\
$\mathbb{Z}_2\text{-}0\text{-}(0,1,0)$ & $\IM{b^{p}}$, $\IM{d^{p}}$ & $\RE{a^{h}}$ & \\
$\mathbb{Z}_2\text{-}0\text{-}(0,1,1)$ & $\IM{c^{p}}$ & $\RE{a^{h}}$  &  \\
$\mathbb{Z}_2\text{-}0\text{-}(1,0,0)$ &  & $\IM{b^{h}}$, $\IM{d^{h}}$ & $U(1)$ at NN \\
$\mathbb{Z}_2\text{-}0\text{-}(1,0,1)$ & $\RE{a^{p}}$ & $\IM{b^{h}}$, $\IM{d^{h}}$  &  \\
$\mathbb{Z}_2\text{-}0\text{-}(1,1,0)$ & $\RE{a^{p}}$ & $\IM{b^{h}}$, $\IM{d^{h}}$   & \\
$\mathbb{Z}_2\text{-}0\text{-}(1,1,1)$ &  & $\IM{b^{h}}$, $\IM{d^{h}}$ & $U(1)$ at NN \\
$\mathbb{Z}_2\text{-}1\text{-}(0,0,0)$ & $\IM{b^{p}}$, $\IM{d^{p}}$ & $\RE{a^{h}}$ & \\
$\mathbb{Z}_2\text{-}1\text{-}(0,0,1)$ & $\IM{c^{p}}$ & $\RE{a^{h}}$ & \\
$\mathbb{Z}_2\text{-}1\text{-}(0,1,0)$ & $\IM{c^{p}}$ &$\RE{a^{h}}$  & \\
$\mathbb{Z}_2\text{-}1\text{-}(0,1,1)$ & $\IM{b^{p}}$, $\IM{d^{p}}$ & $\RE{a^{h}}$ &  \\
$\mathbb{Z}_2\text{-}1\text{-}(1,0,0)$ & $\RE{a^{p}}$ & $\IM{b^{h}}$, $\IM{d^{h}}$  & \\
$\mathbb{Z}_2\text{-}1\text{-}(1,0,1)$ &  & $\IM{b^{h}}$, $\IM{d^{h}}$ & $U(1)$ at NN\\
$\mathbb{Z}_2\text{-}1\text{-}(1,1,0)$ &  & $\IM{b^{h}}$, $\IM{d^{h}}$ & $U(1)$ at NN\\
$\mathbb{Z}_2\text{-}1\text{-}(1,1,1)$ & $\RE{a^{p}}$  & $\IM{b^{h}}$, $\IM{d^{h}}$ & \\
\hline
$U(1)\text{-}n_1 \text{-}0$ &  & $\RE{a^{h}}$   & \\
$U(1)\text{-}n_1 \text{-}1$ &  & $\IM{b^{h}}$, $\IM{d^{h}}$ & \\
\hline
\hline
\end{tabular}
\end{table}

Now that all operations that map the representative bond to itself have been identified, the $O(4)$ transformations in Eq. \eqref{eq: O(4) transformations matrix}, and the phase equations of Eqs. \eqref{eq: Z2 PSG classification} and \eqref{eq: U(1) PSG classification} can be used to deduce all corresponding constraints. From the definitions \eqref{eq: definitions bond operators} and the bosonic commutation relations, we also note that under position exchange the singlet pairing parameter $\Delta_{\vec{r}_{\mu},\vec{r}_{\nu}'}$ is odd  ($\Delta_{\vec{r}_{\mu},\vec{r}_{\nu}'}=-\Delta_{\vec{r}_{\nu}',\vec{r}_{\mu}}$), the triplet pairing $D_{\vec{r}_{\mu},\vec{r}_{\nu}'}^a$  is invariant ($D^a_{\vec{r}_{\mu},\vec{r}_{\nu}'}=D^a_{\vec{r}_{\nu}',\vec{r}_{\mu}}$), and both hopping MF parameters $\chi_{\vec{r}_{\mu},\vec{r}_{\nu}'}$ and $E^a_{\vec{r}_{\mu},\vec{r}_{\nu}'}$ satisfy $A_{\vec{r}_{\mu},\vec{r}_{\nu}'}=A_{\vec{r}_{\nu}',\vec{r}_{\mu}}^*$.

We  have the following three constraints on the MF pairing parameters of the $\mathbb{Z}_2$ PSG classes
\begin{subequations}
\begin{align}
    &(a^{p} , b^{p} , c^{p} , d^{p} )\nonumber \\
    =&  (-1)^{n_1+n_{\overline{C}_6}+n_{\overline{C}_6 S}+n_{S T_1}} (a^{p},  -b^{p}, c^{p} , -d^{p} ) \\
    =& (-1)^{n_1+n_{\overline{C}_6}+n_{S T_1}} (a^{p},  -b^{p}, c^{p} , -d^{p} ) \\
      =& (-1)^{n_{\overline{C}_6 S}} (a^{p},  b^{p}, c^{p} , d^{p} ), 
\end{align}
and the following equations for the hopping MF parameters of both $\mathbb{Z}_2$ and $U(1)$ PSG classes
\begin{align}
    &(a^{h} , b^{h} , c^{h} , d^{h} ) \nonumber \\
    =& (-1)^{n_{\overline{C}_6 S}} (a^{h} , - b^{h},  c^{h} , -d^{h} ) \\
    =& (a^{h*} , - b^{h*},  c^{h*} , -d^{h*} ) \\
    =& (-1)^{n_{\overline{C}_6 S}} (a^{h*} , b^{h*} , c^{h*} , d^{h*} ).
\end{align}
\end{subequations}
The results are summarized in table \ref{tab: independant non-zero MF parameters} where the independent non-zero MF parameters for each PSG class are presented.

\subsection{Relation between MF parameters on different bonds in the unit cell}

\begin{table}
\caption{\label{tab: Relation between the mean-field parameters on different bonds}
Relation between the MF parameters on the different bonds in the $n_1=0$ unit cell. $A_{IGG}$ and $B_{IGG}$ are prefactors to the hopping and pairing terms respectively, for the specified IGG. 
}
\begin{ruledtabular}
\begin{tabular}{c|c|c|c}
Bond              & MF parameters & \multicolumn{2}{c}{Prefactor} \\ \hline
\multirow{2}{*}{$\vec{r}_1 \hspace{-1mm} \to\hspace{-1mm}  \vec{r}_2$} & \multirow{2}{*}{} &  $A_{\mathbb{Z}_2}$ & $B_{\mathbb{Z}_2}$     \\ \cline{3-3} \cline{4-4}
                  &                         &  $A_{U(1)}$  &   $B_{U(1)}$   \\ \hline
\multirow{2}{*}{$0_+ \hspace{-1mm} \to\hspace{-1mm}  1_+$} & \multirow{2}{*}{$ \left( a, b, c, d \right)$}    &  $1$       & $1$     \\ \cline{3-3} \cline{4-4}
                  &                         &  $1$                    &   $1$   \\ \hline
\multirow{2}{*}{$0_+\hspace{-1mm} \to\hspace{-1mm}  2_+$} & \multirow{2}{*}{$ \left( a, b, c, d \right)$}    &  $1$   & $1$     \\ \cline{3-3} \cline{4-4}
                  &                         &  $1$         &   $1$   \\ \hline
\multirow{2}{*}{$0_+ \hspace{-1mm} \to\hspace{-1mm}  3_+$} & \multirow{2}{*}{$ \left( a, b , c, d \right)$}    &  $1$   &  $1$     \\ \cline{3-3} \cline{4-4}
                  &                         &  $1$          &   $1$   \\ \hline
\multirow{2}{*}{$1_+ \hspace{-1mm} \to\hspace{-1mm}  2_+$} & \multirow{2}{*}{$ \left( a, -b, c, -d \right)$}    &  $(-1)^{n_{1} +n_{S T_1}}$  &  $(-1)^{n_{\overline{C}}}$      \\ \cline{3-3} \cline{4-4}
                  &                         &  $(-1)^{n_1}$   &  $1$  \\ \hline
\multirow{2}{*}{$1_+ \hspace{-1mm} \to\hspace{-1mm}  3_+$} & \multirow{2}{*}{$ \left( a, -b, c, -d \right)$}    &   $(-1)^{n_{\overline{C} S} +n_1 + n_{S T_1}}$    &  $(-1)^{n_{\overline{C}} + n_{\overline{C} S}}$   \\ \cline{3-3} \cline{4-4}
                  &                         &   $(-1)^{ n_{\overline{C} S} + n_1 }$   &   $(-1)^{ n_{\overline{C} S} }$  \\ \hline
\multirow{2}{*}{$2_+ \hspace{-1mm} \to\hspace{-1mm}  3_+$} & \multirow{2}{*}{$ \left( a, -b, c, -d \right)$}    &  $(-1)^{n_1+n_{ST_1}}$   & $(-1)^{n_{\overline{C}}}$    \\ \cline{3-3} \cline{4-4}
                  &                         &  $(-1)^{n_1}$  &   $1$   \\ \hline
\multirow{2}{*}{$0_+ \hspace{-1mm} \to\hspace{-1mm}  1_-$} & \multirow{2}{*}{$ \left( a, b,  c, d \right)$}    &  $(-1)^{n_1 + n_{S T}}$  &   $(-1)^{n_1 + n_{\overline{C}} + n_{S T}}$     \\ \cline{3-3} \cline{4-4}
                  &                         &  $(-1)^{n_1}$  &  $(-1)^{n_1}$   \\ \hline
\multirow{2}{*}{$0_+ \hspace{-1mm} \to\hspace{-1mm}  2_-$} & \multirow{2}{*}{$ \left( a, b  , c, d \right)$}    &   $(-1)^{n_1+n_{S T}}$  & $(-1)^{n_1 + n_{\overline{C}} + n_{S T}}$      \\ \cline{3-3} \cline{4-4}
                  &                         &  $(-1)^{n_1}$  &    $(-1)^{n_1}$   \\ \hline
\multirow{2}{*}{$0_+ \hspace{-1mm} \to\hspace{-1mm}  3_-$} & \multirow{2}{*}{$ \left( a, b , c, d \right)$}    &  $(-1)^{n_1+n_{ST}}$   &  $(-1)^{n_1 + n_{\overline{C}} + n_{S T}}$    \\ \cline{3-3} \cline{4-4}
                  &                         &  $(-1)^{n_1}$  &   $(-1)^{n_1}$   \\ \hline
\multirow{2}{*}{$1_+ \hspace{-1mm} \to\hspace{-1mm}  2_-$} & \multirow{2}{*}{$ \left( a, -b, c, -d \right)$}    &  $(-1)^{n_1+n_{ST}}$   &   $1$    \\ \cline{3-3} \cline{4-4}
                  &                         &  $(-1)^{n_1}$  &   $1$   \\ \hline
\multirow{2}{*}{$1_+ \hspace{-1mm} \to\hspace{-1mm}  3_-$} & \multirow{2}{*}{$ \left( a, -b  , c, -d \right)$}    &  $(-1)^{n_{1}+n_{ST}+n_{\overline{C}S}}$    &  $(-1)^{n_{\overline{C} S}}$    \\ \cline{3-3} \cline{4-4}
                  &                         & $(-1)^{n_1 +  n_{\overline{C}S}}$  &  $(-1)^{n_{\overline{C} S}}$  \\ \hline
\multirow{2}{*}{$2_+ \hspace{-1mm} \to\hspace{-1mm}  3_-$} & \multirow{2}{*}{$ \left(  a, -b ,  c, -d \right)$}    &  $(-1)^{n_1 +n_{ST}}$   & $1$     \\ \cline{3-3} \cline{4-4}
                  &                         &  $(-1)^{n_1}$    &    $1$ \\
\end{tabular}
\end{ruledtabular}
\end{table}

To get the MF parameters on all bonds in the unit cell, we first need to identify a SG operations that map the representative bond to the other ones. In this appendix, we specify those relations for all bonds in the $n_1=0$ unit cell (see fig. \ref{fig: primitve cell 0-flux case}). The SG operations that map the representative bond to all others in the unit cell are
\begin{subequations}
\begin{align}
E:& (\vec{0}_0 \to \vec{0}_1) \mapsto (\vec{0}_0 \to \vec{0}_1) \\
\overline{C}_6^{-2}:& (\vec{0}_0 \to \vec{0}_1) \mapsto (\vec{0}_0 \to \vec{0}_2)  \\
\overline{C}_6^{2}:& (\vec{0}_0 \to \vec{0}_1) \mapsto (\vec{0}_0 \to \vec{0}_3)  \\
\overline{C}_6 \circ S^{-1}:& (\vec{0}_0 \to \vec{0}_1) \mapsto (\vec{0}_1 \to \vec{0}_2)  \\
\overline{C}_6^{-2} \circ S \circ \overline{C}_6 :& (\vec{0}_0 \to \vec{0}_1) \mapsto (\vec{0}_1 \to \vec{0}_3)  \\
\overline{C}_6^{-1} \circ S^{-1} :& (\vec{0}_0 \to \vec{0}_1) \mapsto (\vec{0}_2 \to \vec{0}_3).
\end{align}
\end{subequations}
For the six other bonds that connect an up-pointing tetrahedron to another one, the SG operations can be deduced by composing the above transformation with inversion $I=\overline{C}_6^3$. The relations between MF parameters in the $\mathbb{Z}_2$ and $U(1)$ case on all these bonds are summarized in table \ref{tab:  Relation between the mean-field parameters on different bonds}.

\section{\label{appendix: Diagonalization of the mean-field Hamiltonian} Diagonalization of the mean-field Hamiltonian}

To diagonalize the MF Hamiltonian, we can first use the translation symmetry of the ansätz by Fourier transforming the spinon operator
\begin{equation} \label{eq: Fourier transform of annihilation operator}
    b_{\vec{r}_{\mu},\alpha} = \frac{1}{\sqrt{N_{u.c.}}} \sum_{\vec{k}\in\text{BZ}} b_{\vec{k},\mu,\alpha} e^{i\vec{k}\cdot \vec{r}}
\end{equation}
to bring our Hamiltonian \eqref{eq: MF Hamiltonian generic form} to the form
\begin{align} \label{eq: mean-field Hamiltonian, Fourier transformed, not diagonalized}
    H_{MF} =& \sum_{\vec{k}\in BZ} B_{\vec{k}}^\dag \mathcal{H}(\vec{k}) B_{\vec{k}}  \\
    &- N_{u.c.} N_{SL} \lambda (1+\kappa)+ E_0\left(\left\{A\right\}\right) \nonumber, 
\end{align}
where $N_{u.c}$ denotes the number of unit cell, $BZ$ the first Brillouin zone, and $B_{\vec{k}}^\dag = \left( \vec{b}_{\vec{k},1}^\dag, ... \vec{b}_{\vec{k},N_{SL}}^\dag, \vec{b}_{-\vec{k},1}, ...  \vec{b}_{-\vec{k},N_{SL}} \right)$ is a $4\times N_{SL}$ components vector with $N_{SL}$ being the number of sublattices (4 for $n_1=0$ and 16 for $n_1=1$ respectively). With this rewriting, the matrix $\mathcal{H}(\vec{k})$ has the standard Bogoliubov form \cite{colpa1978diagonalization}
\begin{equation}
    \mathcal{H}(\vec{k}) = 
    \begin{pmatrix}
    H_{h}(\vec{k}) & H_{p}(\vec{k}) \\
    H_{p}^\dag(\vec{k}) & H_{h}^T(-\vec{k})
    \end{pmatrix}.
\end{equation}
It can be diagonalized by a Bogoliubov transformation $B_{\vec{k}} = P( \vec{k} ) \Gamma_{\vec{k}}$, such that 
\begin{align} \label{eq: diagonalization of Hamiltonian with Bogoliubov trfs}
   P^\dag (\vec{k}) \mathcal{H}(\vec{k})  P(\vec{k}) =  \Lambda(\vec{k}),
\end{align}
where $\Lambda(\vec{k})$ is a diagonal $4N_{SL} \times 4N_{SL}$ matrix, 
$\Gamma_{\vec{k}}^\dag = \left( \vec{\gamma}_{\vec{k},1}^\dag, ... \vec{\gamma}_{\vec{k},N_{SL}}^\dag, \vec{\gamma}_{-\vec{k},1}, ... \vec{\gamma}_{-\vec{k},N_{SL}} \right)$ is the Bogoliubov spinor and $P(\vec{k})\in \text{SU}(2 N_{SL}, 2 N_{SL})$ must satisfy
\begin{equation} \label{eq: constraint Bogoliubov trf from CCR}
    P^\dag(\vec{k}) J P(\vec{k}) = J, \hspace{6mm} J = \left( \sigma^z \otimes \mathds{1}_{2 N_{SL}, 2 N_{SL}} \right)
\end{equation}
to ensure that the bosonic commutation relations are preserved for the $\gamma_{\vec{k},\mu,\alpha}$ operators. Multiplying Eq. \eqref{eq: diagonalization of Hamiltonian with Bogoliubov trfs} on the left by $P(\vec{k})J$, it can be noted that this problem is strictly equivalent to solving the generalized eigensystem
\begin{equation}
    J \mathcal{H}(\vec{k})\mathbf{a}_{\vec{k},i} = \lambda_{\vec{k},i} \mathbf{a}_{\vec{k},i}. 
\end{equation}
Accordingly, the columns of the Bogoliubov transformations $P(\vec{k})$ are the eigenvectors of the non-hermitian matrix $J \mathcal{H}(\vec{k})$ and the corresponding eigenvalues $\lambda_{\vec{k,i}}$ are the diagonal elements of $J \Lambda(\vec{k})$. With this change of basis, the MF Hamiltonian takes the form
\begin{align} \label{eq: mean-field Hamiltonian, Fourier transformed, diagonalized}
H_{MF} =& \sum_{\vec{k}\in BZ} \sum_{i=1}^{2 N_{S L}} \omega_{i}(\vec{k}) \hat{\gamma}_{k, i}^{\dagger} \hat{\gamma}_{k, i}+\frac{1}{2} \sum_{\vec{k}\in BZ} \sum_{i=1}^{2 N_{S L}} \omega_{i}(\vec{k})\nonumber  \\
&- N_{u.c.} N_{SL} \lambda (1+\kappa)+ E_0\left(\left\{A\right\}\right)
\end{align}

After this diagonalization, all MF ansätz averages can be evaluated in terms of Bogoliubov transformation matrix elements by considering the ground state to be the vacuum for $\gamma_{\vec{k}}$ excitations (i.e., $\gamma_{\vec{k},\mu,\alpha} \ket{0} = 0$). These averages are then compared to the initial MF parameters to solve the self-consistency equations \eqref{eq: self-consistency equations for MF parameters}. The MF energy per unit cell is given by 
\begin{align}
    \frac{\expval{H_{MF}}}{N_{u.c.}} =& \frac{1}{2 N_{u.c.}} \sum_{i=1}^{2 N_{SL}} \sum_{\vec{k}\in BZ}  \omega_{i}(\vec{k}) \\
    &- N_{SL} \lambda (1+\kappa)+ \frac{E_0\left(\left\{A\right\}\right)}{N_{u.c.}},  \nonumber
\end{align}
and can equivalently be minimized with respect to all MF parameters to find the ones that solve the MF self-consistency equations. The MF energy reduces to
\begin{align}
    \expval{H_{MF}}=& - E_0\left(\left\{A\right\}\right)
\end{align}
after all self-consistency equations have been solved.

\section{\label{subsubsec: 0-flux Z2 PSG classes -> condensation pattern} Condensation pattern}

The possible spin ordering of the gapless $\mathbb{Z}_2\text{-}0\text{-}000$ and $\mathbb{Z}_2\text{-}0\text{-}010$ phase can be evaluated by considering the nullspace of the Hamiltonian at the critical momentum. When the spinon dispersion becomes gapless at a given momentum $\vec{k}_{c}$, the spinons condense at that point, thereby leading to a macroscopic occupation that can be treated as the order parameter. The condensation momentum corresponds to the ordering wavevector. To get the explicit spin ordering within one unit cell, the operator is replaced by the c-number $\langle\gamma_{\vec{k}_c}\rangle$, and the corresponding spin pattern is evaluated by using the initial definition of spin in the Schwinger boson representation \eqref{eq: Schwinger boson rep. for spins}. When a Bogoliubov transformation $P(\vec{k})$ satisfying both Eqs. \eqref{eq: diagonalization of Hamiltonian with Bogoliubov trfs} and \eqref{eq: constraint Bogoliubov trf from CCR} exists, this procedure simply amounts to taking an arbitrary linear combinations of the zero eigenmodes of $J\mathcal{H}(\vec{k}_c)$ \cite{sachdev1992kagome}.

Nonetheless, in many cases of interest, the algebraic multiplicity of the $\lambda_{\vec{k}_c}=0$ eigenvalue is larger than its geometric multiplicity, thereby implying that the system cannot be diagonalized by a canonical transformation. As rightfully noted by Liu et al. \cite{liu2019competing}, this problem can be resolved by decomposing our complex operator into two real fields 
\begin{equation}
    B_{\vec{k}} =\frac{1}{\sqrt{2}}\left( X_{\vec{k}} + i P_{\vec{k}} \right)    
\end{equation}
and finding a new basis
\begin{equation}
    \left(\begin{array}{c}
    \hat{X}_{\vec{k}} \\
    \hat{P}_{\vec{k}}
    \end{array}\right)=W(\vec{k})\left(\begin{array}{c}
    \hat{Y}_{\vec{k}} \\
    \hat{Q}_{\vec{k}}
    \end{array}\right)
\end{equation}
where the MF Hamiltonian is in a diagonal form 
\begin{align}
    H_{MF} = \sum_{i,\vec{k}\in BZ} \left( \alpha_{\vec{k},i} \hat{y}_{\vec{k},i}^2 + \beta_{\vec{k},i} \hat{q}_{\vec{k},i}^2  \right)
\end{align}
and the transformation $W(\vec{k})$ is symplectic (i.e., $W(\vec{k})(i\sigma^y\otimes \mathds{1}_{16\times16}) W^T(\vec{k})= (i\sigma^y\otimes \mathds{1}_{16\times16})$) to preserve canonical commutation relations. Such a procedure is equivalent to finding the generalized eigenvectors and rewriting our initial system in a Jordan normal form \cite{bronson1991matrix}. The eigenvalues in this new basis are related the initial ones by $\lambda_{\vec{k}, i}^{2}=\alpha_{\vec{k}, i} \beta_{\vec{k}, i}$. In this framework, the critical Hamiltonian is diagonalizable by a Bogoliubov transformation if all eigenvalues in new basis are zero ($\alpha_{\vec{k}_{c}, i}=\beta_{\vec{k}_{c}, i}=0$), and non-diagonalizable if $\alpha_{\vec{k}_{c}, i}=0, \beta_{\vec{k}_{c}, i} \neq 0$ or $\alpha_{\vec{k}_{c}, i} \neq 0$, $\beta_{\vec{k}_{c}, i}=0$.

Using this procedure and considering the corresponding the low effective field theory, it can be shown that the diagonalizability of a system is related to its dynamical critical exponent $\omega\propto |\vec{k}-\vec{k}_c|^z$: systems with linear dispersions  ($z=1$) are non-diagonalizable, whereas a canonical transformation can always be found in the $z=2$ case \cite{liu2019competing}. Looking at the dispersion relation of the $\mathbb{Z}_2\text{-}0\text{-}000$ and $\mathbb{Z}_2\text{-}0\text{-}010$ PSG classes in fig. \ref{fig: dispersion relation 0-flux Z2 states}, we see that our system is non-diagonalizable at the critical momentum for both. 

\begin{figure}[b!]
\includegraphics[width=0.75\linewidth]{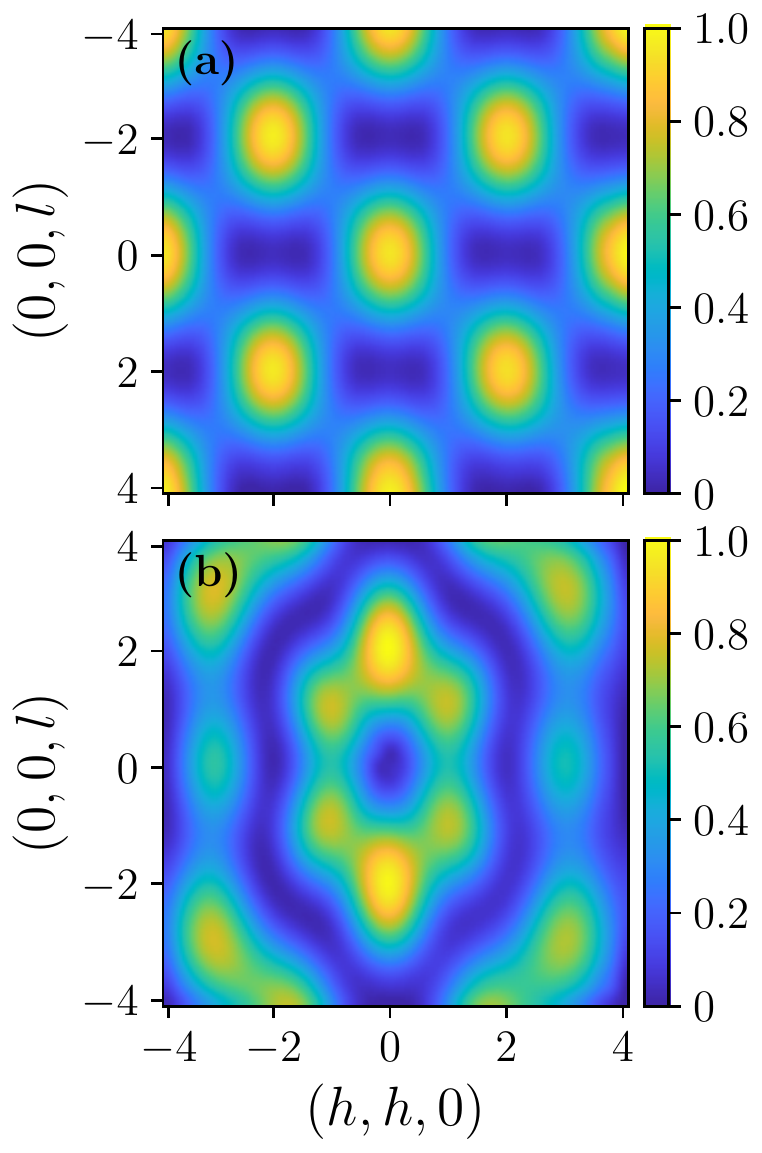}
\caption{ Separation of the two contributions to the total equal-time static neutron scattering amplitude for $\mathbb{Z}_2$-0-101 and $\mathbb{Z}_2$-0-110. (a) The top panel shows the contribution from $S_1^{NS}(\vec{q})$, and (b) the low panel the contribution from $S_2^{NS}(\vec{q})$ in eq. \ref{eq: splitting contribution INS}. \label{fig: single tetrahedron}}
\end{figure}

Specifically, for both states we find four zero eigenvalues, but the null-space of $J \mathcal{H}(\vec{k_c}=\vec{0})$ is only two dimensional and spanned by the time-reversal partners $\mathbf{a}$ and $U_{\mathcal{T}}\mathbf{a}^{*}$, where $U_{\mathcal{T}}=\mathds{1}_{8\times 8}\otimes (i\sigma^y)$. In the diagonal real basis $(\hat{y}, \hat{q})$, the critical Hamiltonian has two gapless modes and a low-energy form
\begin{equation}
    H=\hat{q}_{1}^{2}+\hat{q}_{2}^{2}+0 \cdot \hat{y}_{1}^{2}+0 \cdot \hat{y}_{2}^{2}. 
\end{equation}
The energy is minimized if the gaped modes are unoccupied $\left\langle\hat{q}_{i}\right\rangle=0$, or equivalently if $\hat{y}_{i}$ fluctuate maximally by the uncertainty principle. The spin order is then given by a real linear combination of the two generalized eigenvectors of rank two: $\langle B_{\vec{0}} \rangle = y_1 \mathbf{b} + y_2 U_{\mathcal{T}} \mathbf{b}^{*}$ with $y_{i}=\left\langle\hat{y}_{i}\right\rangle$. Explicitly, the spin pattern in a unit unit cell in terms of the 12-components vector $\mathbf{S}=(\vec{\tau}_0, \vec{\tau}_1, \vec{\tau}_2, \vec{\tau}_3)$ is 
\begin{align}
    \mathbf{S} &\propto \left( S_x, 0, S_z, S_x, 0, S_z, S_x, 0, S_z, S_x, 0, S_z \right)   \\
    &\begin{cases}
    S_x = \sin(\rho) \nonumber \\
    S_z = \cos(\rho)\nonumber
    \end{cases}
\end{align}
with $\rho$ being a function of $y_1$ and $y_2$. Thus, condensation at the $\Gamma$ point for the $\mathbb{Z}_2\text{-}0\text{-}000$ and $\mathbb{Z}_2\text{-}0\text{-}010$ PSG classes describes $X$ and $Z$ all-in-all-out order, which is consistent with classical calculations \cite{patri2020theory}.

\section{\label{appendix: comparison with single tetrahedron calculation} Comparison with the single tetrahedron calculation}

To further compare our results with the single tetrahedron calculation from Castelnovo and Moessner \cite{castelnovo2019rod}, we can split the equal-time neutron scattering amplitude into two different contributions
\begin{align} \label{eq: splitting contribution INS}
    &\widetilde{\mathcal{S}}^{NS}(\vec{q}) = S_1^{NS}(\vec{q}) + S_2^{NS}(\vec{q}) \\
    &\begin{cases}
    S_1^{NS}(\vec{q})  = \left\langle \vec{m}(\vec{q},0)\cdot \vec{m}(-\vec{q},0) \right\rangle   \nonumber \\
    S_2^{NS}(\vec{q})  = - \sum_{ab}  \frac{\vec{q}_{a} \vec{q}_{b}}{|\vec{q}|^2}  \left\langle m^{a}(\vec{q},0) m^{b}(-\vec{q},0) \right\rangle \nonumber
    \end{cases}
\end{align}
These two contributions are displayed in fig. \ref{fig: single tetrahedron}. Our results are indistinguishable from the single tetrahedron calculation (fig. 3 in \cite{castelnovo2019rod}). This confirms the correspondence between the equal-time NS amplitude of a single tetrahedron summed over all classical configurations but the all-in-all-out one, and our two $\mathbb{Z}_2$ QSLs ($\mathbb{Z}_2$-0-101 and $\mathbb{Z}_2$-0-110).

\end{document}